%% file: paper.tex
\documentclass[11pt,a4paper]{article}
\input{macros.tex}

\title{Tailored PDFs for New Physics searches}
\author[a]{Ella Cole}
\author[a]{, Mark N.~Costantini}
\author[b]{, Elie Hammou}
\author[c]{, Luca Mantani}
\author[d]{, \\Francesco Merlotti}
\author[e]{, Manuel Morales-Alvarado}
\author[a]{and Maria Ubiali}
\affiliation[a]{DAMTP, University of Cambridge, Wilberforce Road, Cambridge, CB3 0WA, United Kingdom}
\affiliation[b]{Nikhef Theory Group, Science Park 105, 1098 XG Amsterdam, The Netherlands}
\affiliation[c]{Instituto de F\'isica Corpuscular (IFIC), Universidad de Valencia-CSIC, E-46980 Valencia, Spain}
\affiliation[d]{Institute for Theoretical Physics, ETH, Wolfgang-Pauli-Strasse 27, CH-8093 Z\"{u}rich, Switzerland}
\affiliation[e]{INFN, Sezione di Trieste, SISSA, Via Bonomea 265, 34136, Trieste, Italy}
\emailAdd{ehammou@nikhef.nl}
\emailAdd{m.ubiali@damtp.cam.ac.uk}

\abstract{
Given the non-negligible interplay between parton distribution functions (PDFs) at large $x$ and potential New Physics (NP) effects 
in the high-energy tails of hadron collider observables, a central question is which PDFs can be reliably employed in beyond-the-Standard-Model (BSM) 
analyses. In this work, we examine the fine balance between using PDF sets with small uncertainties in the large-$x$ 
region -- crucial for maximising BSM sensitivity -- and adopting conservative PDF fits that exclude high-energy data potentially contaminated by unaccounted NP contributions.
We systematically assess a range of conservative PDF fitting strategies designed to mitigate such biases and provide a recommendation for 
the class of PDFs best suited for robust BSM searches. In addition, we investigate the alternative approach of performing 
simultaneous fits of Standard Model Effective Field Theory (SMEFT) Wilson coefficients and PDFs, thereby consistently accounting for their mutual correlations.
Starting from a toy model to illustrate the underlying mechanisms, we then analyse two realistic NP scenarios: one modifying high-mass 
Drell–Yan production and another affecting the high-invariant-mass tail of top-quark pair production. Both cases are representative of 
measurements that will be probed with high precision during the High-Luminosity phase of the LHC.
}

\arxivnumber{2602.20235}

\begin{document}

\maketitle
\flushbottom

\input{sections/sec1-intro}

\input{sections/sec2-toy_model}

\input{sections/sec3-simufit}

\input{sections/sec4-recommendations}

\input{sections/sec5-conclusion}
\section*{Acknowledgments}
We thank Fabio Maltoni and Michelangelo Mangano for useful discussions that inspired this work and for their comments that improved it. 
We are immensely grateful to James Moore, who gave a substantial contribution to shape the 
questions that we address here. 
E. H., M. N. C., and M. U.
are supported by the European Research Council under the European Union’s Horizon
2020 research and innovation Programme (PBSP, Grant agreement n.950246). M.~U.~ is partially 
supported by the STFC consolidated grant ST/X000664/1. E. H. is 
partially supported by the Swiss National Science Foundation.
L.~M. acknowledges support from the European Union under the MSCA fellowship (Grant agreement N. 101149078) Advancing global SMEFT fits in the LHC precision era (EFT4ward).
E.~C. is supported by Newnham College, Cambridge and by the STFC Centre for
Doctoral Training (CDT) in Data Intensive Science.
F.~M.~ is supported by Swiss National Science Foundation (SNF) under grant number 200021-231259.

\newpage

\appendix

\input{appendices/app2_new_physics_CT}
\input{appendices/app3_data.tex}

\input{appendices/app4_fit_quality}

\renewcommand{\em}{}
\bibliographystyle{UTPstyle}
\bibliography{references}

\end{document}

%% file: macros.tex
\usepackage{jheppub,multirow}
\usepackage{mathtools}
\usepackage{physics}
\usepackage{float}
\usepackage{graphicx}
\usepackage{url}
\usepackage{cancel}
\usepackage{slashed}
\usepackage{placeins}
\usepackage{feynman}
\usepackage{soul} 
\usepackage{bm}
\usepackage{amsmath}
\usepackage{amssymb}
\usepackage{listings}
\usepackage[toc, page]{appendix}
\usepackage{color}
\usepackage{booktabs} 
\usepackage{pgf}
\usepackage{empheq}
\usepackage{tikz}

\usepackage{longtable}

\usepackage{lipsum}
\usepackage{tabularx}
 \usetikzlibrary{decorations.pathmorphing}
  \usetikzlibrary{decorations.markings}
\tikzset{snake it/.style={decorate, decoration=snake}}
\usetikzlibrary{shapes.geometric, arrows}
\tikzstyle{process} = [rectangle, minimum width=3cm, minimum height=1cm, text centered, draw=black, fill=orange!30]
\tikzstyle{arrow} = [thick,->,>=stealth]
\usetikzlibrary{arrows.meta,bending}
\usepackage{subfigure}
\usepackage{pgf}
\usetikzlibrary{arrows}
\usetikzlibrary{patterns}
\renewcommand{\vec}[1]{\textbf{#1}}
 \usepackage{braket}
 \usepackage[inline]{enumitem} 
 
\usepackage{pifont}
\usepackage{array}
\newcolumntype{P}[1]{>{\centering\arraybackslash}p{#1}}
\usepackage{amsthm}
\usepackage{hyperref}
 \hypersetup{
    colorlinks=true,
    linkcolor=blue,
    filecolor=black,      
    urlcolor=blue,
}
\usepackage{tensor}
\usepackage{slashed} 

\def\G1{{\bf \gamma^{(1)}_N}}

\definecolor{violet}{cmyk}{0,1,0,0.2}

\usepackage[font=small,labelfont=bf,labelsep=period]{caption}
\usepackage{tabularx}
\newcolumntype{C}[1]{>{\centering\arraybackslash}p{#1}}

\newcommand{\be}{\begin{equation}}
\newcommand{\ee}{\end{equation}}
\newcommand{\bea}{\begin{eqnarray}}
\newcommand{\eea}{\end{eqnarray}}
\newcommand{\bi}{\begin{itemize}}
\newcommand{\ei}{\end{itemize}}
\newcommand{\ben}{\begin{enumerate}}
\newcommand{\een}{\end{enumerate}}

\usepackage{tikz}
\usetikzlibrary{arrows}
\usepackage{tikz-3dplot}

\def\G1{{\bf \gamma^{(1)}_N}}

\renewcommand{\vec}[1]{\textbf{#1}}

\newcommand{\simunet}{\texttt{SIMUnet}}

\newcommand{\smefit}{\texttt{SMEFiT}}

\numberwithin{equation}{section}
\numberwithin{figure}{section}
\numberwithin{table}{section}

%% file: sections/sec1-intro.tex
\section{Introduction}
\label{sec:intro}

Over the past decade, the role of global beyond-the-Standard-Model (BSM) fits has undergone a significant evolution. 
Early studies were primarily focused on forecasting the discovery reach of the LHC and identifying promising channels 
in which new physics (NP) signals might first emerge. However, following several null results from Run-I and Run-II, 
the emphasis has progressively shifted towards interpreting increasingly precise Standard Model (SM) measurements in the 
absence of clear deviations. In this context, global analyses have moved from predominantly parameter-estimation exercises 
within specific BSM frameworks to more comprehensive strategies aimed at model comparison, consistency tests, and the 
systematic assessment of correlated uncertainties~\cite{Morrissey:2009tf,GAMBIT:2025qto}. This shift has sharpened the 
need to understand subtle theoretical systematics—among them, the interplay between BSM effects and Parton Distribution 
Functions (PDFs)—which may critically impact the interpretation of high-energy collider data.

In recent years, such an interplay between PDFs and NP has been extensively 
investigated~\cite{Carrazza:2019sec,Greljo:2021kvv,CMS:2021yzl,Iranipour:2022iak,Gao:2022srd,Ball:2022qtp,McCullough:2022hzr,
Kassabov:2023hbm,Hammou:2023heg,Hammou:2024xuj,Shen:2024sci,Costantini:2024xae,Fiaschi:2022wgl,Fiaschi:2021sin,Fiaschi:2021okg}.
It was explicitly shown that, as experimental precision increases and more statistics are accumulated 
in the high-mass region at the HL-LHC, large-$x$ PDFs can mimic the effects of certain new physics scenarios. 
For example, a heavy SU(2)$_L$ triplet that couples universally to all quark flavours~\cite{Farina:2019UV-model} 
can indirectly alter the tails of Drell–Yan distributions used in PDF fits. These effects can be absorbed 
by the relatively unconstrained antiquark distributions at large $x$~\cite{Hammou:2023heg}. Similar issues might 
arise when considering new physics models that affect the large-$p_T$ tails of inclusive jet distributions or 
the large-$m_{jj}$ tails of the dijet distributions that might be absorbed by the poorly constrained large-$x$ gluon~\cite{prep1,prep2}.  
Looking ahead, data from future facilities such as the Electron-Ion Collider (EIC)~\cite{Aschenauer:2017jsk} and experiments from 
the Forward Physics Facilities at CERN~\cite{FASER:2022hcn} will provide complementary 
information to disentangle genuine new physics effects from large-$x$ PDF uncertainties~\cite{Hammou:2024xuj}, 
while lattice-QCD constraints are also expected to contribute in this direction~\cite{DelDebbio:2020rgv}. 

Given the importance of this phenomenological issue, several computational frameworks have been developed 
to systematically assess the interplay between PDFs and new physics by enabling simultaneous fits of PDFs 
and BSM parameters, often within the model-independent Standard Model Effective Field 
Theory (SMEFT) framework.
Among these, the tool {\tt SIMUnet} allows for the simultaneous determination of PDFs and an arbitrary 
number of SMEFT operators, including their linear effects on observables~\cite{Costantini:2024xae}. Similarly, 
a version of {\tt XFitter}~\cite{Shen:2024sci, Abdolmaleki:2023jvw,Accomando:2019vqt} supports joint fits of PDFs and SMEFT operators, although the 
number of implemented operators is currently limited. Both tools are tightly connected to specific PDF 
parametrisations, NNPDF4.0 in the case of {\tt SIMUnet}, and a polynomial-based form in the case of {\tt XFitter}.
The {\tt Colibri} framework~\cite{Costantini:2025agd,Costantini:2025wxp} provides an even more flexible environment, 
which in the near future will allow for a simultaneous determination of PDFs and SMEFT parameters for any PDF parametrisation.

Despite recent progress, several open questions remain in this field, two of which we aim to address in this work. 
\begin{itemize}
    \item[(i)] Given that the inclusion of data probing the high-energy tails of distributions in PDF fits might be 
problematic, as PDFs may be able to mimic and absorb the effects of certain new physics models, 
should we adopt a more conservative approach by excluding all data above a given energy scale, 
defined by a cut-off $Q_{\text{max}}^2$?
\item[(ii)] How to properly use the results of a joint PDF–SMEFT determination? Specifically, given the output of a simultaneous PDF–SMEFT fit, 
can the resulting “SMEFT PDF” be employed directly as input for precision collider predictions or indirect BSM searches? 
Must the assumptions underlying any subsequent analysis be consistent with those made in the joint PDF–SMEFT 
determination, or can such PDFs be used independently? Furthermore, could the difference between a SM–only PDF fit 
and a simultaneous PDF–SMEFT fit be interpreted as an estimate of the systematic uncertainty associated with 
standard PDF determinations? And can a difference in the fit quality be used to conclude that there is evidence of EFT effects?
\end{itemize}
Answering these questions will be of great value not only to those pursuing precision phenomenology at the LHC 
but also to the broader BSM community, including experimental collaborations engaged in direct and indirect 
searches for new physics. In this paper, we tackle both questions. We will show that, while conservative fits 
may appear to be the most robust and model-independent choice, they have several drawbacks, as they rely heavily on older experimental 
data and on the ability of PDF-fitting  frameworks to accurately handle extrapolations into poorly constrained regions. 

The structure of this work is as follows. In Section~\ref{sec:toy_model}, we present the main mathematical ideas of 
this work in the context of a simple toy model. Whilst this toy model is a gross simplification of the realistic 
phenomenological scenario we face, it presents a number of features which translate well to the numerical study 
we subsequently perform. In Section~\ref{sec:pheno}, we set the HL-LHC scenario that 
we aim to explore and define two BSM models that affect the Drell-Yan and the top sector respectively. 
We then  perform a thorough phenomenological study to probe the usefulness of the `conservative' PDF concept 
in the global fits of PDFs and SMEFT Wilson coefficients, and study the impact of using simultaneously-determined `SMEFT' PDFs to make precise theoretical predictions in searches
for New Physics. In Section~\ref{sec:recommendations} we identify several best-practice recommendations to the HEP community. 
Finally in Section~\ref{sec:conclusion} we summarise our findings and plans for future work.

%% file: sections/sec2-toy_model.tex
\section{A toy model for global PDF and SMEFT fits}
\label{sec:toy_model}

In this section, we introduce a toy model for global PDF and SMEFT fits which motivates the phenomenological 
study we perform in the sequel. Throughout this section we work with two data points, one which is used for PDF fitting (the PDF itself is considered to be a single real variable $f$), and one which is used for SMEFT fitting 
(we work with a single SMEFT Wilson coefficient $c$). This crude model is sufficient to demonstrate some of the 
key behaviour which is investigated in realistic cases later in this work and has the advantage of being analytically tractable to some extent.

In Sect.~\ref{subsec:presentation_of_model}, we present the toy model. Subsequently, in Sect.~\ref{subsec:analytic}, 
we derive analytic expressions for the posterior distributions of PDFs and Wilson coefficients in both separate and simultaneous fits. 
Finally, in Sect.~\ref{subsec:results}, we present and discuss the results of both types of fit in this toy model with a numerical example.

\input{sections/sec2_subsecs/subsec2_1-model.tex}
\input{sections/sec2_subsecs/subsec2_2-analytic.tex}
\input{sections/sec2_subsecs/subsec2_3-results.tex}

%% file: sections/sec2_subsecs/subsec2_1-model.tex
\subsection{Presentation of the model}
\label{subsec:presentation_of_model}

Assuming collinear factorization, we can -- very schematically -- model a theoretical prediction for a collider observable ${\cal O}$ 
that is sensitive to a given SMEFT Wilson coefficient $c$ and to a single parton channel as
\begin{equation}
    t(f,c) = s_{\rm SM}\,(f_{\rm ref} + f)\,(1+\,c \, s_{\rm lin}),
\label{eq:model}
\end{equation}
where $s_{\rm SM}$ represents the SM partonic cross section, $(f_{\rm ref} + \,f)$ represents the given parton luminosity channel that the observable is sensitive to and $s_{\rm lin}$ is the linear SMEFT $K$-factor accounting for the interference between the SM amplitude and the SMEFT amplitude. The parton luminosity entering Eq.~\eqref{eq:model} is decomposed into two contributions. The term $f_{\rm ref}$ represents a fixed reference luminosity taken from an input PDF set, while $f$ parametrises a shift of this luminosity induced by including the observable ${\cal O}$ in a PDF fit. Expanding Eq.~\eqref{eq:model} we obtain
\begin{align}
    t(f,c) &= s_{\rm SM} \, f_{\rm ref} + s_{\rm SM} \,f + s_{\rm SM} f_{\rm ref} s_{\rm lin} \, c + s_{\rm SM} s_{\rm lin} \,f\,c \notag\\
 &= \tilde{t}_0 + \tilde{t}_f \,f  + \tilde{t}_c\,c + \tilde{t}_i\,  f\,c.
\label{eq:model_exp}
\end{align}
The interpretation of the above expression is quite straightforward: the first term $\tilde{t}_0 = s_{\rm SM} f_{\rm ref}$ is the SM prediction for the observable ${\cal O}$ 
obtained with a fixed input PDF set; the second term $\tilde{t}_f\,f$, with $\tilde{t}_f = s_{\rm SM}$, is the part of the SM predictions that varies once the 
observable ${\cal O}$ -- sensitive on a particular luminosity channel -- is included in a PDF fit, and the coefficient $\tilde{t}_f$ can be associated to the 
sensitivity of the observable to the PDFs; the third term $\tilde{t}_c$ with $\tilde{t}_c = s_{\rm SM} f_{\rm ref} s_{\rm lin}$ is the linear SMEFT contribution suppressed 
by two powers of the heavy new physics scale $\Lambda^2$, which varies with $c$ once the observable ${\cal O}$ is included in a SMEFT fit, 
and the coefficient $\tilde{t}_c$ can be associated to the sensitivity of the observable to SMEFT corrections; the last term $\tilde{t}_i\,f\,c$, 
with $\tilde{t}_i = s_{\rm SM} s_{\rm lin}$, captures the interplay between PDF and SMEFT effects. It accounts for the fact that the SMEFT contribution 
itself is modulated by shifts in the parton luminosity, and therefore encodes the strength of the correlation between PDF determination and 
SMEFT inference. This term is typically neglected in pure PDF fits (in which $c=0$) and in pure SMEFT fits (where PDFs are held fixed to some $f_{\rm ref}$, 
hence $f=0$). 

To keep the model analytically tractable we exploit the fact that $\tilde{t}_i$ is typically much smaller than $\tilde{t}_f$ and $\tilde{t}_c$, 
as $\tilde{t}_i$ is doubly suppressed, both by the $\Lambda^2$ in the denominator of $s_{\rm lin}$ and by the size of the change in the PDFs associated 
with $f$, which is much smaller than the PDFs themselves\footnote{The inclusion of a new dataset in a PDF fit typically shifts the PDFs by no more than 10\%  
compared to the value of the PDFs before the inclusion of such dataset.}. We can therefore expand Eq.~\eqref{eq:model_exp} around $(f_0,c_0)$, which are 
the maximum likelihood estimators (MLE) of the theory $t(f,c)$ when $\tilde{t}_i$ is set to 0 -- which we will henceforth refer to as {\it linear} theory.
Consider the first-order Taylor expansion of the full theory Eq.~\eqref{eq:model_exp}, 
\begin{align}
    t(f,c) &\simeq 
    t(f_{0},c_{0})
    + \left.\frac{\partial t}{\partial f}\right|_{(f_{0},c_{0})} (f-f_{0})
    + \left.\frac{\partial t}{\partial c}\right|_{(f_{0},c_{0})} (c-c_{0}).
\end{align}
Given that $\partial t/\partial f\,= \tilde{t}_{f} + \tilde{t}_i \,c$ and $\partial t/\partial c = \tilde{t}_c + \tilde{t}_i \,f$ 
we can define the {\it effective linear sensitivities} as the local slopes of the full model around $(f_{0},c_{0})$,
\begin{align}
    t_{f} &\equiv \tilde{t}_f + \tilde{t}_i c_0, \notag\\
    t_{c} &\equiv \tilde{t}_c + \tilde{t}_i f_0,  \label{eq:effsens}
\end{align}
and the linearised expression becomes
\begin{equation}
    t(f,c) \simeq t_0 + t_f\, f     + t_c\, c ,
    \label{eq:linmodel}
\end{equation}
where the constant term $\tilde{t}_0$ of Eq.~\eqref{eq:model_exp} is shifted to $t_0$ by a term that is suppressed by $\tilde{t}_i$.
The expansion of Eq.~\eqref{eq:linmodel} shows that, in the small-$\tilde{t}_i$ limit, the non-linear PDF--EFT interplay term proportional to $\tilde{t}_i$ in 
the full theory~\eqref{eq:model_exp} can be interpreted as a redefinition of the linear
sensitivities around the MLE of a theory in which the interaction term proportional to $\tilde{t}_i$ is completely neglected. This parametrisation has the advantage that the toy model becomes analytically tractable, giving us some simple parametric understanding of what happens in a PDF-SMEFT fit\footnote{In a realistic fits, however, we account for the interaction term, as implemented in the {\tt SIMUnet} methodology.}. 

We now consider two observables and their respective measurements, that are collected into a vector of two data points $\vec{d} = (d_1, d_2)^T$, which we assume 
are distributed according to a multivariate normal distribution centred around our theoretical model 
\begin{equation}
    \vec{d} = \begin{pmatrix} d_1 \\ d_2 \end{pmatrix} \sim \mathcal{N} (\vec{t}(f,c), \Sigma_{\text{exp}} ),
\end{equation}
where $\vec{t} : \mathbb{R}^2 \rightarrow \mathbb{R}^2$ is the linear theory prediction given by Eq.~\eqref{eq:linmodel}, where now the effective $\mathbf{t_f}$ 
and $\mathbf{t_c}$ sensitivities are a 2-dimensional vector, hence the boldface, each component being defined as in Eq.~\eqref{eq:effsens}.

The experimental covariance matrix $\Sigma_{\text{exp}}$ is a $2 \times 2$ matrix, which for simplicity does not contain correlations between the data points:
\begin{equation}
\Sigma_{\text{exp}} = \begin{pmatrix} \sigma_1^2 & 0 \\ 0 & \sigma_2^2 \end{pmatrix}.
\end{equation}

\noindent The dataset $\vec{d} = (d_1, d_2)^T$ is intended to be a gross simplification of a global particle physics dataset. In particular, this dataset can be split into two sets:

\begin{itemize}[label=-]
\item The first data point $d_1$ can be regarded as the `\textit{PDF fitting dataset}'; that is, it contains the data which is usually used to infer SM PDFs. It may be interpreted as a low energy measurement in $Q^2$, where the effects of heavy new physics are very small or negligible.
\item The second data point $d_2$ can be regarded as the `\textit{SMEFT fitting dataset}'; that is, it contains the data which is usually used by SMEFT fitting collaborations 
to infer the SMEFT Wilson coefficients. It may be interpreted as a high energy measurement in $Q^2$.
\end{itemize}

\noindent There are two main modes of fitting, given this categorisation of the data. 
\begin{enumerate}[label = (\arabic*)]
\item \textsc{Separate fit}. The PDF parameter and the Wilson coefficient are determined in two distinct steps, 
without ever maximising a joint likelihood in $(f,c)$, namely $f$ is inferred under the assumption that the Wilson 
coefficient $c$ vanishes, and the Wilson coefficient $c$ is determined from the relevant high energy data, either 
by fixing the PDF parameter $f$ to the obtained best-fit value or by marginalising over its distribution 
(the latter accounts for PDF uncertainty, as is often done in SMEFT fits). Depending on which data are used in the PDF determination, 
this strategy can take two forms.

If the PDF is extracted exclusively from data whose theory predictions depend only very weakly on the Wilson coefficient 
(e.g. low energy data), the procedure is referred to as a \textit{conservative} fit. 
For example, the \smefit{} collaboration~\cite{Giani:2023gfq, terHoeve:2023pvs, Celada:2024mcf,terHoeve:2025gey,terHoeve:2025omu} uses PDF sets that exclude top data when determining 
top-sector SMEFT Wilson coefficients~\cite{Ethier:2021bye}, thereby avoiding overlap between the dataset entering the PDF fit and the dataset entering the SMEFT fit.

On the other hand, if SMEFT-sensitive data is included in the PDF determination while setting $c=0$, then 
potential BSM effects can be absorbed into the PDF parametrisation. 
In this case, the resulting fit is referred to as \textit{BSM-biased}.

\item \textsc{Simultaneous fit}. The PDFs and Wilson coefficient are jointly determined using both the first 
and second data points and including both linear terms in the theory prediction $\mathbf{t}$. This represents the 
type of fit that has been explored in several works, including \cite{Iranipour:2022iak, Kassabov:2023hbm, Gao:2022srd, Shen:2024uop}, and we dub it as \textit{simultaneous} fit.
\end{enumerate}

\noindent In the next section we explore both these options and compare and contrast the results that we obtain in the simple model that we have defined here.

%% file: sections/sec2_subsecs/subsec2_2-analytic.tex
\subsection{Analytic derivations}
\label{subsec:analytic}

In this subsection we discuss the analytic structure underlying the
separate and simultaneous fitting strategies introduced above.  The
goal is to explore the origin of the PDF-SMEFT interplay in a simple Gaussian model that is analytically tractable. 
Throughout, we work in the linearised theory defined in
Eq.~\eqref{eq:linmodel} and consider the measurements described in the previous subsection, 
in the specific case in which $d_1$ is a low energy measurement that is weakly sensitive to the SMEFT coefficient $c$, while $d_2$ is a high energy measurement that 
is highly sensitive to $c$\footnote{At sufficiently low energy, EFT effects are typically negligible. Here we retain a small but non-zero sensitivity in $d_1$ to explore the differences between the separate and simultaneous fitting methodologies.}. Both measurements are sensitive to the PDFs in the same 
parton channel and in the same large $x-$ region of the PDFs. In what follows, we discuss the separate conservative fit, separate BSM-biased fit, and simultaneous fit. 

\paragraph{Separate conservative fit.}

In the conservative strategy, the PDF parameter $f$ is determined solely from
the low-energy observable $d_1$, assuming $c=0$.  The corresponding
likelihood is
\begin{equation*}
\label{eq:d1_likelihood}
    \mathcal{L}_1(d_1 \,|\, f)
    \propto
    \exp\!\left[
        -\frac{1}{2}
        \frac{\bigl(d_1 - t_{0,1} - t_{f,1}f\bigr)^2}{\sigma_1^2}
    \right],
\end{equation*}
where $t_{0,1}$ is the first component of the $\mathbf{t}_0$ vector, and so on. 
Maximisation yields the best-fit value $\hat f$ and its variance $\sigma_f^2$ such that

\begin{equation}
    \hat f_{\rm cons} = \frac{d_1 - t_{0,1}}{t_{f,1}},
    \qquad
    \sigma_{f_{\rm cons}}^2 = \frac{\sigma_1^2}{(t_{f,1})^2}.
\label{eq:cons_unc}
\end{equation}
Using this PDF determination, the SMEFT coefficient $c$ is extracted
from the high energy observable $d_2$, yielding a best-fit value of
\begin{equation*}
    \hat c_{\rm cons} = \frac{d_2 - t_{0,2} - t_{f,2} \hat f_{\rm cons}}{t_{c,2}}.
\label{eq:cons_c}
\end{equation*}
If the PDF uncertainty is not
propagated, the resulting variance is
\begin{equation}
\label{eq:cons_unc_c_nopdf}
    \sigma_{c_{\rm cons}, \ \text{no\,PDF}}^2
    = \frac{\sigma_2^2}{(t_{c,2})^2}.
\end{equation}
A more conservative treatment marginalises over the PDF uncertainty obtained from $d_1$, which simply increases the variance entering the $\chi^2$.  The resulting uncertainty on the SMEFT coefficient becomes
\begin{equation}
\label{eq:cons_unc_c_pdf}
    \sigma_{c_{\rm cons},\ \text{with PDF}}^2
    =
    \frac{\sigma_2^2 + t_{f,2}^2 \sigma_{f_\text{cons}}^2}{t_{c,2}^2} = \sigma_{c_{\rm cons},\ \text{no PDF}}^2 \left( 1 + t_{f,2}^2 \frac{\sigma_{f_\text{cons}}^2}{\sigma_2^2} \right),
\end{equation}
where we have used Eq. (\ref{eq:cons_unc_c_nopdf}) and factorised a degradation coefficient that is related to the PDF uncertainty. Thus, in the conservative fit, failing to propagate PDF uncertainty can lead to an overestimate of the precision in $c$. The impact is controlled by the combination $t_{f,2}^2 \sigma_{f_\text{cons}}^2$. If $d_2$ is only weakly sensitive to the PDFs or if the PDF is very precisely determined, the effect is negligible. Conversely, when the high energy observable has strong PDF sensitivity and the PDF uncertainty is sizeable, the constraint on $c$ is correspondingly degraded.

\paragraph{Separate BSM-biased fit.}

In the BSM-biased strategy, the PDF parameter $f$ is extracted from both $d_1$ and $d_2$ under the SM assumption $c=0$. In this case, the PDF best-fit value is given by
\begin{equation}
\label{eq:sep_bias_f}
    \hat f_{\rm bias}
    =
    \frac{ t_{f,1} (d_1 - t_{0,1})/\sigma_1^2
          +t_{f,2} (d_2 - t_{0,2})/\sigma_2^2 }{D_f},
    \qquad
    D_f \equiv
    \frac{t_{f,1}^2}{\sigma_1^2}
    +
    \frac{t_{f,2}^2}{\sigma_2^2},
\end{equation}
with a corresponding variance
\begin{equation*}
\label{eq:bias_unc}
    \sigma_{f_{\rm bias}}^2 = \frac{1}{D_f}.
\end{equation*}
Notably, since $D_f > \left( t_{f,1} / \sigma_1 \right)^2$, we find that 
\begin{equation}
    \label{eq:sep_f_unc}
    \sigma_{f_{\rm bias}}^2 <  \sigma_{f_{\rm cons}}^2,
\end{equation}
where we have used the variance of Eq. (\ref{eq:cons_unc}). In this way, the PDF uncertainty is reduced with respect to the conservative case. Notice that if the true underlying SMEFT coefficient satisfies $c_{\rm true}\neq 0$, this procedure induces a systematic bias in the extracted PDF,
\begin{equation}
\label{eq:biased_shift_f}
    \mathrm{bias}(\hat f_{\rm bias})
    \equiv
    \mathbb{E}[\hat f_{\rm bias}] - f_{\rm true}
    =
    c_{\rm true}\,
    \frac{
      t_{f,1} t_{c,1}/\sigma_1^2 + t_{f,2} t_{c,2}/\sigma_2^2
    }{
      t_{f,1}^2/\sigma_1^2 + t_{f,2}^2/\sigma_2^2
    },
\end{equation}
which is, intuitively, proportional to the true underlying SMEFT coefficient $c_{\rm true}$. This bias also increases with the SMEFT sensitivity of the datapoints so, even if the low-energy observable is only weakly sensitive (or not sensitive at all) to the SMEFT coefficient, the high energy measurement (where SMEFT effects are usually enhanced) induces a non-vanishing bias.

Fitting $c$ solely from $d_2$ using the biased PDF result leads to
\begin{equation}
\label{eq:sep_bias_c}
    \hat c_{\rm bias}
    =
    \frac{d_2 - t_{0,2} - t_{f,2} \hat f_{\rm bias}}{t_{c,2}},
\end{equation}
with a propagated bias from the PDF sector which translates into a SMEFT bias of
\begin{equation}
\label{eq:biased_shift_c}
    \mathrm{bias}(\hat c_{\rm bias})
    \equiv
    \mathbb{E}[\hat c_{\rm bias}] - c_{\rm true}
    =
    -\,\frac{t_{f,2}}{t_{c,2}}\,\mathrm{bias}(f),
\end{equation}
where we see that the bias in $c$ is therefore directly proportional in size to the bias in $f$,
with a coefficient controlled by the relative sensitivities $t_{f, 2}$ and
$t_{c,2}$ at high energy. Intuitively, an overestimate (underestimate) of
the PDF contribution at high energy is compensated by an underestimate
(overestimate) of the EFT contribution. Notice that by propagating PDF uncertainty the SMEFT uncertainty becomes
\begin{equation}
\label{eq:bias_unc_c_pdf}
    \sigma_{c_{\rm bias},\ \text{with PDF}}^2
    =
    \frac{\sigma_2^2 + t_{f,2}^2\, \sigma_{f,{\rm bias}}^2}{t_{c,2}^2},
\end{equation}
where comparison with Eq. (\ref{eq:cons_unc_c_nopdf}) yields
\begin{equation}
    \label{eq:sep_c_unc}
    \sigma_{c_{\rm bias}}^2 <  \sigma_{c_{\rm cons}}^2,
\end{equation}
where we see that the BSM-biased fit therefore produces both smaller quoted
uncertainties and a systematic bias whenever $c_{\rm true}\neq 0$.

\paragraph{Simultaneous fit.}

In the simultaneous strategy, the PDF parameter $f$ and the EFT
coefficient $c$ are determined jointly from the two measurements
$(d_1,d_2)$. In this case, it is convenient to write the problem in matrix form.  Defining the data vector with constant pieces subtracted,
\begin{equation*}
    \mathbf{d}
    \equiv
    \begin{pmatrix}
        d_1 - t_{0,1} \\[2pt]
        d_2 - t_{0,2}
    \end{pmatrix},
\end{equation*}
the parameter vector
\begin{equation*}
    \boldsymbol{\theta}
    \equiv
    \begin{pmatrix}
        f \\[2pt]
        c
    \end{pmatrix},
\end{equation*}
and the $2\times2$ feature (design) matrix
\begin{equation*}
    X
    \equiv
    \begin{pmatrix}
        t_{f,1} & t_{c,1} \\[2pt]
        t_{f,2} & t_{c,2}
    \end{pmatrix},
\end{equation*}
the Gaussian log-likelihood is equivalent to the weighted
least-squares objective
\begin{equation}
\label{eq:chi2_sim}
    \chi^2(\boldsymbol{\theta})
    =
    (\mathbf{d} - X\boldsymbol{\theta})^{T}
    V^{-1}
    (\mathbf{d} - X\boldsymbol{\theta}),
\end{equation}
with $V=\mathrm{diag}(\sigma_1^2,\sigma_2^2)$. The maximum-likelihood estimator admits (assuming basic invertibility conditions) the closed-form solution
\begin{equation}
\label{eq:theta_hat_sim}
    \hat{\boldsymbol{\theta}}
    =
    (X^{T}V^{-1}X)^{-1}\,X^{T}V^{-1}\mathbf{d}.
\end{equation}
Writing
$X^{T}V^{-1}X=\begin{pmatrix}A&B\\B&C\end{pmatrix}$ with
\begin{align}
    A &\equiv \frac{t_{f,1}^2}{\sigma_1^2} + \frac{t_{f,2}^2}{\sigma_2^2},\\
    B &\equiv \frac{t_{f,1}t_{c,1}}{\sigma_1^2} + \frac{t_{f,2} t_{c,2}}{\sigma_2^2},\\
    C &\equiv \frac{t_{c,1}^2}{\sigma_1^2} + \frac{t_{c,2}^2}{\sigma_2^2},
\end{align}
and $\Delta\equiv AC-B^2>0$, the individual estimators are
\begin{align}
    \hat f_{\rm sim} &= \frac{C\,u - B\,v}{\Delta}, &
    \hat c_{\rm sim} &= \frac{A\,v - B\,u}{\Delta},
\end{align}
where $u$ and $v$ are the corresponding weighted data combinations given by
\begin{align*}
    u &\equiv \frac{t_{f,1} (d_1 - t_{0,1})}{\sigma_1^2} + \frac{t_{f,2} (d_2 - t_{0,2})}{\sigma_2^2},\\
    v &\equiv \frac{t_{c,1} (d_1 - t_{0,1})}{\sigma_1^2} + \frac{t_{c,2} (d_2 - t_{0,2})}{\sigma_2^2}.
\end{align*}
Note that the covariance matrix of the estimator is
\begin{equation}
\label{eq:cov_sim}
    \mathrm{Cov}_{\rm sim}(f,c)
    =
    (X^{T}V^{-1}X)^{-1}
    =
    \frac{1}{\Delta}
    \begin{pmatrix}
        C & -B\\
        -B & A
    \end{pmatrix},
\end{equation}
implying correlated uncertainties for $f$ and $c$, \textit{even} if $d_1$ and $d_2$ are uncorrelated (or if their correlations are neglected, in a real dataset). Also (although trivially) by construction the MLE values in the simultaneous fit are unbiased. The data have been generated according to the same underlying model used in the simultaneous fit, 
as in a closure test setting in the PDF nomenclature.

%% file: sections/sec2_subsecs/subsec2_3-results.tex
\subsection{Results}
\label{subsec:results}

In this subsection we illustrate the analytic results derived above by
applying the three fitting strategies to a concrete toy example.  The
purpose is to visualise the impact of PDF-EFT interplay on both central
values and uncertainties, and to contrast conservative, BSM-biased, and
simultaneous fits in a controlled setting.

We assume that the pseudodata are generated from the linearised model, 
Eq.~\ref{eq:linmodel}, with true underlying parameters
\begin{equation}
    f_{\rm true} = 10.0,
    \qquad
    c_{\rm true} = 0.1 .
\end{equation}
The theory coefficients are fixed to
\begin{equation}
    \mathbf{t}_0 = (0.1,\,0.1),
    \qquad
    \mathbf{t}_f = (0.1,\,0.1),
    \qquad
    \mathbf{t}_c = (0.1,\,1.0),
\end{equation}
so that the first observable $d_1$ is dominantly sensitive to the PDF
parameter $f$, while the second observable $d_2$ exhibits enhanced
sensitivity to the EFT coefficient $c$.  Experimental uncertainties
are taken to be uncorrelated and proportional to the true central
values, with a relative size of $1\%$.  For clarity, the pseudodata
are set equal to their expectation values, so that all differences
between the fits arise solely from modelling assumptions rather than
statistical fluctuations.

\begin{figure}[ht!]
    \centering
    \includegraphics[width=0.49\linewidth]{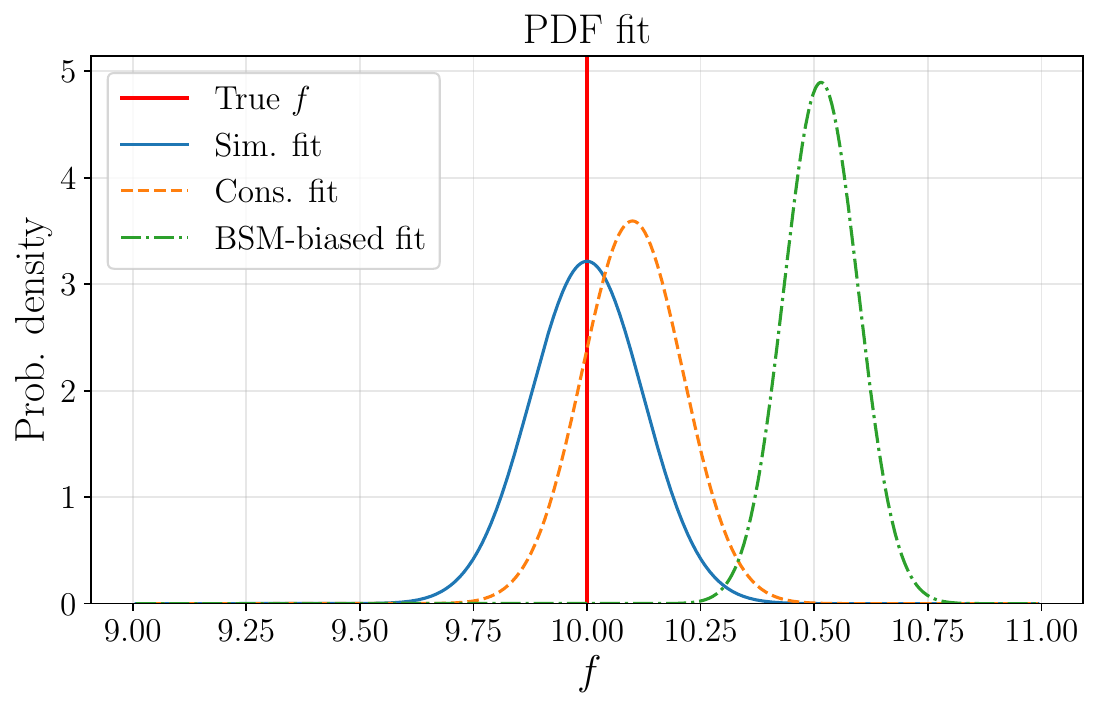}
    \includegraphics[width=0.49\linewidth]{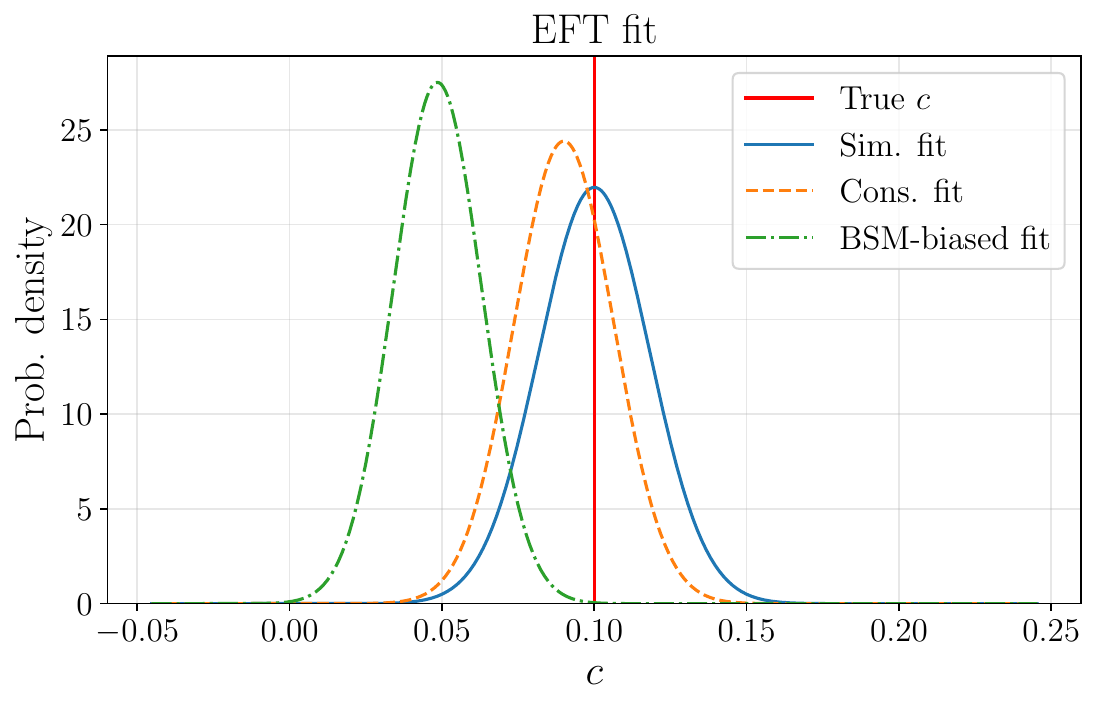}
    \caption{
    One--dimensional posterior distributions for the PDF parameter $f$
    (left) and the EFT coefficient $c$ (right) obtained in the toy model.
    The red vertical lines indicate the true values
    $(f_{\rm true},c_{\rm true})$.  The conservative fit is unbiased but
    exhibits inflated uncertainties, the BSM-biased fit yields
    artificially precise yet shifted posteriors, and the simultaneous
    fit correctly recovers both parameters in this closure--test
    configuration.
    }
    \label{fig:toy_comparison}
\end{figure}

Figure~\ref{fig:toy_comparison} compares the one-dimensional posterior
distributions for the PDF parameter $f$ (left) and the EFT coefficient
$c$ (right) obtained using the three fitting strategies discussed in
Sec.~\ref{subsec:analytic}.  We begin with the \emph{simultaneous fit},
which serves as the reference or ``golden standard''.  In this case,
$f$ and $c$ are inferred jointly, fully exploiting the information
content of both observables (and their correlations in a more realistic dataset, as we will see later in the text).  As expected from
the analytic results, the simultaneous fit yields unbiased estimates
for both parameters.

The \emph{conservative fit} provides the next best alternative.  Here,
the PDF parameter $f$ is determined using only the low energy observable
$d_1$, and the resulting PDF uncertainty is consistently propagated
into the determination of $c$ from the high energy observable $d_2$.
This strategy sacrifices sensitivity relative to the simultaneous fit,
leading to broader posteriors for both $f$ and $c$, but it remains
robust: the extracted values are close to the true parameters
$(f_{\rm true},c_{\rm true})$.  The loss
of precision can be directly traced to the deliberate exclusion of
$d_2$ from the PDF determination.

Finally, the \emph{BSM-biased fit} illustrates the failure mode
identified analytically in Sec.~\ref{subsec:analytic}.  In this case,
both $d_1$ and $d_2$ are used to constrain the PDF under the incorrect
assumption $c=0$.  Because the high energy observable $d_2$ has strong
sensitivity to the EFT coefficient, its constraining power artificially
reduces the PDF uncertainty, producing a posterior for $f$ that is both
shifted away from $f_{\rm true}$ and spuriously narrow.  This
overconstrained and biased PDF determination is then propagated into
the EFT fit, yielding a posterior for $c$ that is likewise displaced
from $c_{\rm true}$ (in the opposite direction) and exhibits a smaller uncertainty than in the conservative case.  The result is an incorrect determination: the apparent gain in precision is entirely spurious and arises from absorbing genuine EFT effects into the PDF fit.

These toy model results provide a controlled illustration of how PDF-EFT interplay, bias, and uncertainty propagation arise in separate and simultaneous fits; in the next section, we extend this analysis to realistic new physics scenarios in the context of the HL-LHC.

%% file: sections/sec3-simufit.tex
\section{Disentangling new physics signals and PDFs at the HL-LHC}
\label{sec:pheno}

In this section, we start by presenting the methodological setup adopted in this analysis, 
including the simultaneous closure-test framework and the full set of data and HL-LHC projections considered (Sect.~\ref{subsec:settings}). 
We then examine two benchmark BSM scenarios separately: one primarily affecting the quark sector at large~\(x\) via Drell-Yan high-mass measurements 
(Sect.~\ref{subsec:DY}) and one impacting the gluon sector at large~\(x\) via measurements of top pair production at large invariant mass (Sect.~\ref{subsec:top}). 
For each scenario, we demonstrate how the corresponding BSM signal can be partially absorbed into the PDFs, in the quark and gluon sectors respectively,
thereby introducing a source of bias. We then perform a comparative assessment of the extent to which the use of conservative PDFs, or alternatively a 
simultaneous fit of PDFs and SMEFT coefficients mitigates the resulting BSM-induced bias in the extracted constraints.

\input{sections/sec3_subsecs/subsec1_settings.tex}
\input{sections/sec3_subsecs/subsec2_wmodel.tex}
\input{sections/sec3_subsecs/subsec3_zmodel.tex}

%% file: sections/sec3_subsecs/subsec1_settings.tex
\subsection{Analysis settings}
\label{subsec:settings}

Our analysis is based on the \textit{simultaneous closure test} framework~\cite{Iranipour:2022iak, Costantini:2024xae}, 
which extends the PDF closure-test methodology introduced in Refs.~\cite{NNPDF:2014otw, DelDebbio:2021whr, Barontini:2025lnl}. 
The framework consists of a three-step procedure:
\begin{itemize}
\item[(i)] In the first step, a PDF set $\mathbf{f}_{\rm true}$ is selected and identified as the \emph{true} underlying PDFs, 
together with a BSM model defining the \emph{true} law of nature. 
When applicable, BSM effects are parametrised in terms of a set of SMEFT Wilson coefficients $\mathbf{c}_{\rm true}$; 
synthetic data, referred to as Level-0 ($L_0$) data, are then generated according to
\[
  \mathbf{D}^{0} 
    = T[\mathbf{f}_{\rm true},\mathbf{c}_{\rm true}] 
    = \hat{\boldsymbol{\sigma}}_{\rm NNLO}^{\rm SM} \otimes \mathcal{L}_{\rm true}
      \left[1+\sum_i c_i^{\rm true} R_i\right],
    \label{eq:L0data}
\]
where $\hat{\boldsymbol{\sigma}}_{\rm NNLO}^{\rm SM}$ denotes the vector of partonic cross sections computed at NNLO in perturbative QCD within the SM, 
and $\mathcal{L}_{\rm true}$ are the parton luminosities constructed from the {\it true} PDFs. We remind the reader that the integrated luminosity for the parton pair $i,j$ is defined as:
\begin{equation}
\label{eq:integrated_luminosity}
  {\cal L}_{ij}(m_X,\sqrt{s}) = \frac{1}{s}\int\limits_{-y}^{y}\,d\tilde{y}\,
  \left[
  f_{i}\left(\frac{m_X}{\sqrt{s}}e^{\tilde{y}},m_X\right)\, 
  f_j\left(\frac{m_X}{\sqrt{s}}e^{-\tilde{y}},m_X\right)
  +
  (i \leftrightarrow j)
  \right]
  ,
  \end{equation}
 where $f_i \equiv f_i(x,Q)$ is the PDF corresponding to the parton flavour $i$ 
 , and the integration limits are defined by:
 \begin{equation}
 \label{eq:rapidity}
  y=\ln\left(\frac{\sqrt{s}}{m_X}\right).
  \end{equation}
In particular, we focus on the luminosities that are most constrained by the NC and CC Drell-Yan data respectively, namely   
  \begin{align}
    {\cal L}^{\rm NC}(m_X,\sqrt{s}) &= {\cal L}_{u\bar{u}}(m_X,\sqrt{s}) + {\cal L}_{d\bar{d}}(m_X,\sqrt{s}),\label{eq:zlumi}\\[1.5ex]
      {\cal L}^{\rm CC}(m_X,\sqrt{s}) &= {\cal L}_{u\bar{d}}(m_X,\sqrt{s}) + {\cal L}_{d\bar{u}}(m_X,\sqrt{s}).\label{eq:wlumi}
    \end{align}

The term in brackets accounts for BSM contributions through the corresponding linear SMEFT $K$-factors $R_i$, 
computed with the {\it true} PDFs; the synthetic observables $\mathbf{D}^{0}$ are assigned uncertainties 
and correlations according to the experimental covariance matrices of the measurements or projections included in the fit.
\item[(ii)]
In the second step, a simultaneous fit to the synthetic data is performed, in which both the PDFs and 
the Wilson coefficients parametrising the BSM model are treated as free parameters and determined concurrently.
\item [(iii)]
In the third step, the fitted PDFs and Wilson coefficients are compared to the corresponding \emph{true} PDF and {\it true} 
SMEFT coefficient values used as input in the generation of the synthetic data. The simultaneous closure test is 
deemed successful if the fitted results are statistically compatible with the underlying law of nature within the quoted uncertainties.
\end{itemize}
%
%
Simultaneous closure tests probe the ability of a joint PDF--SMEFT fitting procedure to disentangle distortions induced by potential BSM effects 
from genuine PDF features. This framework is particularly well suited to isolating the interplay between PDFs and BSM contributions. 
By construction, the synthetic data are internally consistent and their theoretical description is exact. As a result, additional 
effects arising from experimental tensions between datasets~\cite{Barontini:2025lnl} or from missing higher-order corrections in the SM 
predictions~\cite{NNPDF:2024nan,NNPDF:2024dpb} are absent.

\begin{figure}[ht!]
    \centering
    \includegraphics[width=\textwidth]{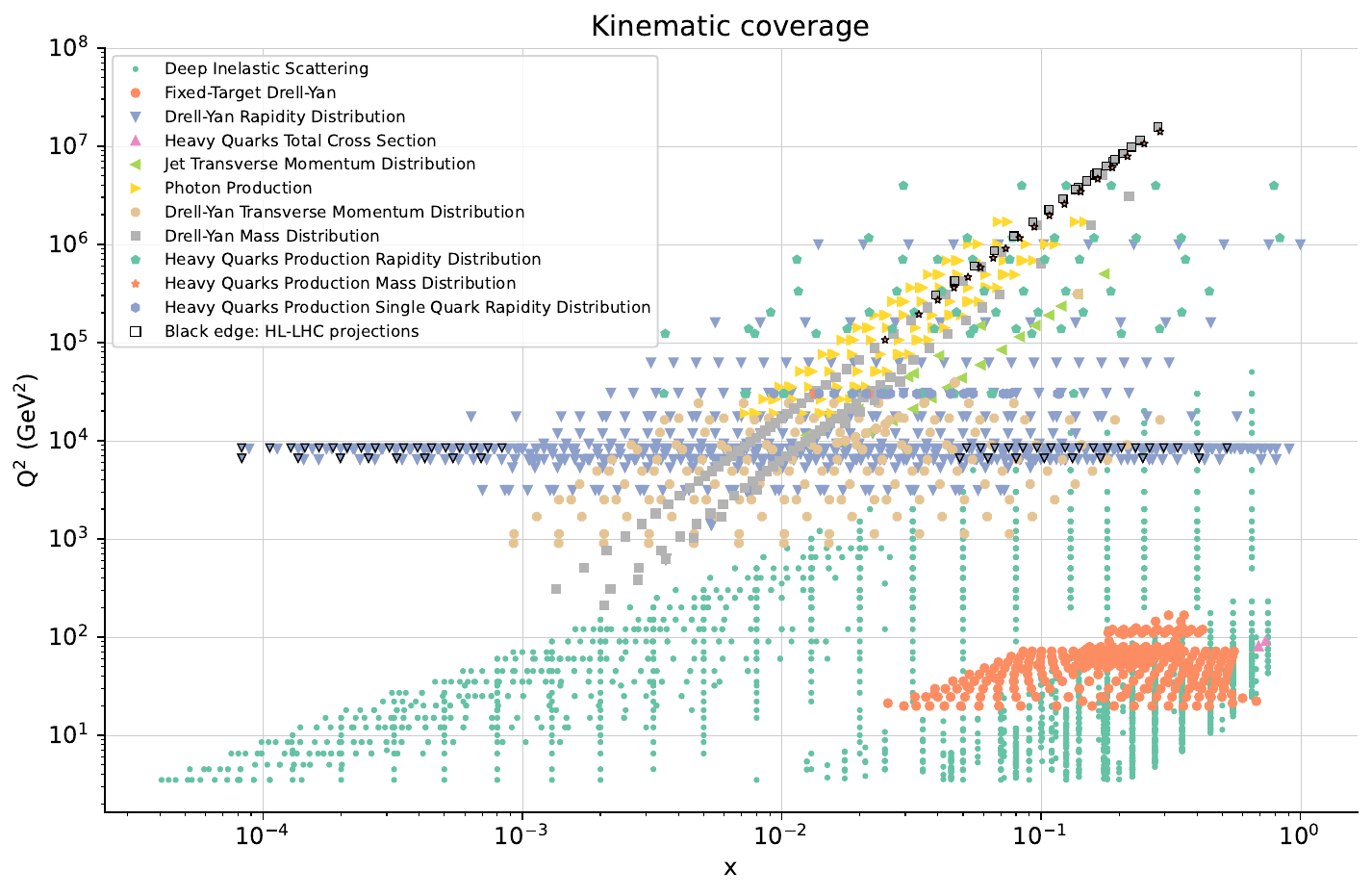}
    \caption{Kinematic coverage of the data points included in the 
    PDF fit. The points highlighted with a black
    edge are our HL-LHC projections, while the other data points correspond to existing measurements.
    The values of $x$ have been computed using a leading order approximation.}
    \label{fig:xq2}
  \end{figure}

From an experimental perspective, we generate a synthetic dataset comprising a total of 4363 data points, covering a broad range of processes. 
The $L_0$ data constructed according to Eq.~\eqref{eq:L0data} are drawn from two sources:
\begin{enumerate}
  \item Observables currently included in the NNPDF4.0 analysis~\cite{Ball:2021leu}, excluding all jet observables so that we 
  can isolate the top data and use them as the main source of constraints for the large-$x$ gluon. Moreover we supplement the NNPDF4.0 dataset 
  by the Drell--Yan and top-quark measurements incorporated in Refs.~\cite{Greljo:2021kvv,Iranipour:2022iak} and~\cite{Kassabov:2023hbm}, respectively. 
  After applying kinematic cuts, this dataset comprises $n^{\rm current}_{\rm data} = 4271$ data points.
  \item Projections for the HL-LHC in high-mass Drell--Yan, forward Drell--Yan, and high-mass top-pair production, 
  yielding a total of $n^{\rm HL}_{\rm data} = 92$ data points. The uncertainties associated with the first category are taken directly from the corresponding experimental publications and therefore reflect realistic experimental conditions. In contrast, the uncertainties assigned to the HL-LHC projections are constructed following dedicated projection scenarios, which are described in detail later in this section. 
\end{enumerate}
The kinematic coverage of the data points used in this study is shown in Fig.~\ref{fig:xq2}. The points
are shown in \((x,Q^2)\) space with the HL-LHC projections highlighted with a border.
The high-mass Drell-Yan, forward on-shell Drell-Yan and top quark pair production HL-LHC projections 
that we generate are summarised in Tab.~\ref{tab:hmdy_hllhc}, Tab.~\ref{tab:lhcb_dy_hllhc} and Tab.~\ref{tab:ttbar_hllhc}, 
and illustrated below.

\paragraph{HL-LHC high-mass Drell-Yan (HMDY) projections} The invariant mass distribution
projections are generated at \(\sqrt{s} = 14\) TeV, assuming an
integrated luminosity of \(\mathcal{L} = 6 \text{ ab}^{-1}\) (
\(3 \text{ ab}^{-1}\) collected by ATLAS and  \(3
\text{ ab}^{-1}\) by CMS).
Both in the case of NC and CC Drell-Yan cross sections, the pseudodata
were generated using the {\tt MadGraph5\_aMCatNLO} NLO Monte Carlo event
generator~\cite{Frederix:2018nkq} with additional $K$-factors to include the NNLO QCD and
mass-dependent NLO EW corrections. The pseudodata consist of four datasets
(associated with NC/CC distributions with muons/electrons in the final
state), each comprising respectively 12 (NC) and 16 (CC) bins in the $m_{ll}$
invariant mass distribution or transverse mass $m_T$ distributions
with both $m_{ll}$ and $m_T$ greater than 500 GeV , with the highest energy bins reaching $m_{ll}=4$
TeV ($m_T=3.5$ TeV) for NC (CC) data.
The rationale behind the choice of number of bins and the width of each bin was outlined 
in Ref.~\cite{DYpaper}, and stemmed from the requirement that the expected number of events
per bin was big enough to ensure the applicability of Gaussian
statistics. The choice of binning for the $m_{ll}$ ($m_T$) distribution at the
HL-LHC is displayed in Fig.~5.1 of Ref.~\cite{DYpaper}.
%
\begin{table}[ht!]
    \centering
    \footnotesize
    \begin{tabularx}{\textwidth}{Xccc}
    \toprule
    Process & N$_\text{dat}$ & $m$ [GeV] & Ref. \\
    \midrule
    High-mass Drell--Yan (NC) & 12 & $m_{\ell\ell} \in [\, 500, 4600 \, ]$   &\cite{DYpaper} \\
    High-mass Drell--Yan (CC) & 16 & $m_T \in [\, 500, 3500 \, ]$            & \cite{DYpaper} \\
    \bottomrule
    \end{tabularx}
    \caption{High-mass Drell--Yan projections at the HL-LHC already presented in \cite{DYpaper} and already included in \cite{Hammou:2023heg}.}
    \label{tab:hmdy_hllhc}
\end{table}

\paragraph{HL-LHC forward Drell-Yan projections} HL-LHC projections for LHCb are generated assuming an integrated luminosity of $0.3~\mathrm{ab}^{-1}$ as a benchmark, and focus on forward $W$ and $Z$ production.
The predictions for each process were generated using the
{\tt MadGraph5\_aMCatNLO} NLO Monte Carlo event
generator~\cite{Frederix:2018nkq} interfaced with
Pineline \cite{Barontini:2023vmr} with additional $K$-factors to 
include the NNLO QCD correction produced using NNLOJET \cite{NNLOJET:2025rno}.
The $Z$ boson is 
produced on-shell ($60 \text{ GeV} < m_{ll} < 120 \text{ GeV}$). In both processes we impose requirements 
on the transverse momentum ($p^l_T > 20 \text{ GeV}$) and on the rapidity to study the forward region ($2 < |y| < 4.5$)
following the selections used in Ref.~\cite{LHCb:2015mad}.
In Ref.~\cite{Hammou:2023heg}, we showed that these data primarily constrain the $u$ and $d$ quark PDFs, while providing 
only limited sensitivity to the light antiquark distributions. This can be understood from the kinematics of forward lepton 
production at LHCb: for the final-state leptons to populate the forward region, one of the initial-state partons must carry 
a significantly larger longitudinal momentum fraction than the other. Since quarks are far more likely than antiquarks to carry 
a large fraction of the proton momentum, the resulting constraints predominantly affect the quark sector. 
%
\begin{table}[ht!]
    \centering
    \footnotesize
    \begin{tabularx}{\textwidth}{Xccc}
    \toprule
    Process & N$_\text{dat}$ & $y$ & Ref. \\
    \midrule
    Forward Drell--Yan (NC) & 18 & $y_Z \in [\, 2.0, 4.5 \, ]$& \cite{LHCb:2015mad} \\
    Forward Drell--Yan (CC) & 8 & $y_\mu \in [\, 2.0, 4.5 \, ]$& \cite{LHCb:2015mad} \\
    \bottomrule
    \end{tabularx}
    \caption{Forward on-shell Drell--Yan HL-LHC projections generated for this study and inspired 
    by the existing LHCb measurements of Ref.~\cite{LHCb:2015mad}.}
    \label{tab:lhcb_dy_hllhc}
\end{table}

\paragraph{HL-LHC top quark pair production} The LHC operates as a top-quark factory, producing very large samples of $t\bar{t}$ events as well 
as sizeable datasets for rarer top-quark processes. The prospects for measurements at the HL-LHC, corresponding to an integrated luminosity of $3~\mathrm{ab}^{-1}$, are 
taken from Ref.~\cite{Durieux:2022cvf}, which in turn were obtained through an extrapolation of current Run~II results.

In the projection scenario considered in Ref.~\cite{Durieux:2022cvf}, statistical uncertainties and a subset of experimental 
systematic uncertainties are assumed to scale with the inverse square root of the integrated luminosity, while the remaining 
experimental systematics are reduced by an overall factor of five. In this work, we focus on differential measurements of the 
$t\bar{t}$ production cross section as a function of the invariant mass of the $t\bar{t}$ system. Such measurements at large 
invariant mass already play a central role in constraining four-fermion operators. To fully exploit the HL-LHC potential, 
the kinematic reach of the projections in the $t\bar{t}$ invariant mass is extended from the current upper limit of approximately $1.5~\mathrm{TeV}$ to $3~\mathrm{TeV}$.
Theoretical predictions for these observables are computed at NNLO in QCD for the SM, using \texttt{fastNLO} grids~\cite{Czakon:2017Dynamical} supplemented 
by NNLO/NLO $K$-factors computed with the {\tt HighTea} tool~\cite{Czakon:2023hls}, while the SMEFT contributions are included at leading order through the 
corresponding $K$-factors.

\begin{table}[ht!]
    \centering
    \footnotesize
    \begin{tabularx}{\textwidth}{Xccc}
    \toprule
    Process & N$_\text{dat}$ & $m_{t\bar{t}}$ [GeV] & Ref. \\
    \midrule
    High-mass $t\bar t$ production & 18 & $m_{t\bar{t}} \in [\, 250, 3000\, ]$ & \cite{Durieux:2022cvf} \\
    \bottomrule
    \end{tabularx}
    \caption{Top-pair projections at the HL-LHC, from Ref.~\cite{Durieux:2022cvf}.}
    \label{tab:ttbar_hllhc}
\end{table}

\begin{figure}[ht!]
    \centering
    \includegraphics[width=0.49\textwidth]{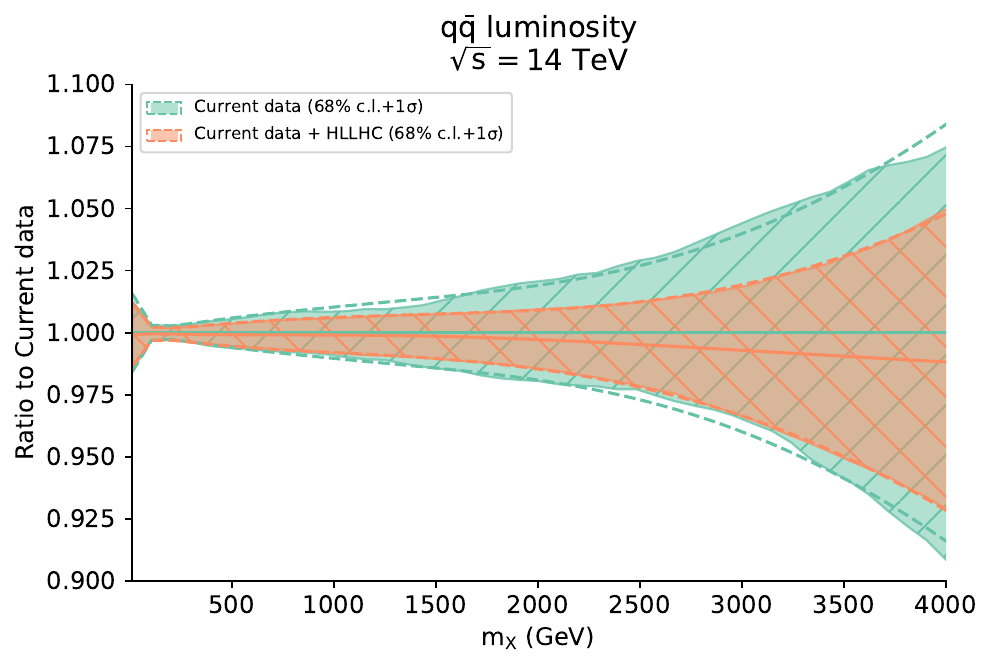}
    \includegraphics[width=0.49\textwidth]{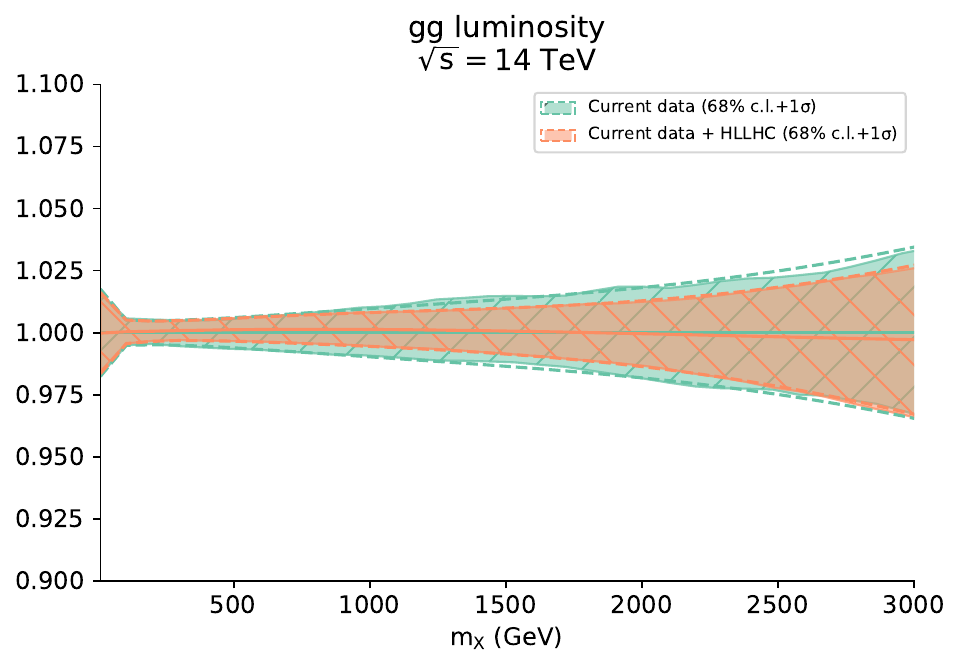}
    \caption{Impact of the HL-LHC projections on the uncertainties of the quark-antiquark (left) 
    and gluon-gluon (right) PDF luminosities.}
    \label{fig:impact_HLLHC}
  \end{figure}

\noindent The impact of the HL-LHC projections on the PDFs is assessed in Fig.~\ref{fig:impact_HLLHC} by 
comparing a fit on $L_0$ data one including the HL-LHC projections (orange band) and one excluding them (green band). 
A substantial reduction in the uncertainties of the quark--antiquark luminosity is observed at invariant masses above 
$1~\mathrm{TeV}$, indicating that the HL-LHC Drell--Yan projections play a dominant role in constraining the quark and antiquark PDFs 
in the large-$x$ region. 
%
By contrast, the effect of the HL-LHC top-quark projections on the gluon--gluon luminosity is significant, but less stringent in comparison to the 
high-mass Drell-Yan distributions. This suggests that the current extensive top-quark data already included in the fit,
which was implemented in~\cite{Kassabov:2023hbm} remain competitive with the HL-LHC projections in constraining the gluon PDF at large~$x$.

%% file: sections/sec3_subsecs/subsec2_wmodel.tex
\subsection{BSM and PDF interplay in the Drell-Yan sector}
\label{subsec:DY}

In this section we consider two BSM scenarios, that can be parametrised in the SMEFT, impacting the Drell-Yan sector: the oblique parameters $\hat{W}$ 
(corresponding to a heavy universally coupled $W'$ field) and $\hat{Y}$
(corresponding to a heavy universally coupled $Z'$ field), playing a role similar to the $c$ coefficient introduced in Sect.~\ref{subsec:presentation_of_model}. 
More details on the models are given in App.~\ref{app:wmodel} and 
App.~\ref{app:ymodel} respectively. These models have already 
been explored in the context of PDF fitting in Refs.~\cite{Greljo:2021kvv,Iranipour:2022iak,Hammou:2023heg,Hammou:2024xuj}, 
where it was shown that -- within a global PDF fit -- synthetic data generated with a non-zero $\hat{Y}$ injected would be flagged as incompatible 
with the bulk of the data included in the fit, whereas the injection of a non-zero $\hat{W}$ in the high--energy tails of the HL-LHC 
measurements could be completely absorbed by the PDFs if those were fitted by assuming the SM, hence producing {\it BSM--biased} PDFs. 
Throughout this section, we analyse the results in the context of a closure test in which the {\it true} underlying law that 
we use to generate the synthetic data is given by the {\tt NNPDF4.0 NNLO} PDF set~\cite{NNPDF:2021njg} -- which we assume to be the \emph{true} 
PDFs -- and the \emph{true} theory of nature is given by the SM augmented by new heavy universally--coupled boson(s) that can be parametrised at the LHC energies 
in the SMEFT by a combination of non-zero $\hat{Y}$ and $\hat{W}$ Wilson coefficients that we will specify in the text. 

\begin{figure}[tbh]
    \centering
    \includegraphics[width=0.51\linewidth]{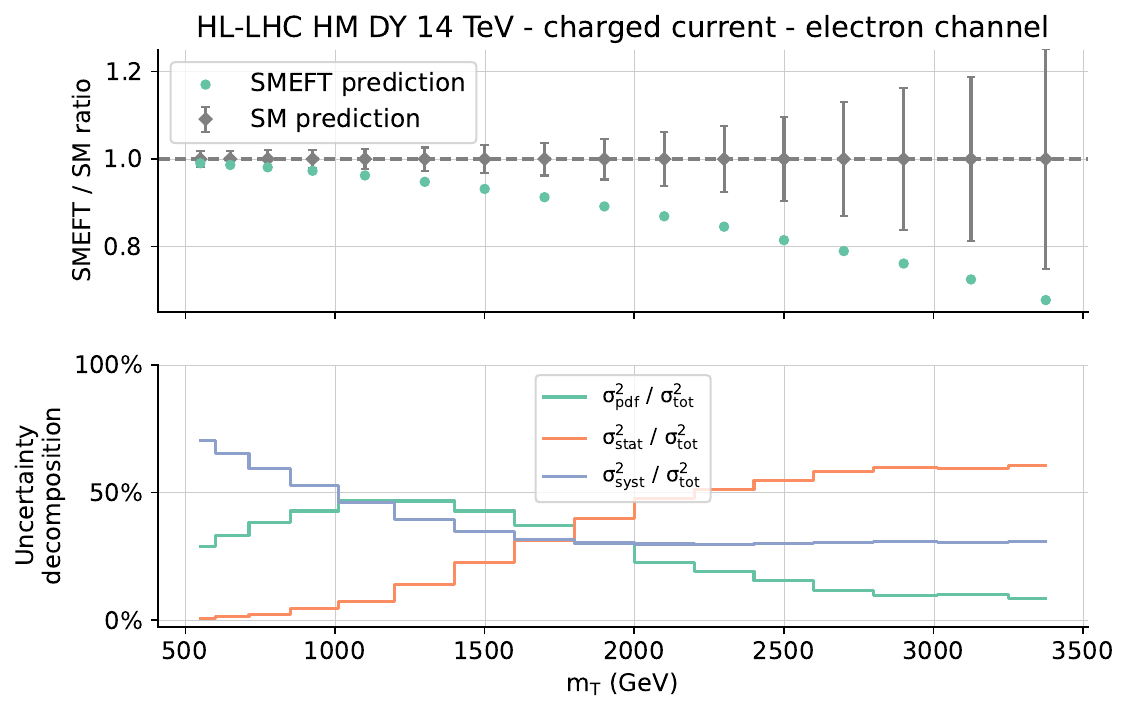}
	\includegraphics[width=0.47\linewidth]{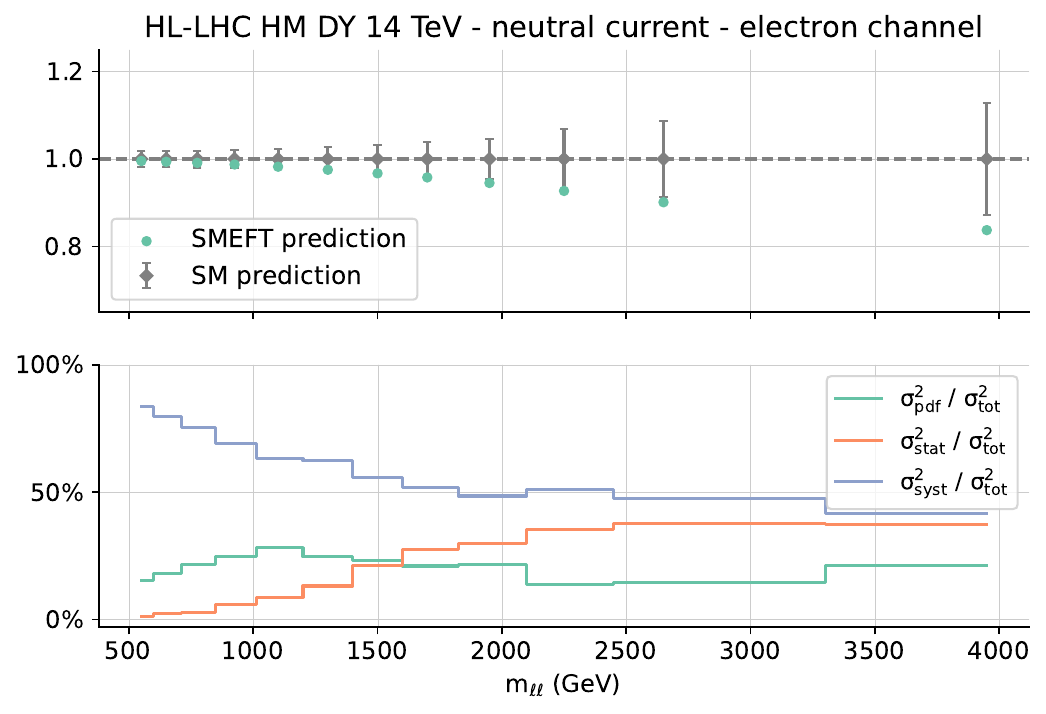}\\
    
    \includegraphics[width=0.51\linewidth]{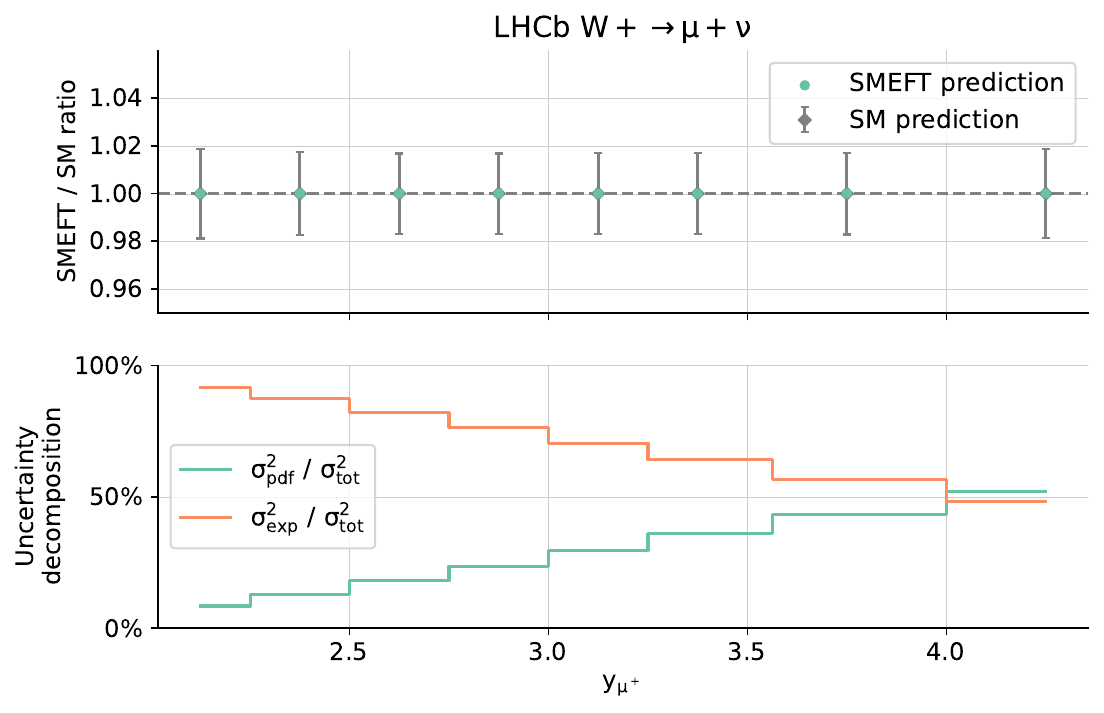}
    \includegraphics[width=0.47\linewidth]{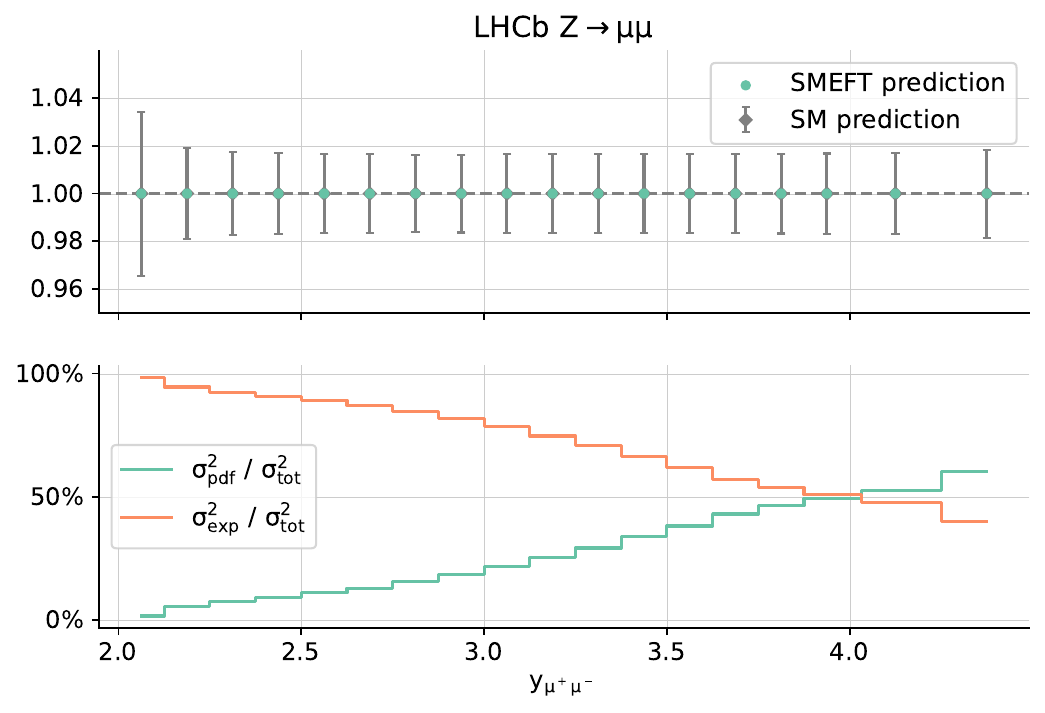}
    
	\caption{Projected impact of a flavour universal $W'$ with $m_{W'} = 13.8$ TeV ($\hat{W} = 8 \times 10^{-5}$) 
    on high-mass Drell-Yan (DY) distributions
	measured by ATLAS and CMS in the HL-LHC phase (top row) and in the forward bosons production measured by LHC in the HL-LHC phase(bottom row) for 
    charged current DY (left column) and neutral current DY (right column). 
    The BSM signal is compared to the SM prediction and its total uncertainty, which is split into component due to PDF uncertainty and a component due to experimental uncertainties  
    in the lower insets. In the case of high-mass DY the experimental uncertainty is split into a statistic and a systematic components, while 
    in the case of forward DY distributions the statistic uncertainty is subdominant and is not displayed separately. 
	}
	\label{fig:W_impact}
\end{figure}

The current Run I and Run II high-mass Drell--Yan measurements that we include in our analysis, comprising neutral-current Drell--Yan measurements 
performed by ATLAS at $\sqrt{s}=7$ and $8$~TeV~\cite{ATLAS:2013xny, ATLAS:2016gic} and by CMS at $\sqrt{s}=7$, $8$ and 
$13$~TeV~\cite{CMS:2013zfg,CMS:2014jea, CMS:2018mdl}, are sensitive to the BSM scenario considered here~\cite{Farina:2016rws}. 
The HL-LHC high--mass projections enhance by a large factor the current sensitivity. In Fig.~\ref{fig:W_impact} we show the deviation from the 
SM predictions that the HL-LHC projections would display if a new heavy universally--coupled $W'$ boson with mass 
$M_{W'} = 13.8$ TeV was present in nature (corresponding to a SMEFT Wilson coefficient $\hat{W} = 8 \times 10^{-5}$). 
We observe that the distortion in the tails is larger than (comparable to) the uncertainty of the SM theoretical predictions for masses above 1~TeV in the 
case of charged (neutral) current. 
In the case of charged currents the uncertainty -- for invariant mass around 500~GeV -- is dominated by the systematic experimental uncertainties 
while above 2~-TeV the statistical uncertainty dominates over systematics. The PDF uncertainty becomes larger than experimental uncertainties in the 
1 TeV -- 1.8 TeV region.
In the case of neutral current PDF uncertainties are 
smaller than the experimental ones across the whole $m_{\ell\ell}$ spectrum, and are dominated by systematic uncertainties. 
On the other hand distortion effects associated to this model 
are very suppressed in the HL-LHC on-shell LHCb forward Drell-Yan measurements, as shown in the bottom row of Fig.~\ref{fig:W_impact}, as the 
SMEFT predictions sit on top of the SM predictions, which in turn are dominated by systematic experimental uncertainties, 
apart from the most forward bins in which the PDF uncertainty becomes comparable to the experimental one.

\begin{figure}[htb]
    \centering
    \includegraphics[width=0.8\linewidth]{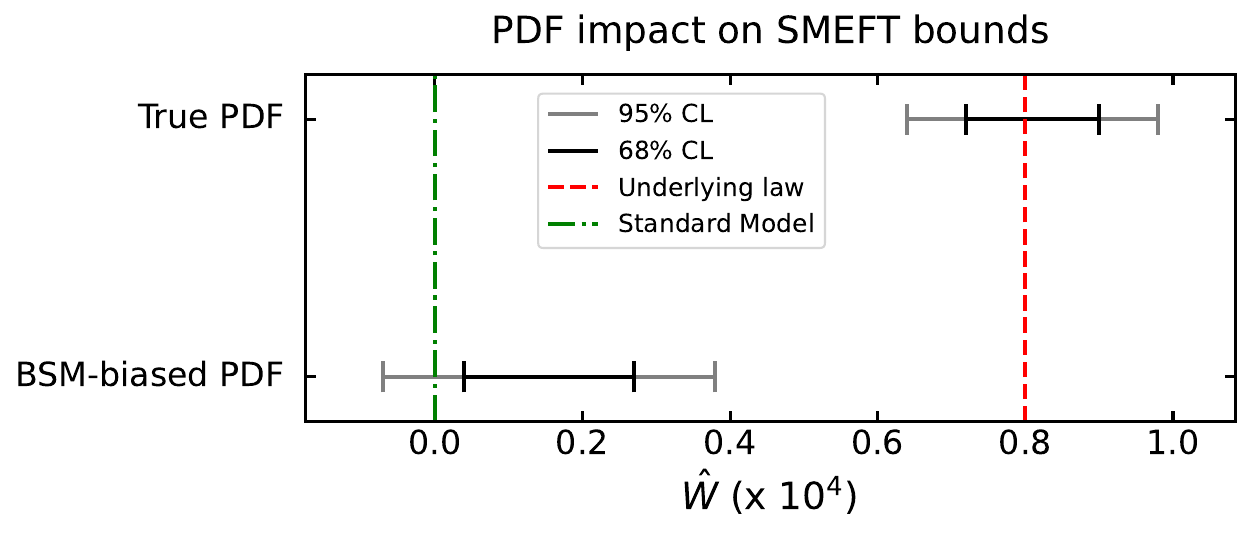}
	\caption{SMEFT fit performed on the synthetic Drell-Yan high-mass data and HL-LHC pseudodata 
    presented in Sect.~\ref{subsec:settings} with $\hat{W} = 8 \times 10^{-5}$ injected. The 95\% C.L. (grey) and 68\% C.L. bands 
    are displayed for a SMEFT fit in which the {\it true} PDFs are used as input set (top) and a fit in which the {\it BSM-biased} 
    PDFs are used (bottom). 
}
	\label{fig:DY_PDF_cont}
\end{figure}

In Fig.~\ref{fig:DY_PDF_cont}, we show that a fit of the $\hat{W}$ Wilson coefficient can yield completely unreliable 
results if {\it BSM-biased} PDFs are used as input in the SMEFT fit. When the correct {\it true}  
PDFs are employed instead, the injected {\it true} value of $\hat{W}$ is accurately recovered, differing from the SM 
expectation by more than  $5\sigma$. By contrast, using a BSM-biased PDF set in the SMEFT fit leads to an apparent 
compatibility with the SM, while simultaneously excluding the {\it true} underlying theory with a significance exceeding $5\sigma$. 
This behaviour qualitatively reproduces the pattern already observed in the toy-model study shown in Fig.~\ref{fig:toy_comparison}.

We now assess a set of strategies aimed at mitigating this undesirable behaviour and achieving a more robust determination of 
both PDFs and SMEFT coefficients, following the prescriptions outlined in Sect.~\ref{subsec:analytic}. First, we perform a conservative 
PDF fit in which all high-mass Drell–Yan observables are excluded, and we subsequently use the resulting PDF set as input to a SMEFT fit. 
Second, we carry out a fully simultaneous fit of the PDFs and SMEFT coefficients.

\begin{figure}[!ht]
    \centering
	\includegraphics[width=0.49\linewidth]{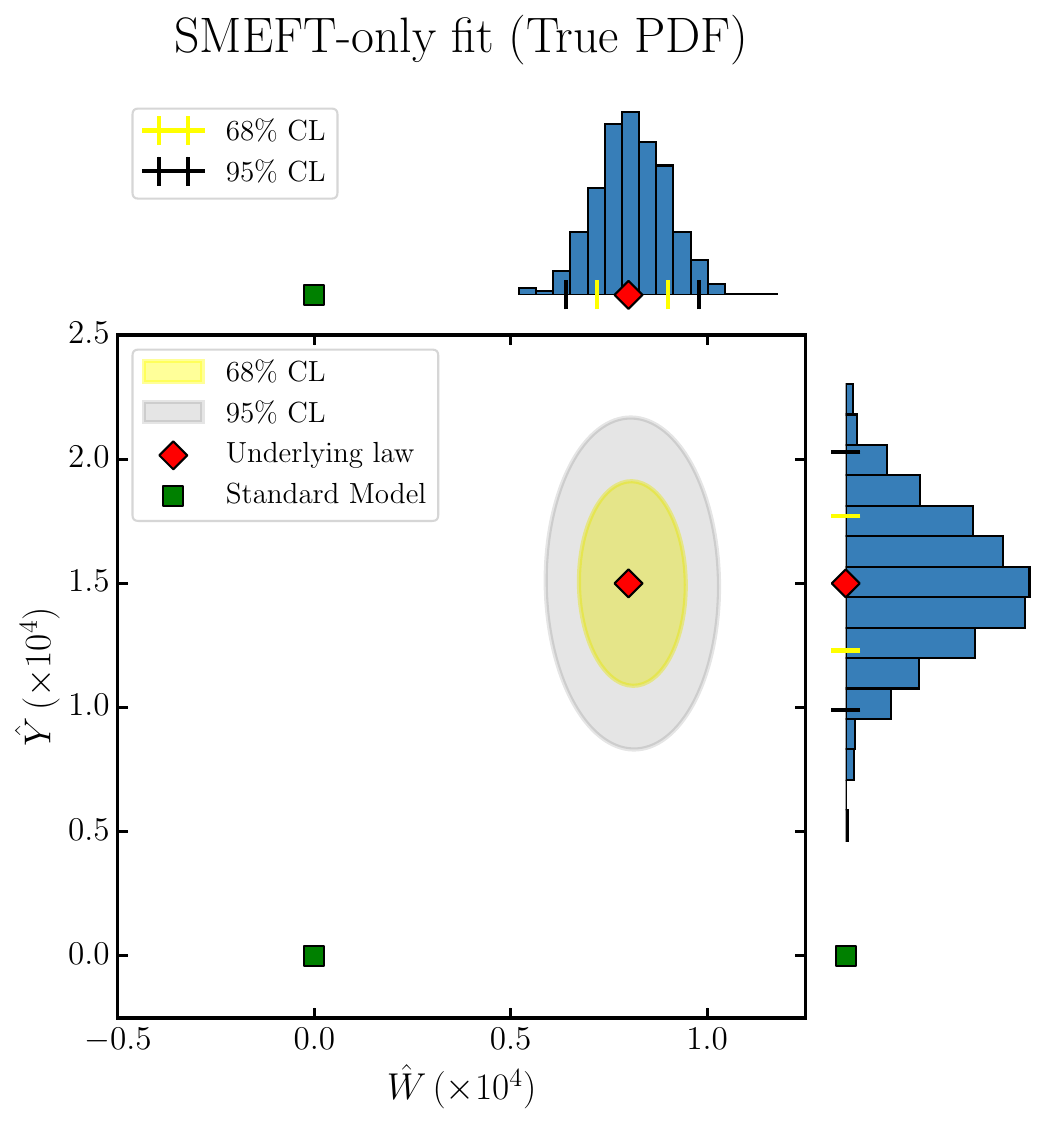}
	\includegraphics[width=0.49\linewidth]{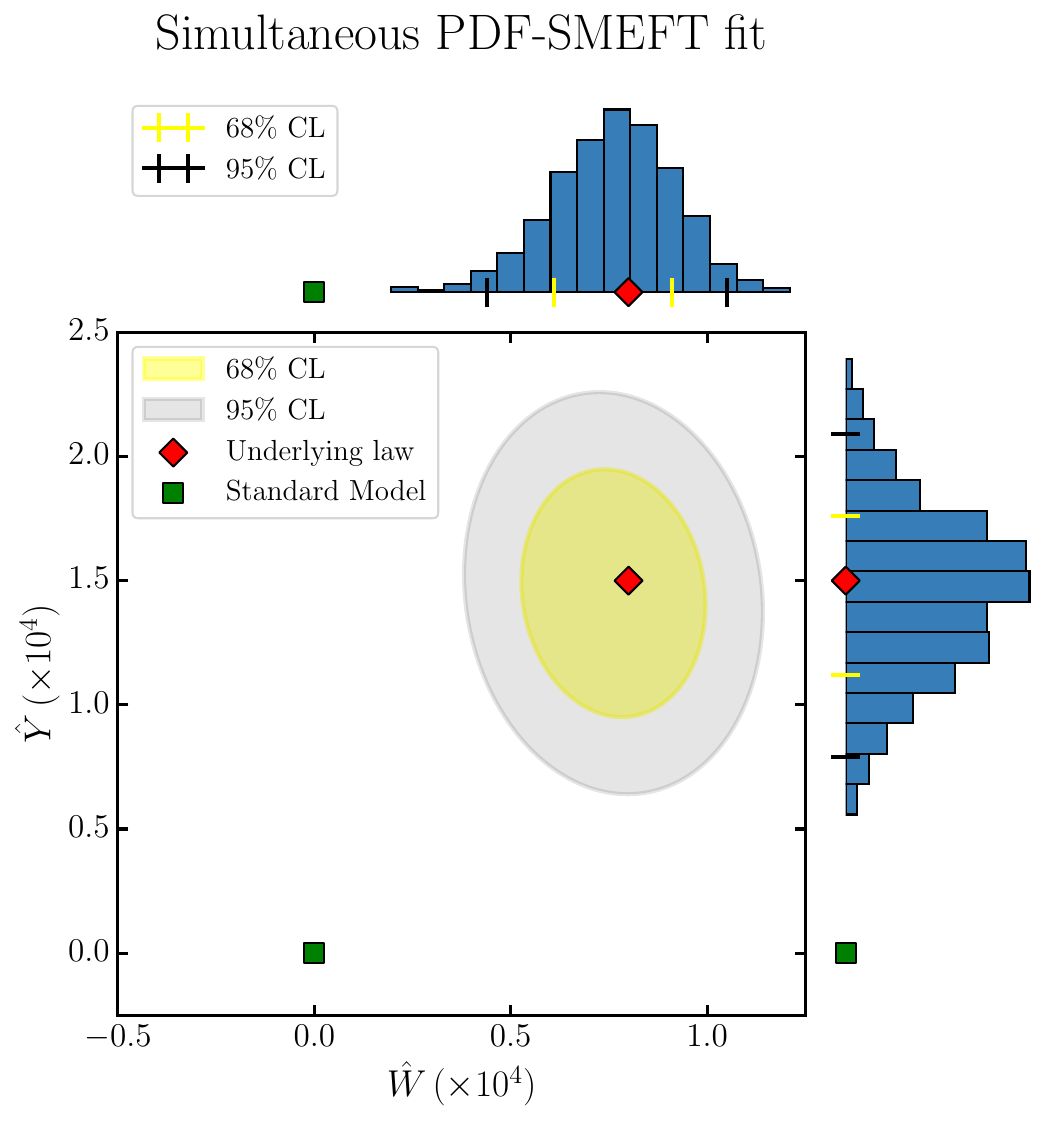}\\
	\includegraphics[width=0.49\linewidth]{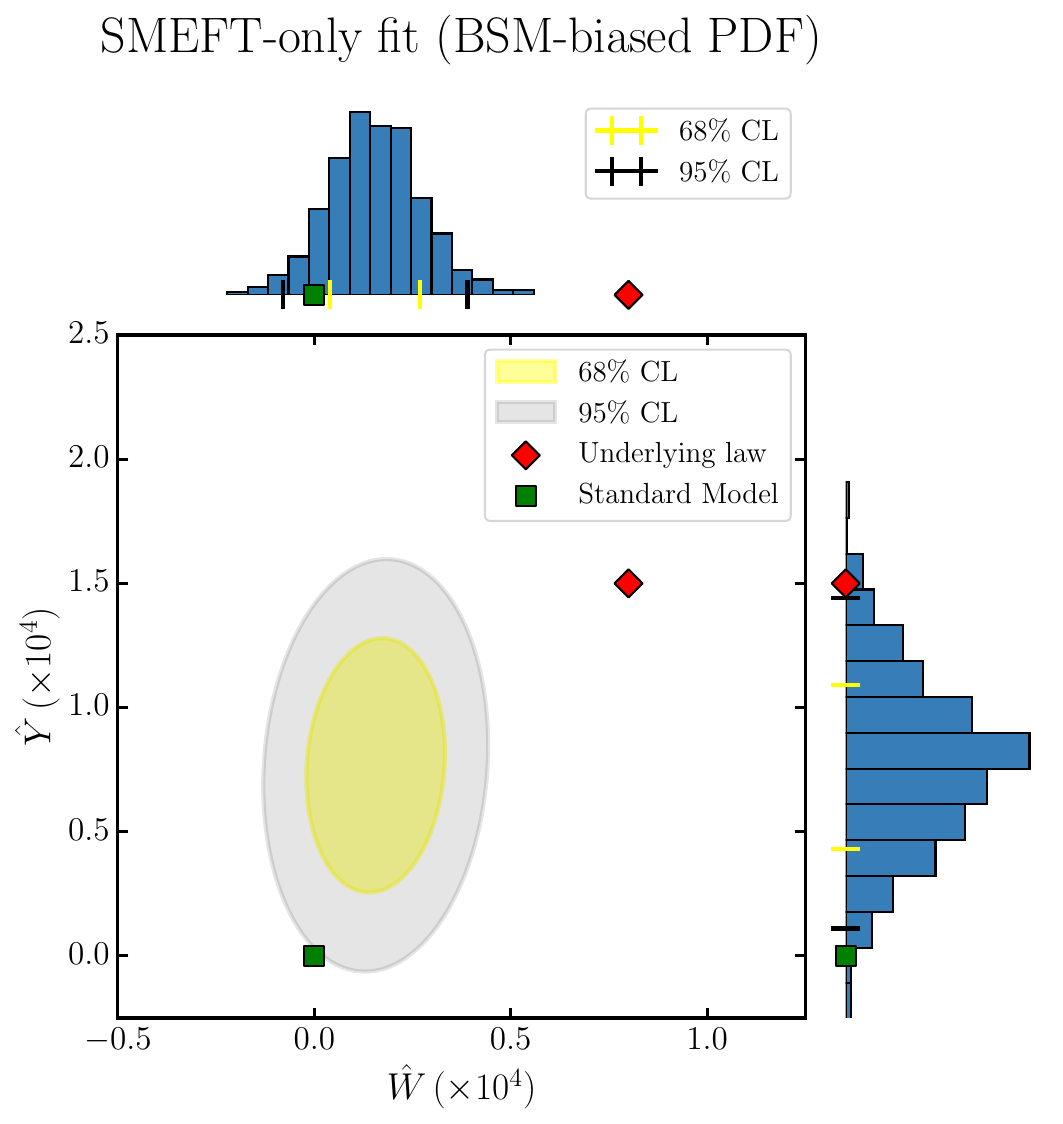}
	\includegraphics[width=0.49\linewidth]{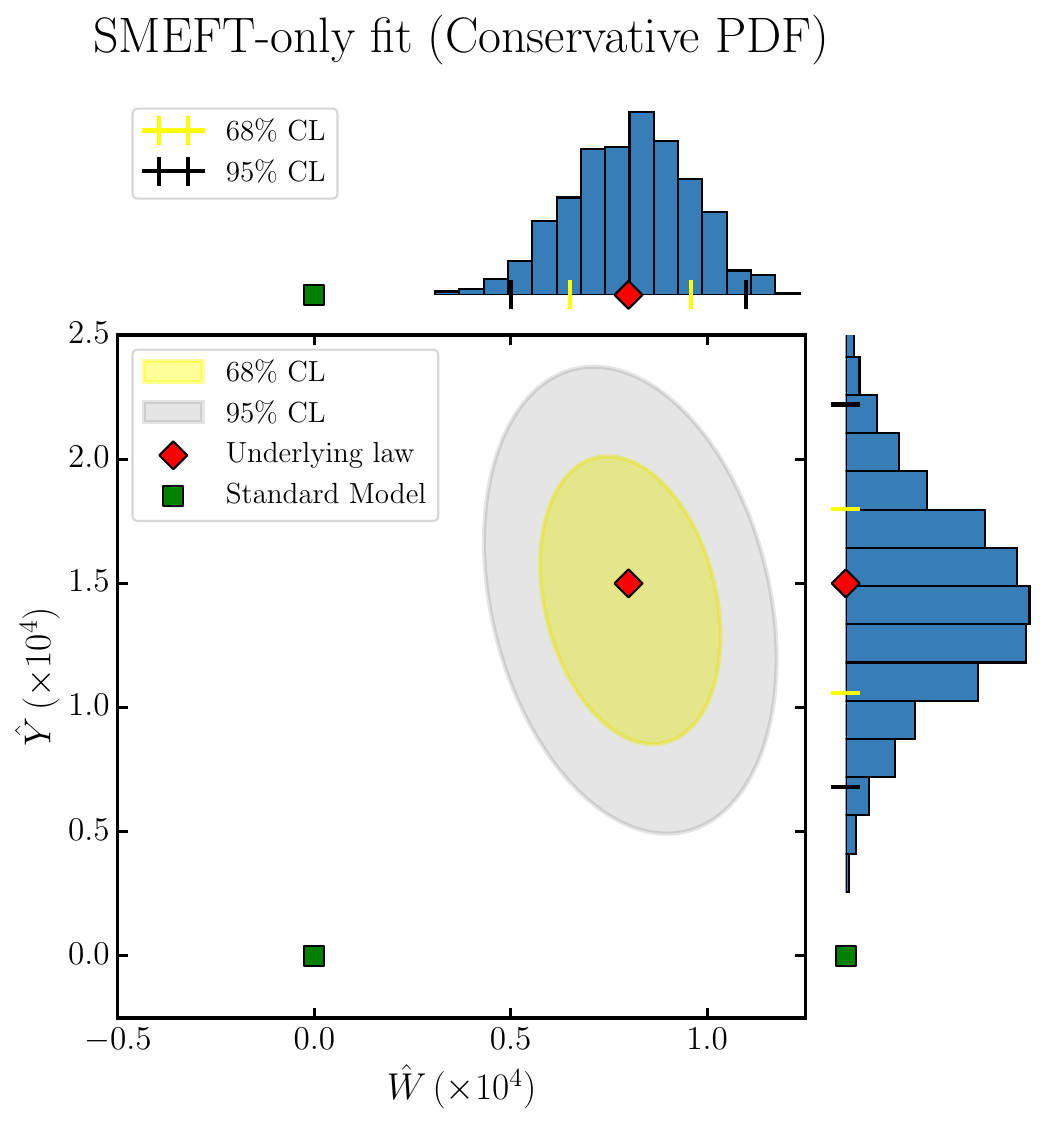}
	\caption{Comparison of SMEFT fits alongside the $\hat{W}$ and $\hat{Y}$ directions on Drell-Yan observables whose 
	underlying law is $\hat{W} = 8 \times 10^{-5}$ and $\hat{Y} = 1.5 \times 10^{-4}$. Upper left: Fixed-PDF SMEFT fit 
	using the true PDFs used to generate the pseudodata. Bottom left: Fixed-PDF SMEFT fit using BSM-biased PDFs.
	Upper right: Simultaneous fit of the SMEFT and PDFs. Bottom right: Fixed-PDF fit using the conservative PDFs.}
	\label{fig:DY_SMEFT_comparison}
\end{figure}

\begin{figure}[!ht]
    \centering
	\includegraphics[width=0.7\linewidth]{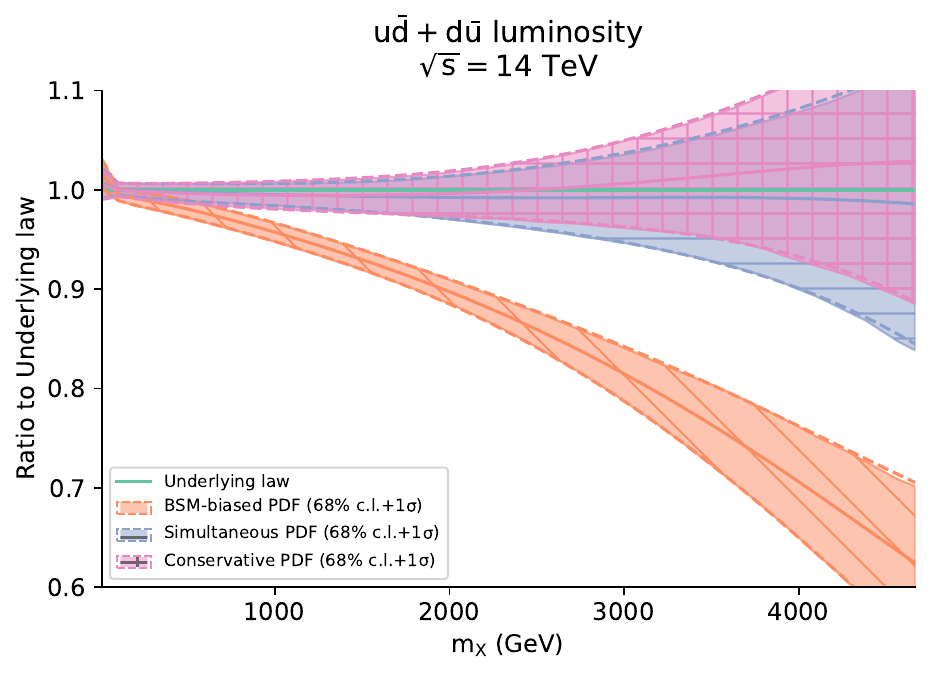}
	\caption{Comparison between the $u\bar{d} + d\bar{u}$ PDF luminosities defined in this section and the 
    \emph{true} PDFs of the underlying law: the PDF luminosity obtained in a conservative fit, where all the 
    high-mass Drell-Yan data are excluded (pink), the PDF luminosity obtained in a simultaneous fit of the PDFs with the 
    SMEFT $\hat{W}$ and $\hat{Y}$ coefficients (blue) and the PDF luminosity obtained assuming the SM, hence BSM-biased (in orange). 
	The error bands represent the 68\% C.~L. uncertainty of the resulting PDFs. }
	\label{fig:DY_PDF_lumi_comparison}
\end{figure}

To ensure the generality of our conclusions and to test the ability of these approaches to disentangle correlated effects, we generate 
synthetic data assuming an underlying scenario in which both $\hat{W}$ and $\hat{Y}$ are non-zero. 
We retain the same value of $\hat{W}=8\cdot10^{-5}$ as before and additionally 
inject $\hat{Y}=1.5\times10^{-4}$, corresponding to an additional $Z'$ with mass $M_{Z'}=18.7$~TeV.
%
The resulting SMEFT constraints are shown in Fig.~\ref{fig:DY_SMEFT_comparison}, while the corresponding PDF 
luminosities are presented in Fig.~\ref{fig:DY_PDF_lumi_comparison}.

The SMEFT bounds shown in Fig.~\ref{fig:DY_SMEFT_comparison} demonstrate that both the simultaneous fit and the 
conservative fixed-PDF SMEFT fit successfully recover the underlying theory and clearly exclude the SM, in contrast to the 
BSM-biased fixed-PDF SMEFT fit. This behaviour closely mirrors that observed in the toy-model study presented in Fig.~\ref{fig:toy_comparison}.
For comparison, we also perform a fixed-PDF SMEFT fit using the true PDF set constrained by the full dataset. 
Although this does not represent a realistic scenario -- since such PDFs are not accessible in 
practice -- it provides an idealised benchmark corresponding to the best achievable constraints 
on the SMEFT coefficients. As expected, both the simultaneous and conservative fits yield wider uncertainties 
than this true fixed-PDF reference. 
The origin of the uncertainty broadening differs between the two approaches. In the conservative fit, the 
looser SMEFT bounds arise directly 
from the increased PDF uncertainties in the large-$x$ region, which result from the removal of high-mass data 
that otherwise constrain large-$x$ 
quark and antiquark distributions. In the simultaneous fit, by contrast, the dataset is identical to that of the 
reference fit; the broader uncertainties 
instead reflect the much larger number of fitted parameters. In addition to the two SMEFT degrees of freedom, 
the fit simultaneously optimises the 
hundreds of parameters of the NNPDF parametrisation. Despite this significant increase in model complexity, the resulting uncertainty bands remain well under control.
This indicates that the available dataset is sufficiently constraining to disentangle the marginal distributions 
of the SMEFT coefficients and the PDFs. 
Numerical values for the SMEFT confidence interval (CI) marginal bounds 
and the corresponding change in the $\chi^2$ (compute on all experimental points that are sensitive to 
$\hat{W}$ and $\hat{Y}$ are reported in Tab.~\ref{tab:statistics_DY_SMEFT_fit_methods}.
We observe there that the marginalised bounds on the $\hat{W}$ parameter are slightly tighter for the conservative than 
for the simultaneous one, while it is the opposite for the $\hat{Y}$ parameter.
\input{tables/statistics_DY_SMEFT_fit_methods.tex}

We further observe that the simultaneous fit reduces the correlation between $\hat{W}$ and $\hat{Y}$ with respect to the conservative fit, reflecting the fact that the additional PDF-sensitive data help to better disentangle these two parameters. The Pearson correlation coefficient between $\hat{W}$ and $\hat{Y}$ is defined as
\begin{equation}
\label{eq:pearson}
    \rho = \frac{\langle\hat{W}\,\hat{Y}\rangle \, - \, \langle\hat{W}\rangle \, \langle \hat{Y}\rangle}{\sigma_{\hat{W}} \sigma_{\hat{Y}}},
\end{equation}
where the average $\langle\rangle$ and the standard deviation $\sigma$ are computed over the 1000 replicas of the fit of the $\hat{W}$ and $\hat{Y}$ coefficients. Indeed, the correlation coefficient obtained in the simultaneous fit, $\rho_{\rm simu} = -0.09$, is closer to zero than that of the conservative fit, $\rho_{\rm cons} = -0.25$, indicating that the joint determination of PDFs and Wilson coefficients allows the fit to better identify the two parameters independently.

As far as the global $\chi^2$ is concerned, we observe that the global $\chi^2$ is basically unchanged with respect to the baseline if conservative PDFs or if the PDFs 
obtained from a simultaneous PDF and SMEFT fit are used. On the other hand, if the BSM-biased PDF are used the global $\chi^2$ deteriorates only marginally, less than 0.5$\sigma$ 
away, compared to the baseline, hence making the datasets with new physics injected look compatible with the SM. 

In Fig.~\ref{fig:DY_PDF_lumi_comparison} we present the $u\bar{d}+d\bar{u}$ PDF luminosity -- which is mostly relevant for the 
charged--current Drell--Yan observables. The corresponding $u\bar{u}+d\bar{d}$ luminosity for the neutral--current channel exhibits a very similar behaviour. 
We observe that the BSM--biased PDF luminosity effectively absorbs the energy--growing BSM signal present in the synthetic data (with BSM injected) 
shown in Fig.~\ref{fig:W_impact}, thereby failing to reproduce the true underlying PDF law. This mechanism is responsible for the bias observed 
in the SMEFT analyses, which consequently appear compatible with the SM. 
In contrast, both the simultaneous and conservative PDF determinations robustly recover the underlying law, with comparable uncertainty bands. 

To summarise, we have shown that, in this case, both the simultaneous fit of PDFs and SMEFT coefficients and the conservative 
fixed-PDF SMEFT fit are able to disentangle the PDF–SMEFT mixing in the Drell–Yan sector first identified in~\cite{Greljo:2021kvv}. 
Within the scope of this study, neither approach appears to outperform the other. It is worth noting, however, that the conservative 
PDFs could be safely determined by excluding only the high-mass Drell–Yan observables. In scenarios where additional sectors must be regarded 
as not “SM-safe” and consequently removed from the PDF fit, the performance of the conservative approach would likely deteriorate, 
whereas the simultaneous fit would remain largely unaffected.

%% file: tables/statistics_DY_SMEFT_fit_methods.tex
\begin{table}[!ht]
    \centering
     \begin{tabular}{|l|c|c|c|}
        \hline
        \textbf{Fit} & \textbf{$\hat{W} \times 10^4$}  & 
        \textbf{$\hat{Y} \times 10^4$} &
        \shortstack{$\Delta\chi^2$} \\
        \hline
        True PDF       & [0.64, 0.98]  & [0.99, 2.03] &  - \\
        BSM-biased PDF & [-0.08, 0.39] & [0.11, 1.44] & +32 (+48) \\
        Conservative PDF      & [0.50, 1.10]  & [0.68, 2.22] & -0.1 (+0.0)\\
        Simultaneous fit      & [0.44, 1.05]  & [0.79, 2.09] & -0.7 (-1.0)\\      
        \hline
    \end{tabular}
    \caption{Statistical summary of the $\hat{W}$ and $\hat{Y}$ fits to Drell--Yan observables shown in Fig.~\ref{fig:DY_SMEFT_comparison}. 
    The quoted bounds should be compared to the injected BSM signal used to generate the synthetic data, corresponding to 
    $\hat{W}=8\times10^{-5}$ and $\hat{Y}=1.5\times10^{-4}$. 
    For each fit we show the 95\%  confidence interval (CI) marginal bounds on the  $\hat{W}$ and $\hat{Y}$ fits obtained 
    using different input PDFs and how much the 
    total $\chi^2$ (computed over all data-points included in the PDF fits, $N_{\rm dat} = 4363$) 
    improves/deteriorates with respect to the baseline closure test in which 
    the \emph{true} PDFs are used as 
    an input including or excluding (in brackets) the PDF uncertainty in the computation of the $\chi^2$.}
    \label{tab:statistics_DY_SMEFT_fit_methods}
\end{table}

%% file: sections/sec3_subsecs/subsec3_zmodel.tex
\subsection{BSM and PDF interplay in the top sector}
\label{subsec:top}

In this section, we assess for the first time in the literature whether a BSM-induced bias may appear when the top data included in 
a PDF fit feature signs of new physics. In this case the gluon PDFs in the large-$x$ region 
could potentially absorb such signs.
Like the large-$x$ anti-quark, the large-$x$ gluon is poorly constrained by pre-LHC data and 
its knowledge heavily relies on LHC data that are measured at large energy scale $Q$ and/or large 
rapidity $y$. The strongest constraints come from high-$p_T$ inclusive jets, high invariant mass dijets, 
and the tails of the $t\bar{t}$ invariant mass distributions~\cite{Beneke:2012wb,Czakon:2013tha,Nocera:2017zge}.  
If there was a BSM model producing energy-growing effects on such distributions, the effect could be possibly mimicked and absorbed by the gluon.
Here we focus on the top sector, excluding the jet data from the input dataset for our PDF and SMEFT fits, leaving the latter to future studies~\cite{prep1,prep2}. 


The model that we explore here is the $\hat{Z}$ parameter, which belongs to the class of so-called 
\emph{universal theories}~\cite{Barbieri:2004qk}. It arises from the dimension-6 SMEFT matching of the 
Coloron model~\cite{Chivukula:1996yr,Simmons:1996fz}, featuring a massive ${\rm SU}(3)_c$ color-octet vector field. 
A detailed description of the UV and SMEFT models is provided in App.~\ref{app:zmodel}.

In order to assess whether the gluon PDF can absorb signs of such BSM model, we replicate the study of Ref.~\cite{Hammou:2023heg} 
in the $\hat{Z}$ scenario. We inject several non-zero values of the $\hat{Z}$ parameter in top 
invariant--mass distributions and identify what is the largest value of $\hat{Z}$, 
corresponding to the strongest BSM-induced deviation, which can be absorbed in the gluon PDF \emph{without} visibly worsening the fit quality 
of the data as compared to the baseline SM fit. 
A poor fit quality would indicate that the data are not consistent with the SM predictions, and point out the possible presence of 
NP in them. We use the NNPDF criteria to tag a dataset as inconsistent, namely $\chi^2 \geq 1.5$ and $n_\sigma \geq 2$ \cite{NNPDF:2024dpb}. 
We find that the threshold for maximal absorption 
of the NP signal in the gluon PDF is $\hat{Z} = 4 \times 10^{-4}$. In the UV Coloron model such a value of $\hat{Z}$ 
corresponds to a coupling of $g_{\mathcal{G}} = 2$ and $\alpha_{\mathcal{G}} = g_{\mathcal{G}}^2/4\pi \leq 1$ and 
$M_{\mathcal{G}} = 8.5$ TeV, which is large enough for the EFT approach 
to be valid even in the highest energy bins of the HL-LHC $t\bar{t}$ pseudo-data that we consider here.

We only consider the linear SMEFT corrections in this study,
both in the generation of the pseudodata and in the subsequent 
fits for consistency. We investigated the impact of the 
quadratic corrections and found that they amount to a maximum 
of a 12\% deviation in the highest energy bins of the HL-LHC 
projections. While they should be considered in a real data 
precision study, neglecting them does not change the qualitative 
behaviour we are investigating, as they are always noticeably 
smaller than the experimental uncertainties.


\begin{figure}[ht!]
    \centering
    \includegraphics[width=0.8\linewidth]{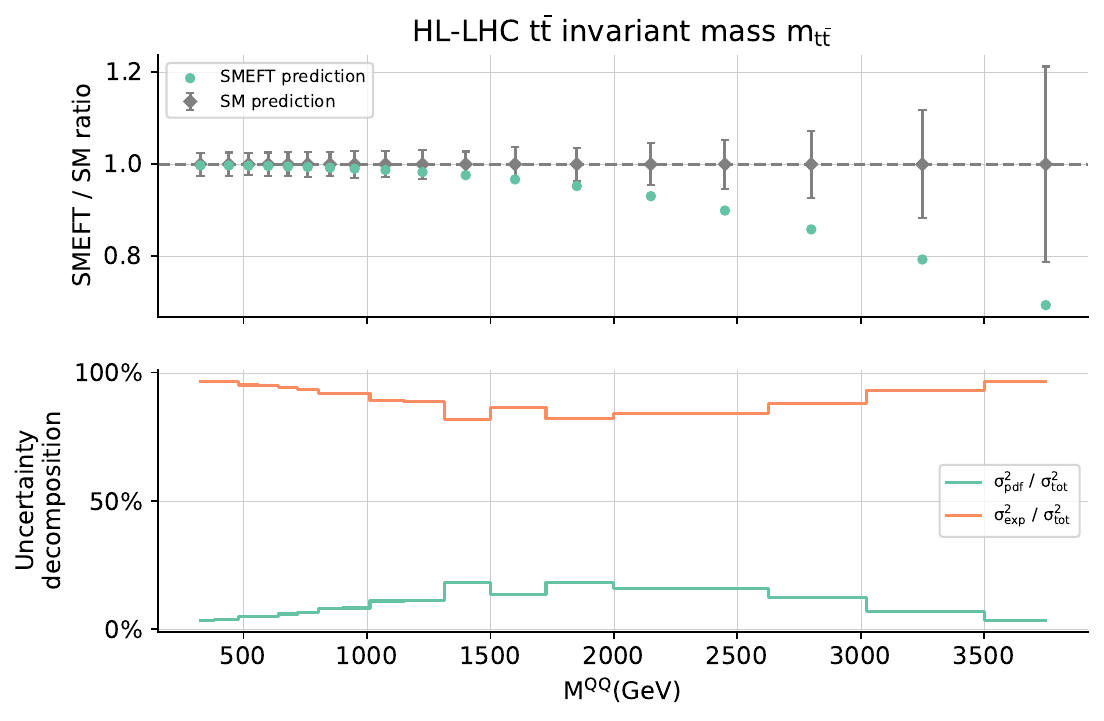}
    \caption{Same as Fig.~\ref{fig:W_impact}, here showing the impact of the $\hat{Z} = 4 \times 10^{-4}$ model on the HL-LHC projection for the $t\bar{t}$ 
    invariant mass distribution.}
    \label{fig:impact_z4}
\end{figure}

In the top-quark sector, in addition to the inclusive and differential $t\bar{t}$ cross sections and $t$-channel single-top production 
already included in {\tt NNPDF4.0}, we incorporate the additional observables introduced in Ref.~\cite{Kassabov:2023hbm}. 
These include $t\bar{t}$ production asymmetries, $W$-boson helicity fractions, associated top-pair production with electroweak gauge 
bosons and heavy quarks—such as $t\bar{t}Z$, $t\bar{t}W$, $t\bar{t}\gamma$, $t\bar{t}t\bar{t}$, and $t\bar{t}b\bar{b}$—as well as $s$-channel 
single-top production and associated single-top plus vector-boson processes. A non-zero value of $\hat{Z}$ would affect all these distributions 
in the high--invariant mass bins, but the effect is enhanced in the HL-LHC scenario that we consider in our work.
The impact on the $t \bar{t}$ projections at the HL-LHC binned 
in the invariant mass of the top pair is displayed in Fig.~\ref{fig:impact_z4}. We see that the effect remains below 10\% until 
the top pair invariant mass reaches 2 TeV, beyond which the impact reaches the 20-30\% level exceeding the PDF uncertainty. 

\begin{figure}[ht!]
    \centering
    \includegraphics[width=0.9\linewidth]{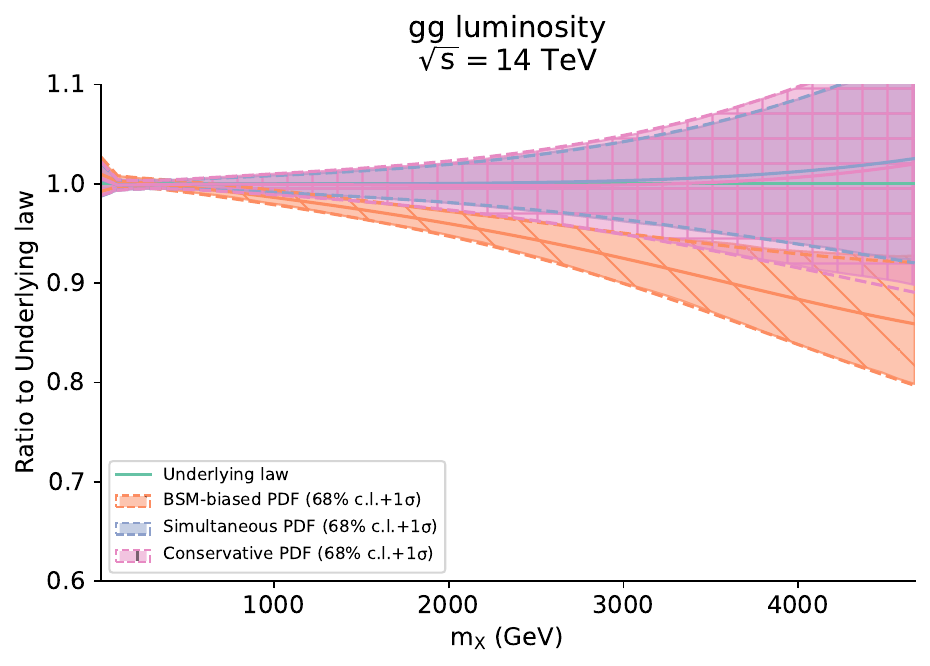}
    \caption{The \emph{true} baseline gluon-gluon luminosity at $\sqrt{s}=14$ TeV in the central 
    rapidity region (green line) is compared to the BSM-biased luminosity (orange) obtained by fitting 
    synthetic data in which $\hat{Z} = 4 \times 10^{-4}$ has been injected by assuming the SM, to a conservative luminosity (orange) obtained by 
    excluding all top data from the PDF fit and to the result of a simultaneous fit of PDFs and SMEFT Wilson coefficients (blue). 
    The results are normalised to the baseline SM luminosities and the 68\% C.L. 
    bands are displayed.}
    \label{fig:gg_lumi_Z4_comp}
\end{figure}

We fit the PDFs on the synthetic data presented in Sect.~\ref{subsec:settings} -- excluding all jets and dijets distributions -- 
injecting the BSM signal in the data and assuming the SM in our theory prediction. For $\hat{Z}=4\times 10^{-4}$ we 
obtain a fit-quality that is statistically equivalent to the 
fit performed on SM synthetic data using a SM theory, thus indicating that the BSM signal gets absorbed by the PDFs. 
The effect on the gluon-gluon luminosity is visible in Fig.~\ref{fig:gg_lumi_Z4_comp}, where we see that the gluon luminosity (pink) 
is mimicking the new physics signature present in the data and fails to recover the underlying law. 
In comparison both the conservative PDFs in which all high-mass 
top observables are excluded and the result of a simultaneous fit of PDF and SMEFT
recover the underlying law. The PDF 
uncertainty bands associated with the conservative fit is systematically larger than the uncertainty of the 
PDFs obtained in a the simultaneous fit. The difference increases with the energy scale.

\begin{figure}[ht!]
    \centering
    \includegraphics[width=0.9\linewidth]{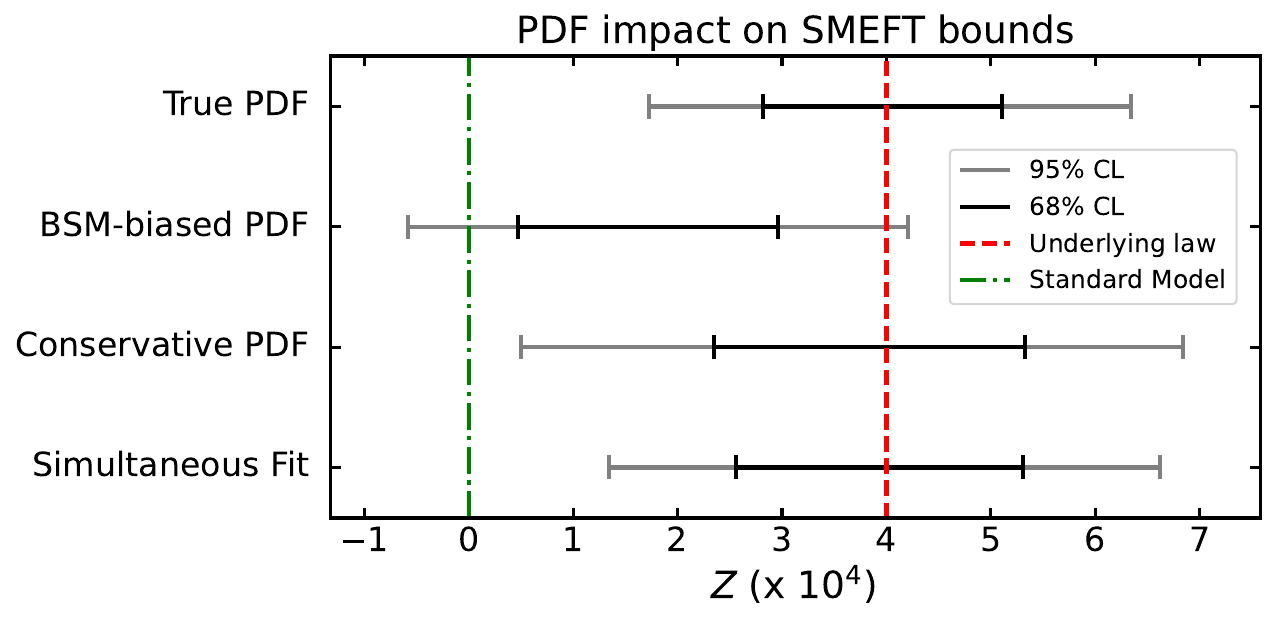}
    \caption{
    SMEFT fit performed on the $t\bar{t}$ productions pseudodata
    mirroring existing measurements and augmented by HL-LHC 
    projections, presented in Sect~\ref{subsec:settings}, 
    with $\hat{Z}=4 \times 10^{-4}$ injected. The 95\% (grey) and 
    68\% (black) C.L. correspond to SMEFT fits using different 
    PDF sets as inputs and to a simultaneous determination of 
    PDFs and SMEFT.}
    \label{fig:ttbar_smeft_fits}
\end{figure}

Finally, in Fig.~\ref{fig:ttbar_smeft_fits} and Table~\ref{tab:statistics_ttbar_SMEFT_fit_methods} we 
compare the bounds on the $\hat{Z}$ Wilson coefficient obtained using: the \emph{true} PDF input set;   
the BSM-biased PDFs, the conservative PDFs and the SMEFT-PDF from a simultaneous fit of the PDFs and $\hat{Z}$. 
Again, using the \emph{true} PDF, we are able to clearly spot a deviation with respect to the SM. 
Inputting the BSM-biased PDF, we see that 
the absorption of NP signals in the gluon PDFs is only partial since we are not excluding the true underlying 
law, contrarily to what we observe in the DY sector. Instead, we obtain SMEFT bounds somewhat compatible with both the SM and the true BSM model.
The result of the mixing between PDF and SMEFT is a somewhat watering down the deviation from the SM, from 
$n_\sigma = 3.43$ to $n_\sigma = 1.41$. On the other hand the conservative fixed-PDF SMEFT fit and the simultaneous fit
are both able to recover robustly the injected value of $\hat{Z}$. However, this time the simultaneous fit
produces noticeably tighter bounds as compared to the bounds obtained by using \emph{conservative} PDFs. 
This means that the benefit of being able to use the 
$t\bar{t}$ observables binned in the invariant mass to constrain the gluon PDF outweigh the uncertainty increase
due to fitting the PDF degrees of freedom alongside the SMEFT.
\input{tables/statistics_ttbar_SMEFT_fit_methods.tex}

To conclude, we have shown that the risk of absorbing signs of new physics inside the PDFs is 
not limited to the DY sector and the $\hat{W}$ model, as it can also happen in the top quark sector. Thus, even 
if it appears to be a model and sector dependent effect since we only observed a partial absorption 
in the top sector while it was a total one for DY, we can assume that this risk is quite general and 
must be taken into account for robust SMEFT analyses. Once again, we have shown that both simultaneous 
and conservative fits are able to break the degeneracy between the PDF and the SMEFT.

%% file: tables/statistics_ttbar_SMEFT_fit_methods.tex
\begin{table}[!htb]
    \centering
\renewcommand{\arraystretch}{1.25}
\begin{tabular}{|l|c|c|}
  \hline
\textbf{Fit}
  & \textbf{$\hat{Z} \times 10^4$}
  & \shortstack{$\Delta\chi^2$} \\
  \hline
  True PDF        & [1.73, 5.11] &  - \\
  BSM-biased PDF  & [-0.49, 4.35] & +5.5 (+7.3) \\
  Conservative PDF       & [0.50, 6.84]  & +0.1 (+0.2) \\
  Simultaneous fit       & [1.34, 6.62]  & +0.0  (+0.0) \\
  \hline
\end{tabular}
    \caption{Same as Table~\ref{tab:statistics_DY_SMEFT_fit_methods}, now for the
     $\hat{Z}$ fits to top observables shown in Fig.~\ref{fig:ttbar_smeft_fits}. 
    The quoted bounds should be compared to the injected BSM signal used to generate the synthetic data, corresponding to 
   $\hat{Z}=4 \times10^{-4}$. }
    \label{tab:statistics_ttbar_SMEFT_fit_methods}
\end{table}

%% file: sections/sec4-recommendations.tex
\section{Practical recommendations}
\label{sec:recommendations}

In this section, we provide practical recommendations to identify a degeneracy between PDFs and potential signs
of new physics. As we saw in Sect.~\ref{sec:pheno}, the fit quality of a BSM-biased PDF can be as good as the one
of a consistent fit, both in the DY and in the top sector. 
This makes it difficult to distinguish between the two without prior knowledge of the true 
underlying law. 
In this section we explore two approaches that can be used in real life -- where the underlying BSM model is unknown -- to identify possible PDF-induced bias.
In Sect.~\ref{subsec:energy_cuts}, 
we explore model-agnostic conservative PDF fits, while in Sect.~\ref{subsec:test_observables}, we discuss more in depth the nature 
of the degeneracy between the NP signals, which scales with energy, and the PDFs, which depend 
on the Bjorken $x$ and design a set of ratio observables at different centre-of-mass energies that could disentangle NP signals and PDF effects.

\input{sections/sec4_subsecs/subsec1_energy_cuts.tex}

\input{sections/sec4_subsecs/subsec2_test_observables.tex}

\input{sections/sec4_subsecs/subsec3_sector_observables.tex}

%% file: sections/sec4_subsecs/subsec1_energy_cuts.tex
\subsection{Implementing energy cut-offs}
\label{subsec:energy_cuts}

In Sect.~\ref{sec:pheno}, we  performed SMEFT fits by using some \emph{ad-hoc} conservative PDFs 
that we produced by removing from the global PDF analysis all the data affected by the \emph{known} BSM model 
that we were injecting in the synthetic data. In particular, in the case of Drell-Yan we removed  
all high invariant and transverse masses DY distributions, while in the case of top we removed all 
high--invariant mass distributions. 

In realistic scenarios, where the specific BSM dynamics responsible for potential distortions in the high-energy tails of 
measured distributions are unknown, it is generally unclear which datasets should be excluded in order to construct a 
conservative PDF fit free from BSM-induced biases. In this section, we adopt an agnostic strategy in which conservative 
PDF sets are obtained by performing fits that exclude all data above a given energy threshold, $Q_{\rm max}$. 
This approach exploits the fact that SMEFT-induced deviations typically grow with energy, allowing us to identify 
characteristic signatures in both the PDFs and the SMEFT constraints obtained using conservative PDFs as input sets.
If PDFs are effectively absorbing BSM-induced deviations in the observables, varying the energy cut is expected to modify 
these distortions in a way that renders the corresponding PDF fits mutually incompatible. By contrast, in the SM (baseline) case, 
the universality of PDFs implies that fits performed with different energy thresholds should remain consistent, with only an  
increase in uncertainties as higher-energy data are progressively removed.
We therefore expect the uncertainty bands to broaden as $Q_{\rm max}$ is lowered and the available dataset is reduced. 
One should be aware that if the energy cuts become too restrictive, extrapolation effects may lead to spurious behaviour, 
as insufficient data remain to constrain the PDFs in the large-$x$ region.
\begin{figure}[htb!]
    \centering
	\includegraphics[width=0.53\linewidth]{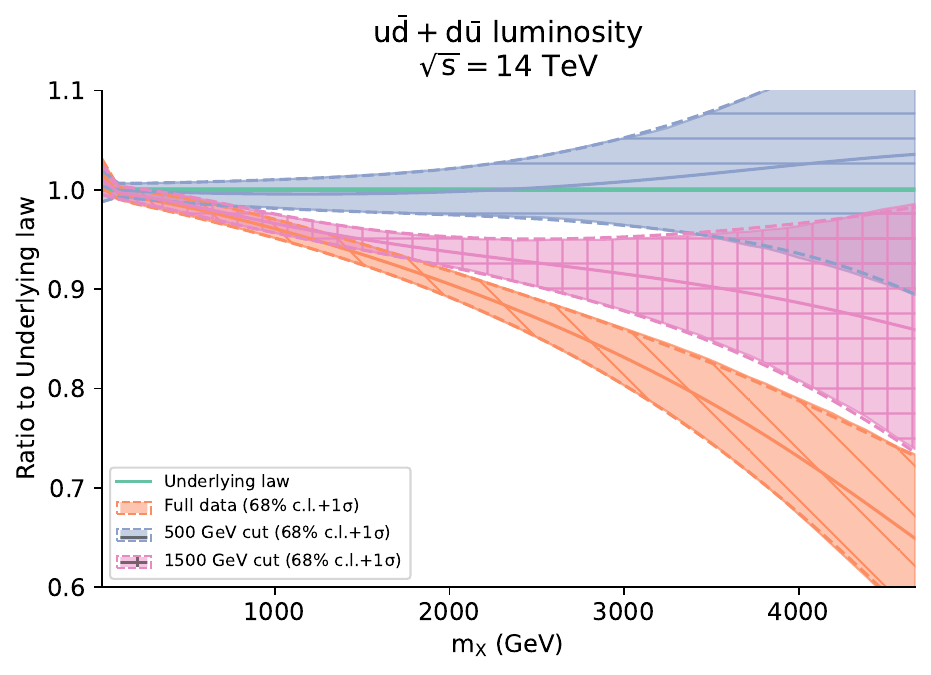}
	\includegraphics[width=0.46\linewidth]{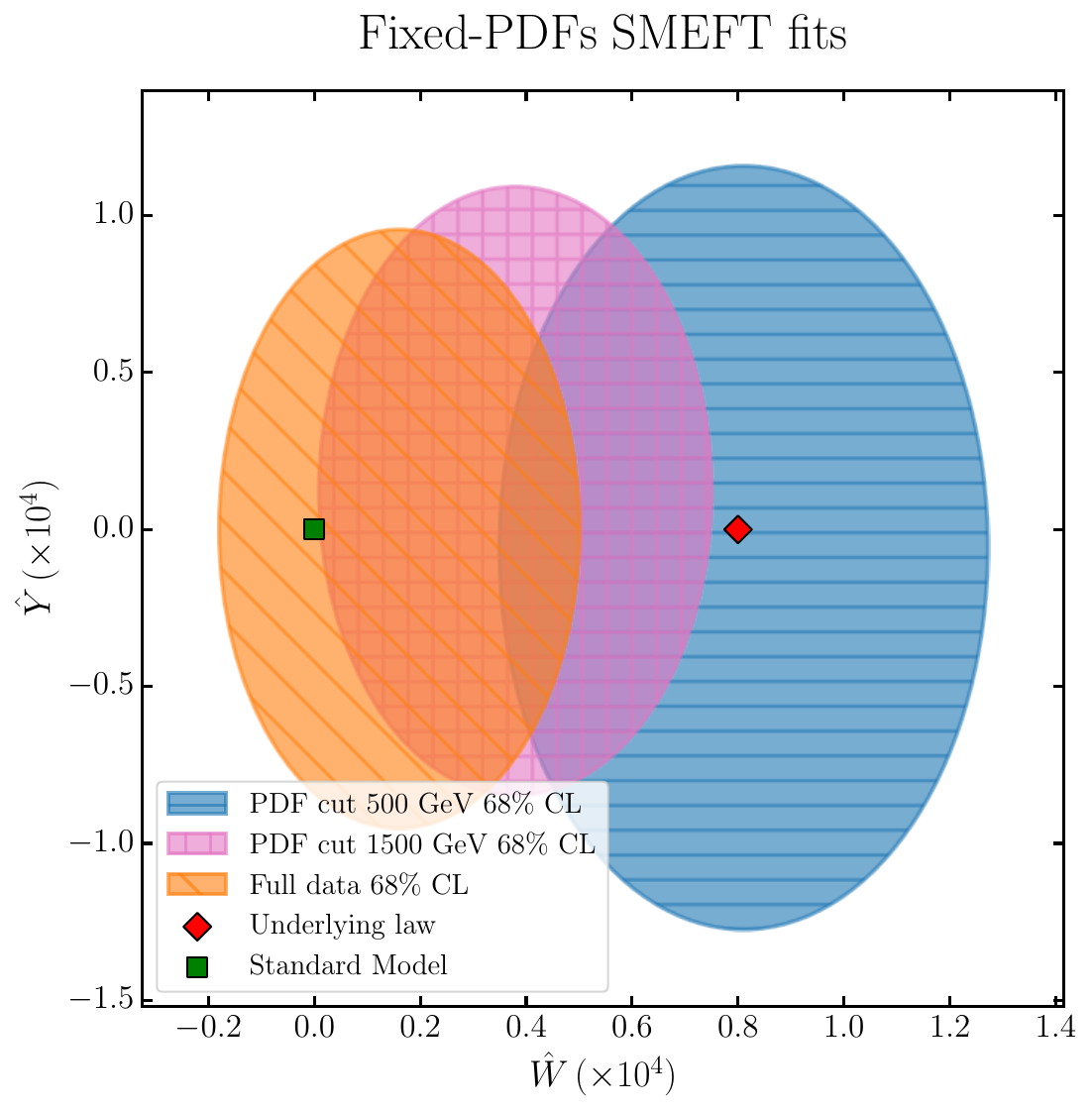} \\
	\includegraphics[width=0.53\linewidth]{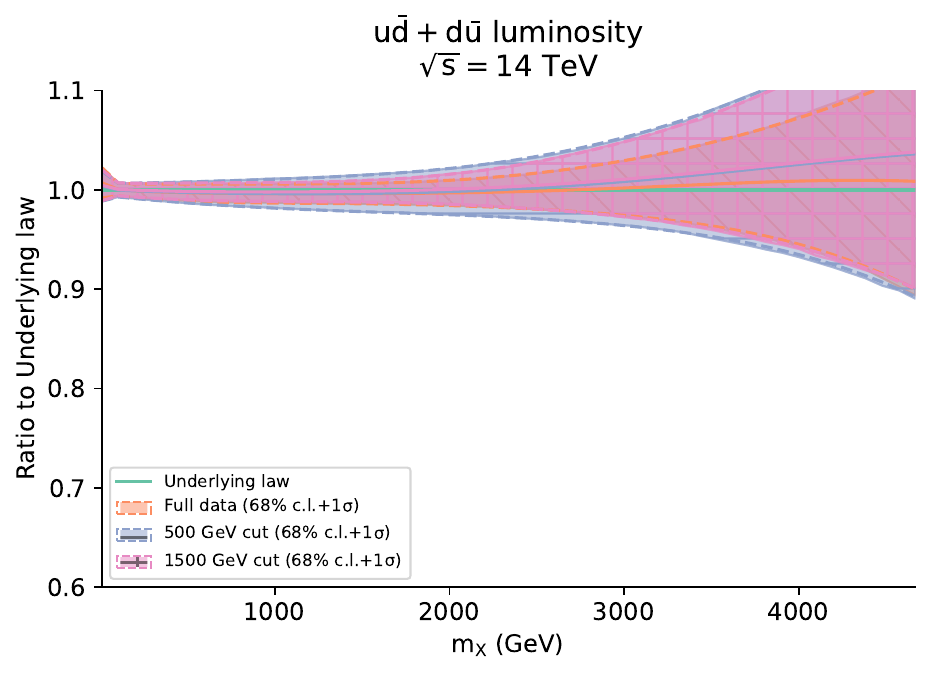}
	\includegraphics[width=0.46\linewidth]{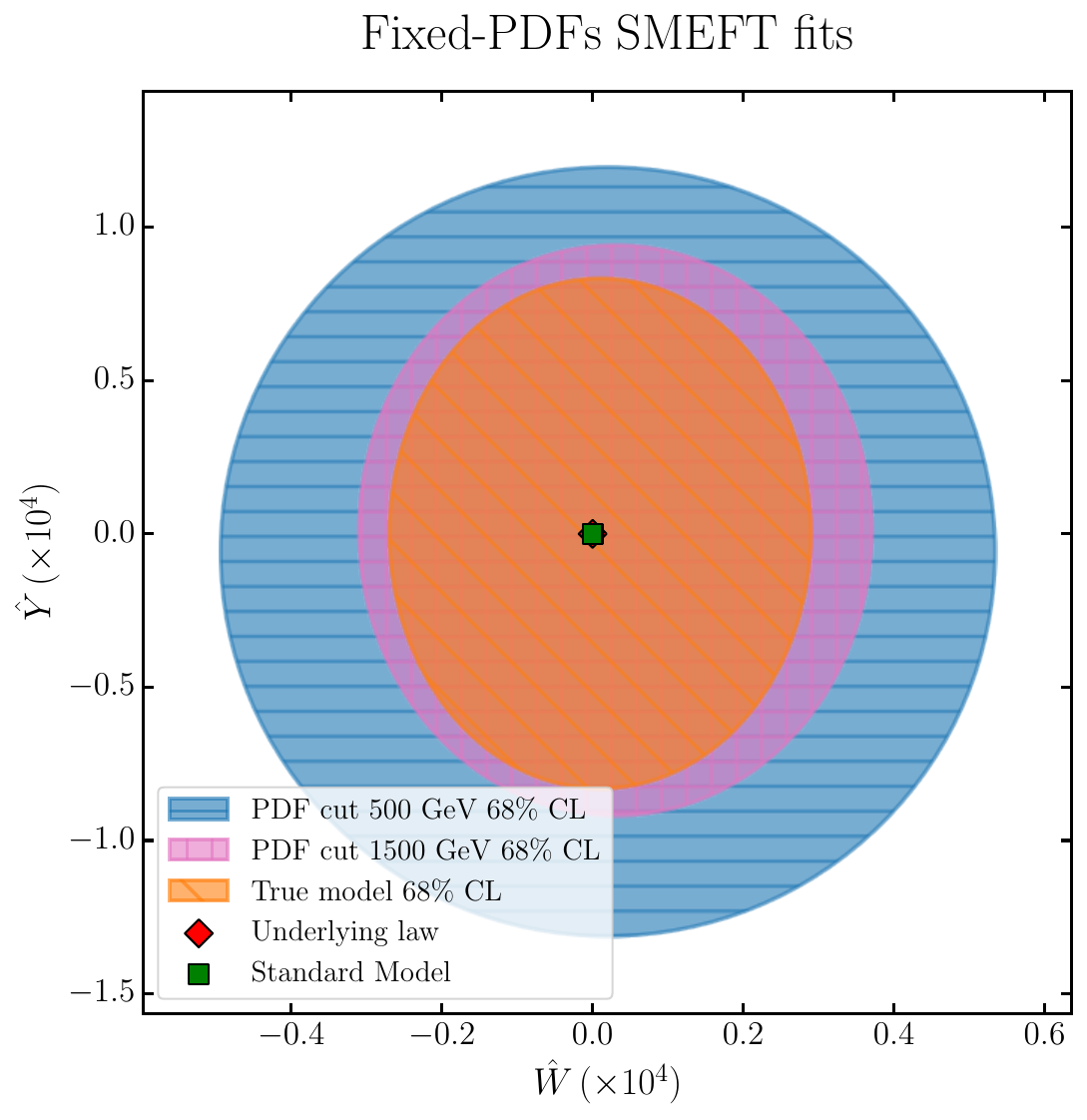}
	\caption{Impact of various energy cut $Q_{\rm max}$ in the input PDF sets. In the top row all synthetic data were generated by injecting 
    the BSM model discussed in Sect.~\ref{subsec:DY} (corresponding to $\hat{W}=8\cdot 10^{-5}$. 
    In the bottom row the synthetic data were generated assuming the SM. 
    On the left panels the relevant PDF luminosities $u \bar{d} + \bar{u} d$ 
    obtained by varying $Q_{\rm max}$ are compared in the two cases. 
    On the right panels the SMEFT bounds obtained in a SMEFT-only fit by using the PDFs with different $Q_{\rm max}$ as an input are compared. 
}
	\label{fig:energy_cuts_cont}
\end{figure}

We test this model-agnostic diagnostic strategy in the Drell--Yan scenario discussed in Sect.~\ref{subsec:DY}, with 
$\hat{W}=8\times10^{-5}$, for which both the injected BSM signal and the resulting PDF bias are known. 
Figure~\ref{fig:energy_cuts_cont} illustrates the impact of imposing global energy cuts at $Q_{\rm max}=500$ and 1500~GeV on 
the extracted PDFs and on the corresponding SMEFT fits obtained using these conservative PDF sets as an input.  
In the top panel, where the synthetic data include a BSM contribution, we observe that imposing a very conservative cut 
of $Q_{\rm max}=500$~GeV (blue band) completely removes the bias present when the full dataset is used (orange band), 
thereby restoring the correct underlying theory. By contrast, a milder cut at $Q_{\rm max}=1500$~GeV (pink band) is insufficient to eliminate the contamination. The progressive departure 
from the underlying law as increasingly NP-sensitive data are included is clearly visible. 
Such behaviour would not arise if the data were generated consistently within the SM, as illustrated in the bottom panel.  
As expected, the PDF uncertainty bands broaden as $Q_{\rm max}$ is lowered and more data are excluded from the fit. 
A similar pattern emerges in the SMEFT-only analyses performed using the increasingly conservative PDF sets, shown 
in the right panel. The SMEFT bounds obtained with the $Q_{\rm max}=500$~GeV PDFs are well centred on the underlying BSM value, 
albeit with substantially larger uncertainties. As the energy cut is relaxed, the bounds progressively shift towards the 
SM---ultimately excluding the correct BSM scenario---while becoming increasingly stringent.  
Finally, in the bottom panels of Fig.~\ref{fig:energy_cuts_cont}, we show that when the synthetic data are generated within the SM, 
the SMEFT bounds remain centred on the true values as $Q_{\rm max}$ is varied, with only a gradual increase in the uncertainties as higher-energy data are removed.

\begin{table}[htb!]
  \begin{center}
    \footnotesize

    \begin{tabular}{lccc}
      \toprule
      SMEFT Fit using & SM data & BSM data & $N_{\rm points}$ \\
      \midrule
      No cut  & ---  & 20.9 & 4363 \\
      $Q_{\rm max} =$ 1500 GeV  & -1.1 & 13.2 & 4339 \\
      $Q_{\rm max} =$ 500 GeV  & -1.00  & -1.2 & 4262 \\
      \bottomrule
    \end{tabular}
    \caption{\small Total $\Delta \chi^2$ of the SMEFT fits presented on the right panels
    of Fig.~\ref{fig:energy_cuts_cont} computed using fixed PDF sets trained on datasets cut at different values of $Q_{\rm max}$.
    Results are shown for fits to SM synthetic data and data with BSM signal injected (corresponding to $\hat{W}=8\times 10^{-5}$).
      \label{tab:global_chi2}
    }
  \end{center}
\end{table}
In Table~\ref{tab:global_chi2}, we report the variation in the global fit quality,
$\Delta\chi^2=\chi^2_{\rm fit}-\chi^2_{\rm baseline}$, obtained using the different PDF sets constructed by 
imposing a maximum energy cut $Q_{\rm max}$. The baseline reference corresponds, as usual, to the fit of SM synthetic data using SM predictions.  
Two distinct patterns emerge when comparing fits performed on synthetic data generated within the SM to those obtained from data containing an 
injected BSM signal. In the SM case, the choice of $Q_{\rm max}$ has essentially no impact on the fit quality. This behaviour is expected, 
since all datasets are mutually consistent by construction, being generated from the same underlying theory.  
By contrast, for synthetic data generated with an injected $\hat{W}=8\times10^{-5}$ signal, the fit performed without any high-energy cuts 
exhibits an increase in the global $\chi^2$ of about 21 units, indicating a mild tension. When an energy cut at 1500 GeV are imposed, 
the corresponding $\Delta\chi^2$ values decrease progressively, but remain positive. This behaviour can be understood from the top-left panel of 
Fig.~\ref{fig:energy_cuts_cont}: as the energy cut becomes more stringent, the PDF luminosity approaches the underlying law, yet still 
fails to reproduce it accurately. This reflects the fact that neither cut is sufficient to fully remove the BSM contamination present in the HL-LHC dataset.  
In contrast, the $Q_{\rm max}=500$~GeV cut successfully eliminates the residual BSM contribution. As a result, the associated PDF and SMEFT fits robustly 
recover the underlying theory, and the corresponding fit quality becomes comparable to that of the baseline case.

In summary, this relatively simple diagnostic test can be used to assess the risk that energy-growing deviations are being absorbed into the PDFs. 
A progressive shift of the type observed in Fig.~\ref{fig:energy_cuts_cont} would act as warning signal, indicating the presence of potential 
inconsistencies associated with energy-dependent effects, and hence possibly of BSM origin, in the dataset used for the PDF fit. 
Such behaviour would motivate the adoption of a conservative $Q_{\rm max}$ cut in order to ensure a robust and unbiased PDF determination.

%% file: sections/sec4_subsecs/subsec2_test_observables.tex
\subsection{Breaking Bjorken-$x$ and $M^2$ degeneracy}
\label{subsec:test_observables}

A second approach to identify a potential BSM-bias in the PDFs is to use test observables not present in the fit which break 
the degeneracy between the PDF and the SMEFT behaviour. In the case of a proton-proton collision producing a final state $A$, 
where $A$ could be for example a DY pair or a $t \bar{t}$ pair, collinear factorisation yields, up to power suppressed terms: 
\begin{equation}
    \sigma_{\rm pp\to A} (M^2) = \sum_{ij} \int  dx_1 \, dx_2 \, f^{(1)}_i  (x_1, \, \mu) f^{(2)}_j (x_2, \,  \mu )  \hat{\sigma}_{ij\to A} 
    (x_1, \, x_2, \, M^2, \, \mu),
\label{eq:pp_factorisation}
\end{equation}
where $f^{(1)}_i$ and $f^{(2)}_j$ are the PDFs of the two incoming protons, $x_1$ and $x_2$ are the Bj\"orken-$x$ of the partons involved
in the perturbative interaction ($ij\to A$), $M$ is the invariant mass of the final states produced in the hard scattering collision, 
and $\mu$ collectively indicate the factorisation and renormalisation scales of the process.

At tree level, since $x_1$ and $x_2$ correspond to the fraction of the momentum of the protons carried by the partons, conservation of energy simply yields
\begin{equation}
	M^2 = x_1  x_2 s,
\label{eq:M2_x1x2}
\end{equation}
where $\sqrt{s}$ is the centre-of-mass energy of the proton-proton collision.

The main point here is that the SMEFT effects grow with the energy scale $M$. The linear corrections stemming from 
dim-6 four fermion operators grow with $M^2 / \Lambda^2$, where $\Lambda$ is around the scale of the new physics, where the SMEFT expansion breaks.  
On the other hand, PDFs depend on $x$. The observed BSM-induced bias in the PDFs corresponds to a degeneracy between 
the $x$ and $M^2$ dependencies of the observables described in Eq.~\ref{eq:pp_factorisation} at a given centre-of-mass energy. 
If $\sqrt{s}$ changes, the same $M^2$ bin would correspond to different Bjorken-$x$ values. 
The argument on the dependence of the PDFs on the $M^2/s$ ratio 
has been considered in Ref.~\cite{Mangano:2012mh}, where examples of cross section ratios taken 
at different centre-of-mass energies were considered and their implication in terms of precision SM measurements 
was discussed.

\begin{figure}[!htb]
    \centering
	\includegraphics[width=0.7\linewidth]{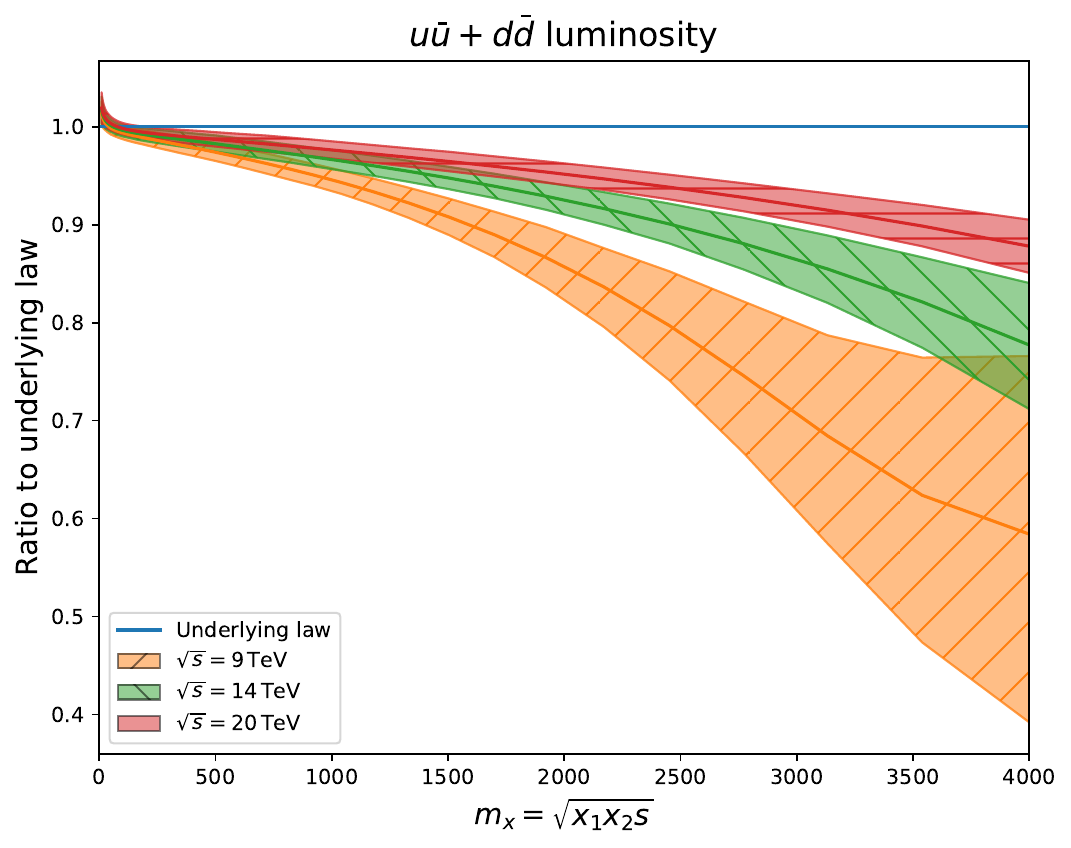}
	\caption{Ratio of the $u \bar{u} + d \bar{d}$ parton luminosity at different centre-of-mass energies with respect to the underlying law.  
    The input PDF set is given by the BSM-biased PDF mimicking the deviations caused by $\hat{W} = 8 \times 10^{-5}$ model in high--mass Drell--Yan observables 
    at $\sqrt{s} = 14$ TeV, discussed in Sect.~\ref{subsec:DY}.
	\label{fig:lumi_DY_different_CoM}
    }
\end{figure}
In Fig.~\ref{fig:lumi_DY_different_CoM}, we compare the $u\bar{u}+d\bar{d}$ parton luminosities obtained from the same BSM-biased 
PDF set, which mimics the deviations induced by a $\hat{W}=8\times10^{-5}$ scenario in high-mass Drell--Yan observables at $\sqrt{s}=14$~TeV, 
as discussed in Sect.~\ref{subsec:DY}. We show the ratios of these luminosities at different hadronic centre-of-mass energies, namely $\sqrt{s}=9$, 14, and 20~TeV, 
to the corresponding underlying law derived from SM synthetic data and SM theoretical predictions. The choices of 9 and 20~TeV are not motivated by specific experimental 
proposals, but are adopted here for illustrative purposes, in order to demonstrate how varying $\sqrt{s}$ can help break the degeneracy between the partonic 
momentum fractions and the invariant mass.  

We observe that the BSM-mimicking deviation, which develops from observables measured at $\sqrt{s}=14$~TeV, exhibits a pronounced dependence on the hadronic 
centre-of-mass energy. As $\sqrt{s}$ decreases, the deviation from the underlying law increases for fixed values of $M^2=x_1x_2s$, while the opposite trend 
is observed when $\sqrt{s}$ is increased.  
This behaviour can be understood from simple kinematic considerations. For a smaller centre-of-mass energy, for instance $\sqrt{s_{\rm small}}=9$~TeV compared 
to $\sqrt{s_{\rm large}}=14$~TeV, the partonic momentum fractions corresponding to a given invariant mass $M$ are larger, since $x_1x_2=M^2/s$, namely
\begin{equation}
	\frac{M^2}{s_{\rm small}} = x^{(s_{\rm small})}_1 x^{(s_{\rm small})}_2 >
	x^{(s_{\rm large})}_1 x^{(s_{\rm large})}_2 = \frac{M^2}{s_{\rm large}} \quad , \qquad \text{for fixed } M.
\end{equation}
Conversely, for fixed values of $x_1x_2$, the corresponding invariant mass rescales with the hadronic centre-of-mass energy as
\begin{equation}
	M^2(s) = x_1x_2\,s 
	\quad \Rightarrow \quad 
	M^2(s_{\rm large}) = M^2(s_{\rm small})\frac{s_{\rm large}}{s_{\rm small}} 
	\quad , \qquad \text{for fixed } x_1x_2.
\end{equation}

\begin{figure}[h!]
    \centering
	\includegraphics[width=0.49\linewidth]{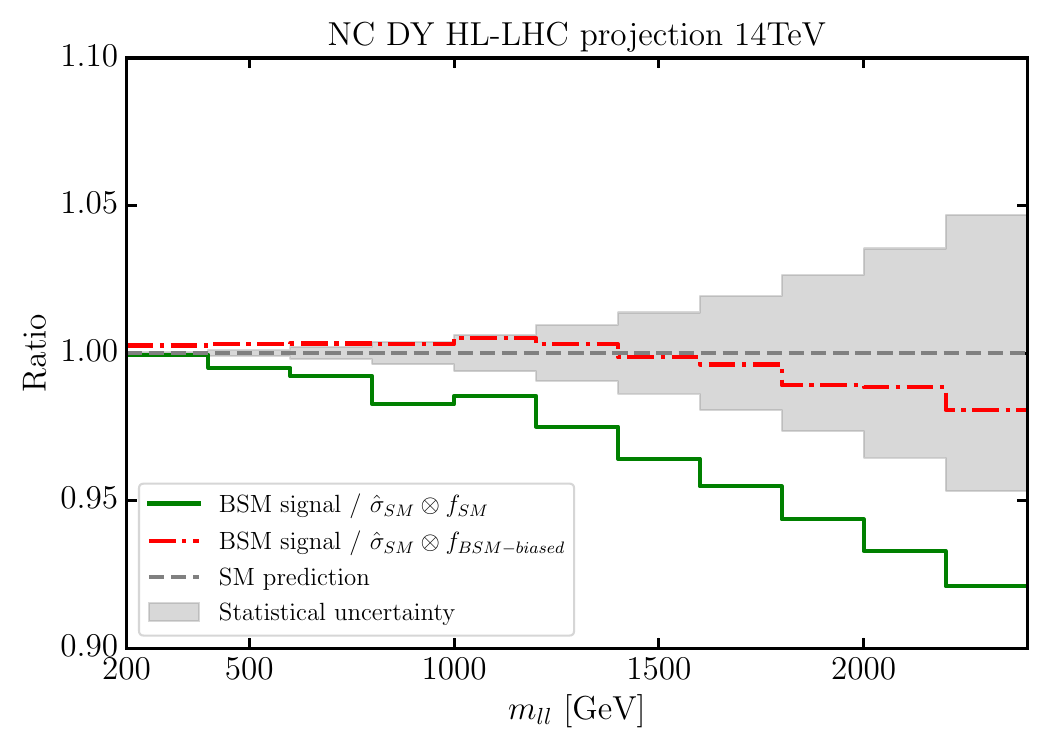}
	\includegraphics[width=0.49\linewidth]{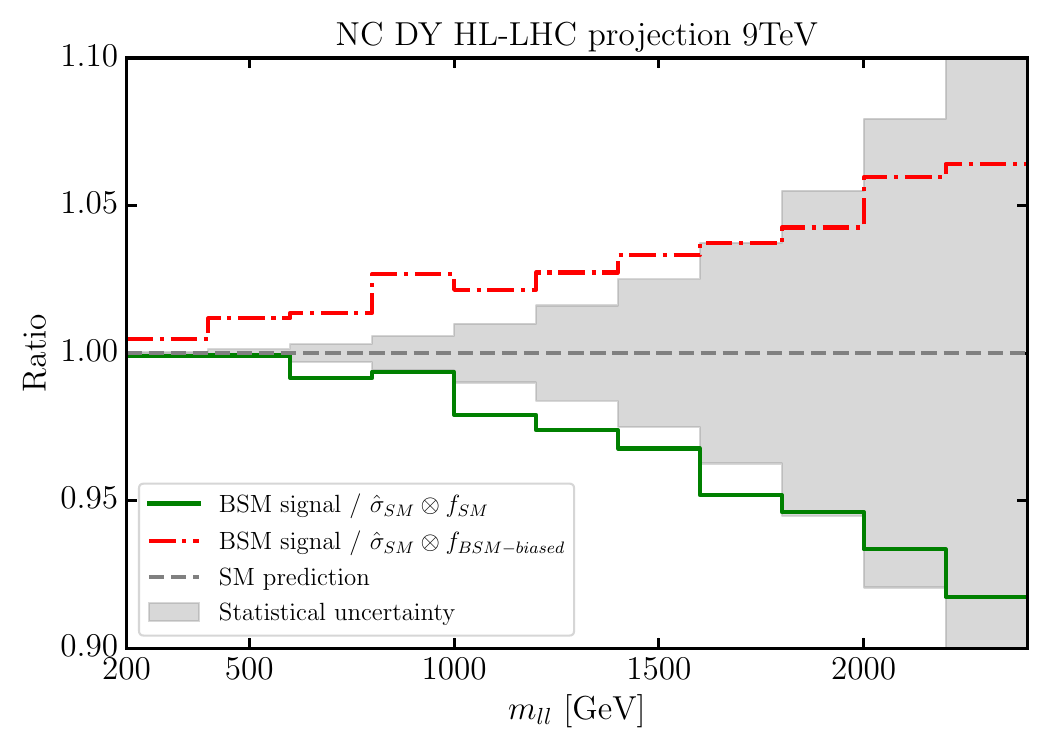}
	\caption{Predictions for NC DY ($pp\to l^- l^+$) inclusive cross sections
	differential in $M_{ll}$, with $l= (e,\mu)$ for $\sqrt{s} = 14$ TeV (left) and $\sqrt{s} = 9$ TeV (right).
	The BSM observables are compared to SM predictions using the true PDFs (green) and the BSM-biased PDFs (red).}
	\label{fig:prediction_DY_different_CoM}
\end{figure}
This feature can be exploited to construct test observables at centre-of-mass energies different from those used in the PDF fit. 
In Fig.~\ref{fig:prediction_DY_different_CoM}, we show the ratio of BSM Drell--Yan projections at the HL-LHC to the corresponding 
SM predictions, evaluated using both the true PDFs and the BSM-biased ones. We present leading-order projections for $\sqrt{s}=14$~TeV in 
the left panel and for $\sqrt{s}=9$~TeV in the right panel.  
For $\sqrt{s}=14$~TeV, the BSM-biased PDFs successfully mimic the genuine new-physics deviations, thereby making the observables 
appear spuriously consistent with the SM, whereas a clear discrepancy is visible when using the true PDFs. For $\sqrt{s}=9$~TeV, 
the SMEFT-induced deviations remain unchanged, as they depend only on the binning in the invariant mass $m_{\ell\ell}^2$. 
Consequently, the ratio of the BSM prediction to the SM expectation obtained with the true PDFs is unaffected. In contrast, 
the comparison based on the BSM-biased PDFs is no longer compatible with the SM. In this case, the PDFs overcompensate 
the SMEFT effects, as already illustrated in Fig.~\ref{fig:lumi_DY_different_CoM}, leading to an apparent deviation from 
the SM in the \emph{opposite direction} to the true one.  

We note that, at $\sqrt{s}=9$~TeV, the number of events at large $m_{\ell\ell}$ is significantly reduced compared to $\sqrt{s}=14$~TeV, 
resulting in larger statistical uncertainties in this region. For simplicity, we have assumed the full HL-LHC integrated luminosity 
for both centre-of-mass energies, although this does not correspond to a realistic experimental scenario. Nevertheless, 
realistic measurements can still disentangle PDF effects from SMEFT contributions. As shown in Ref.~\cite{Hammou:2024xuj}, 
deep-inelastic scattering data from the EIC~\cite{AbdulKhalek:2022hcn} and measurements at the Forward Physics Facility~\cite{FPF:2025bor} 
can help break this degeneracy. Other proposed experimental programmes, such as the LHeC~\cite{Ahmadova:2025vzd}, 
are expected to provide similar benefits. These measurements probe the same partonic channels as high-mass Drell--Yan processes, 
but at much higher rapidities, thereby breaking the alignment between $x$ and $M^2$ that characterises BSM-biased PDFs.

%% file: sections/sec4_subsecs/subsec3_sector_observables.tex
\subsection{Comparing different sectors}
\label{subsec:different_sectors}

A complementary strategy to disentangle PDF and SMEFT effects using test observables consists in exploiting the fact that different processes 
and kinematic sectors can be affected in qualitatively different ways by the same NP scenario. By comparing observables across such sectors, 
one can therefore construct powerful consistency tests that are sensitive to BSM contamination in PDF fits.

\begin{figure}[!ht]
    \centering
	\includegraphics[width=0.7\linewidth]{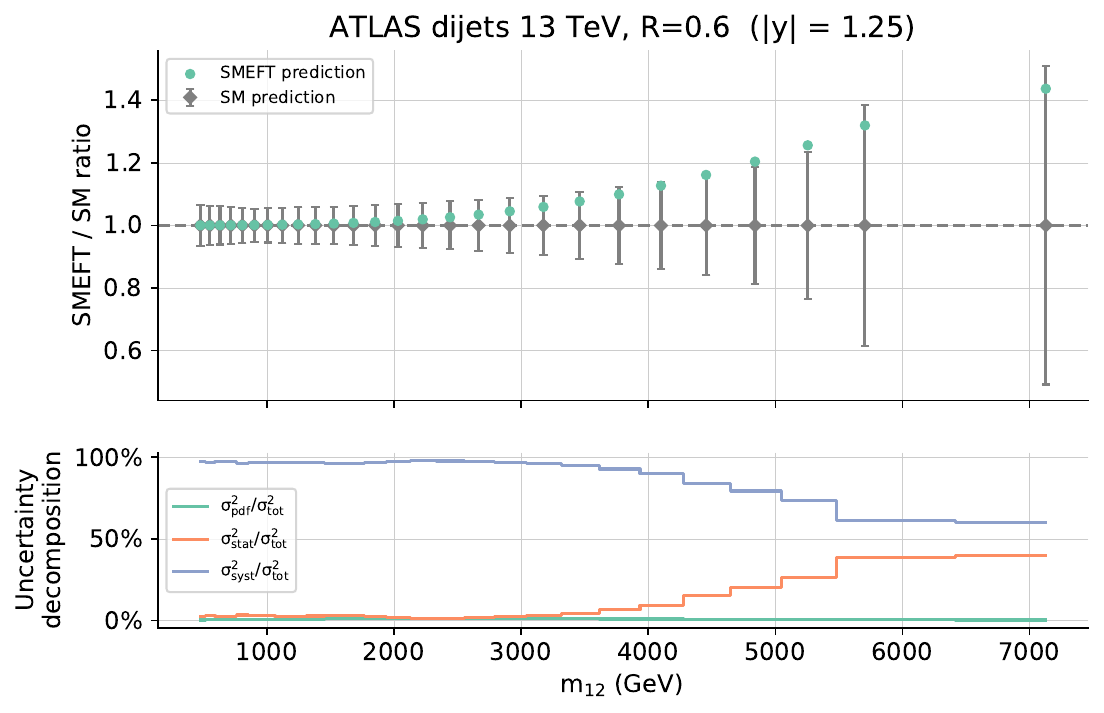}
	\caption{Top: Impact of the $\hat{Z} = 4 \times 10^{-4}$ model on the dijet invariant mass $m_{12}$ distribution in the $1.0 < |y^*| < 1.5$ bin measured by the ATLAS collaboration \cite{ATLAS:2017ble}. Bottom: uncertainty decomposition in terms of PDF (green), statistical (orange) and  systematic (blue) as ratio to the total uncertainty.
    }
	\label{fig:impact_dijet}
\end{figure}

As a representative example, in Fig.~\ref{fig:impact_dijet} we show the impact of the $\hat{Z}=4\times10^{-4}$ scenario discussed in Sect.~\ref{subsec:top} on a dijet observable \cite{ATLAS:2017ble}.
This measurement from ATLAS at $\sqrt{s} = 13$~TeV is presented as double differential cross sections in the invariant mass of the 2-jet system, $m_{12}$, and the absolute rapidity separation $|y^*|$ of the pair of jets in the 2-jet event.
SM theory predictions for these distributions have been released~\cite{Britzger:2022lbf} as \textsc{APPLfast} interpolation grids on the \textsc{Ploughshare} website\footnote{https://ploughshare.web.cern.ch/ploughshare}. They have been computed using the \textsc{NNLOjet} parton level event generator~\cite{NNLOJET:2025rno}. 
In~\cite{Chiefa:2025loi}, the aforementioned grids were converted in the \textsc{PineAPPL} format~\cite{Carrazza:2020gss}.
In contrast to the suppression of events in the $t\bar{t}$ invariant-mass distribution, illustrated in Fig.~\ref{fig:impact_z4}, the injection of the $\hat{Z}$ model in the dijet predictions produces an enhancement in the predicted event yield with respect to the SM expectation. 
This opposite response of the two sectors constitutes a highly non-trivial signature of the underlying BSM dynamics.
To our knowledge, such a direct and systematic comparison between top-quark and jet observables as a diagnostic tool for PDF--BSM interplay has not been explored in previous studies.
The fact that the same BSM scenario induces qualitatively different effects in these channels provides a novel and robust handle to test the consistency of PDF-based interpretations.

\begin{figure}[h!]
    \centering
	\includegraphics[width=0.7\linewidth]{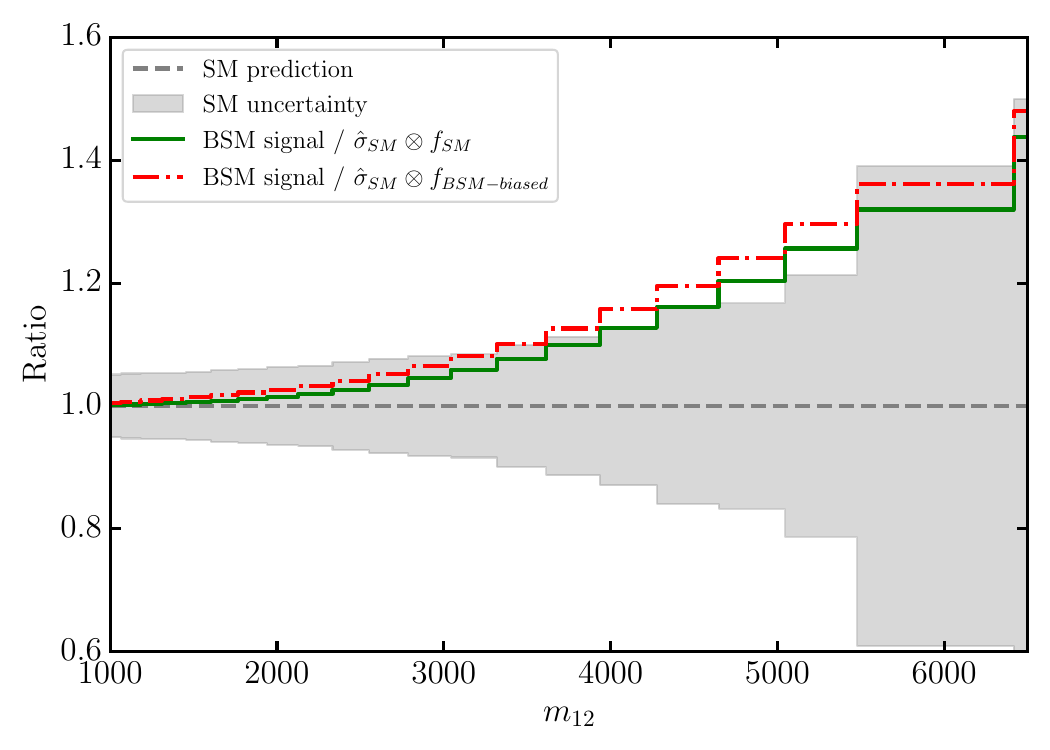}
	\caption{Predictions for dijet inclusive cross sections
	differential in the dijet invariant mass $m_{12}$.
	The BSM observables are compared to SM predictions using the true PDFs (green) and the BSM-biased PDFs (red).}
	\label{fig:prediction_dijet}
\end{figure}

In the present case, dijet observables were not included in the PDF fit, whereas $t\bar{t}$ data are. As a consequence, 
the BSM-biased PDFs absorb the suppression of high-energy events observed in the $t\bar{t}$ channel and effectively encode 
a decreasing behaviour with energy. When such PDFs are subsequently used to analyse dijet distributions, they no longer mask 
the BSM effects. Instead, they amplify the discrepancy with respect to the SM prediction, as shown in 
Fig.~\ref{fig:prediction_dijet}, thereby revealing the underlying inconsistency.

More generally, in Ref.~\cite{Hammou:2023heg} we demonstrated that BSM-biased PDFs can also induce artificial 
deviations in processes involving the same partonic channels as high-mass Drell--Yan production, but which are 
not directly affected by the same SMEFT operators, such as $WW$ production. Since these processes are measured 
at the same centre-of-mass energy, the PDF-induced distortions are comparable, while the SMEFT contributions differ. 
This mismatch leads to observable tensions that can be exploited as additional consistency checks.

Taken together, these examples illustrate that combining information from multiple sectors with distinct 
sensitivities to new physics provides a powerful means of disentangling PDF effects from genuine SMEFT contributions. 
While this strategy is intrinsically model dependent and requires a detailed understanding of how specific BSM scenarios 
impact different observables, it offers a valuable and largely unexplored avenue for improving the robustness of global BSM analyses.

%% file: sections/sec5-conclusion.tex
\section{Conclusions}
\label{sec:conclusion}

To summarise, in this work we have investigated two complementary strategies to disentangle the mixing between PDFs 
and BSM effects first observed in Ref.~\cite{Hammou:2023heg}, considering benchmark scenarios involving a new $W'$ and $Z'$ 
in the Drell--Yan sector, as well as a new $G'$ boson in top-quark pair production. Both approaches proved successful in 
mitigating BSM-induced biases in PDF-based analyses. These consist of the use of \emph{conservative} PDF fits, 
in which observables potentially affected by new physics are excluded by imposing an upper energy threshold $Q_{\rm max}$, 
and of fully \emph{simultaneous} fits of PDFs and SMEFT coefficients.

Conservative fits constrain the PDFs under the assumption of the SM by relying, in principle, only on SM-safe measurements. 
The resulting PDFs are then used as input in a subsequent SMEFT analysis. The main limitation of this approach lies 
in the choice of an energy cut that is sufficiently conservative to suppress BSM-induced biases, while not being 
so restrictive as to excessively inflate PDF uncertainties, which are then propagated to the SMEFT fit. 
In Sect.~\ref{sec:pheno}, we have shown that, in the presence of $W'$ and $Z'$ contributions to Drell--Yan production, 
conservative fits yield constraints that are competitive with those obtained from simultaneous fits. By contrast, 
in the case of a $G'$ affecting $t\bar{t}$ production, conservative fits lead to visibly weaker bounds. This indicates that 
the performance of this strategy is intrinsically dependent on the underlying model and on the sector under consideration.

Simultaneous fits, on the other hand, make use of the full dataset to constrain both PDFs and SMEFT coefficients, 
without assuming the SM during the fitting procedure. In this way, potential BSM-induced distortions of the PDFs are absorbed 
into the extended parameter space rather than biasing the PDF determination. As a result of the substantially larger number 
of trainable parameters, simultaneous fits typically exhibit broader uncertainties than idealised SMEFT-only fits 
based on the true PDFs, which are not accessible in practice. Nevertheless, as demonstrated in both the Drell--Yan and 
$t\bar{t}$ studies presented here, the inclusion of high-energy observables affected by BSM dynamics provides sufficient 
constraining power to compensate for this increase in dimensionality. In all scenarios considered, this approach proved 
effective in recovering both the PDFs and, most importantly, the underlying BSM dynamics.

It is important to stress, however, that simultaneous PDF--SMEFT fits cannot be straightforwardly extended 
to arbitrarily large operator bases, due to the rapidly increasing dimensionality of the parameter space 
and the associated computational and statistical challenges. In this respect, systematic strategies to identify 
the most relevant operators, such as those proposed in Ref.~\cite{Hirsch:2025qya}, play a crucial role in 
reducing the effective dimensionality of the problem and enabling scalable global analyses.

To summarise our findings in reference to the questions posed in Sect.~\ref{sec:intro}:
\begin{itemize}
    \item[(i)] The exclusion of data above a specific energy threshold $Q_{\rm max}$ represents a robust and general solution. 
    While this approach may lead to increased uncertainties in large-$x$ PDFs and, consequently, more conservative SMEFT bounds, it effectively 
    mitigates potential BSM-induced biases to yield reliable results. Furthermore, the availability of multiple PDF sets determined with 
    varying $Q_{\rm max}$ values provides a diagnostic tool to identify energy-growing deviations, which may signal a hidden interplay 
    between PDF evolution and BSM effects.  
    \item[(ii)] Simultaneous PDF--SMEFT fits deliver accurate and precise results for both the PDFs and the resulting SMEFT constraints. 
    However, this strategy relies on an \textit{a priori} selection of the BSM operators likely to be present in the data. In the absence of 
    such indications, using these ``SMEFT-aware'' PDFs directly for general collider predictions is not straightforward. Nevertheless, these 
    sets highlight a frequently overlooked source of theoretical uncertainty: the potential presence of BSM signals within high-energy data. 
    Quantifying this uncertainty and systematically incorporating it into global fits remains a subject for future investigation.
\end{itemize}

Finally, we have proposed three practical recommendations for identifying potential BSM-induced biases in 
PDF determinations, which can otherwise remain hidden, since the absorption of BSM effects does not necessarily 
lead to an observable deterioration of the fit quality. The first consists in performing PDF fits with different agnostic 
energy cuts and testing the mutual compatibility of the resulting PDFs and SMEFT constraints. Incompatibilities associated with 
energy-growing deviations may signal the presence of BSM contamination. Care must be taken, however, as excessively 
stringent cuts may induce spurious tensions through uncontrolled extrapolation in poorly constrained kinematic regions. 
The second recommendation is to exploit test observables measured at different centre-of-mass energies or in distinct rapidity ranges, 
thereby breaking the degeneracy between PDFs and BSM effects that exists at fixed kinematics. Finally, the third recommendation consists in devising 
test observables -- not included in a PDF fit -- so that combining information from different sectors with distinct 
sensitivities to new physics may enable us to disentangle PDF effects from genuine BSM contributions.

Overall, our results highlight the importance of systematically accounting for the interplay between PDFs and SMEFT 
in global analyses aimed at indirect searches for new physics. The methodologies developed in this work have been 
validated through closure tests based on consistent fixed-order synthetic data. 
The continuous development of novel Machine Learning (ML) techniques applied to high energy physics~\cite{Ubiali:2026myh}  
may provide additional tools to identify latent correlations between PDFs and SMEFT directions, and to construct optimised diagnostic observables.

A natural and important extension 
of this programme is the investigation of real experimental measurements that induce substantial shifts in PDF determinations, 
such as those in the jet sector, which is currently under active study. 
A further next step will be the application of 
these methods to full global PDF analyses based on real experimental data, including higher-order theoretical corrections. 
It will also be important to explore how these strategies can be integrated within PDF4LHC-style combinations and future community recommendations. 

%% file: appendices/app2_new_physics_CT.tex
\section{Description of the new physics scenarios}
\label{sec:NP_CT}

\input{appendices/app2_NP_scenarios_subapps/subapp1_wmodel}

\input{appendices/app2_NP_scenarios_subapps/subapp2_ymodel}

\input{appendices/app2_NP_scenarios_subapps/subapp2_zmodel}

%% file: appendices/app2_NP_scenarios_subapps/subapp1_wmodel.tex
\subsection{The $W'$ model and the $\hat{W}$ parameter}
\label{app:wmodel}

We consider an extension of the SM by a heavy $W'$ triplet transforming in the adjoint representation of $\text{SU}(2)_L$ and coupling universally to 
left-handed fermions and the Higgs doublet. The full details of the model are provided in Sect.~3 of \cite{Hammou:2023heg} and in \cite{Wells:2015uba}. 
The complete UV Lagrangian is:

\begin{equation}
	\begin{split}
		\mathcal{L}^{W'}_{\text{UV}} &= \mathcal{L}_{\text{SM}} - \frac{1}{4} {W'}^{a}_{\mu \nu} {W'}^{a, \mu \nu} + \frac{1}{2} M_{W'}^{2} {W'}_{\mu}^{a} {W'}^{a, \mu}\\
		&\qquad - g_{W'} {W'}^{a,\mu} \sum_{\substack{f_L}} \bar{f}_L T^{a} \gamma^{\mu} f_L
		 - g_{W'} ({W'}^{a, \mu} \varphi^{\dagger} T^{a} i D_{\mu} \varphi + \textrm{h.c.}) \, ,
	\end{split}
    \label{eq:Wprime_UV}
\end{equation}

For masses $M_{W'}$ greater than 10 TeV, we find that this model is well described by the dimension-6 SMEFT Lagrangian for 
the HMDY HL-LHC projections presented in Tab.~\ref{tab:hmdy_hllhc}. Furthermore, at the energies at which the BSM physics becomes relevant for the observables, 
the SMEFT four-fermion operators largely dominate the BSM contributions.
Therefore, in practice, the relevant 
SMEFT Lagrangian used throughout this study is:

\begin{equation}
	\mathcal{L}^{W'}_{\text{SMEFT}} = \mathcal{L}_{\text{SM}} -\frac{g_{W'}^{2}}{2 M^2_{W'}} J^{a, \mu}_L J^a_{L, \mu}, \qquad J^{a, \mu}_L = \sum_{\substack{f_L}} \bar{f_L} T^a \gamma^{\mu} f_L. \, 
    \label{eq:Wprime_EFT}
\end{equation}

We can measure the strength of the BSM signal by using the $\hat{W}$ dimensionless parameter:

\begin{equation}
	\mathcal{L}^{W'}_{\text{SMEFT}} = \mathcal{L}_{\text{SM}} -\frac{g^{2} \hat{W}}{2 m^2_W} J^{a, \mu}_L J^a_{L, \mu}, \qquad \hat{W} = \frac{g_{W'}^{2}}{g^{2}} \frac{m_{W}^{2}}{M_{W'}^{2}} \, \propto \mathcal{C}_{lq}^3 \text{ (Warsaw basis)},
\end{equation}

where $g$ and $m_{W}$ are the SM $W$ boson coupling and mass respectively.
If we assume that $g_{W'} = 1$, we have a direct connection between $\hat{W}$ and $M_{W'}$.
The limit $\hat{W} = 0$ corresponds to $M_{W'} \rightarrow \infty$, \emph{i.e.} 
the SM limit. On the other hand, a 
greater value of $\hat{W}$ corresponds to a lower mass of the $W'$, which 
would have a greater impact on the HL-LHC observables. 

%% file: appendices/app2_NP_scenarios_subapps/subapp2_ymodel.tex
\subsection{The $Z'$ model and the $\hat{Y}$ parameter}
\label{app:ymodel}

Similarly, we consider a $Z'$ field associated with a $U(1)_Y$ interaction. The full details of the model are also provided in Sect.~3 of \cite{Hammou:2023heg}
and in \cite{Wells:2015uba}. 
It extends the SM Lagrangian by the UV terms:

\begin{equation} 
\label{eq:Zprime}
	\begin{split}
		\mathcal{L}^{Z'}_{\text{UV}} &= \mathcal{L}_{\text{SM}} - \frac{1}{4} Z'_{\mu \nu} {Z'}^{\mu \nu} + \frac{1}{2} M_{Z'}^{2} Z'_{\mu} {Z'}^{\mu} \\
		&\qquad - g_{Z'} Z'_{\mu} \sum_{\substack{f}} Y_{f} \bar{f} \gamma^{\mu} f - Y_{\varphi} g_{Z'} ( Z'_{\mu} \varphi^{\dagger} i  D^{\mu} \varphi + \textrm{h.c.} ) \, .
	\end{split}
\end{equation}

As in the $W'$ case, we can match it to the SMEFT provided that $M_{Z'}$ is sufficiently larger than the experimental scale, which we have ensured throughout 
the study:

\begin{equation}
\label{eq:zprimeL}
	\mathcal{L}^{Z'}_{\text{SMEFT}} = \mathcal{L}_{\text{SM}} -\frac{g_{Z'}^{2}}{2 M_{Z'}^{2}} J^{\mu}_Y J_{Y, \mu}, \qquad J^{\mu}_Y = \sum_{\substack{f}} Y_{f} \bar{f} \gamma^{\mu} f \, .
\end{equation}

The dimensionless oblique parameter $\hat{Y}$ is used to describe the intensity of the new physics signal:

\begin{equation}
	\mathcal{L}^{Z'}_{\text{SMEFT}} = \mathcal{L}_{\text{SM}} -\frac{{g'}^2 \hat{Y}}{2 m^2_W} J^{\mu}_Y J_{Y, \mu}, \qquad \hat{Y} = \frac{g_{Z'}^{2}}{M_{Z'}^{2}} \frac{m^2_W}{{g'}^2} \, .
\end{equation}

%% file: appendices/app2_NP_scenarios_subapps/subapp2_zmodel.tex
\subsection{The coloron model $\mathcal{G}$ and the $\hat{Z}$ parameter}
\label{app:zmodel}

Finally, we consider a massive $\text{SU}(3)_c$ octet field $\mathcal{G}_a^\mu$, where $a$ is the colour index associated with the generators of the adjoint representation of the $\text{SU}(3)_c$ group.
This is referred to as the \emph{Coloron} model in the literature \cite{Chivukula:1996yr,Simmons:1996fz}. This model was not explored in \cite{Hammou:2023heg}, we will therefore provide some more details about it here.

The Coloron has properties similar to those of the gluon, and we might improperly refer to it as 
a `heavy gluon'. It is assumed that it acquires its mass $M_\mathcal{G}$ via a Higgs-like mechanism, which we do not discuss in detail. It couples to the $\text{SU}(3)_c$ current similarly to 
what the gluon does with a coupling constant $g_\mathcal{G}$. The UV Lagrangian extension reads:
\begin{equation}
    \mathcal{L}^\mathcal{G}_{\text{UV}} = \mathcal{L}_{\text{SM}} - \frac{1}{4} \mathcal{G}_{\mu\nu}^a \mathcal{G}^{a, \mu\nu} + \frac{1}{2} M_\mathcal{G}^2 \mathcal{G}_\mu^a \mathcal{G}^{a, \mu}  - g_\mathcal{G} \mathcal{G}_\mu^a \sum_q \bar{q} \gamma^\mu T^a q,
    \label{eq:coloron-lagrangian}
\end{equation}
where $q$ are the SM quarks, and $T^a$ the generators of the fundamental representation of $\text{SU}(3)_c$. The field strength tensor reads
\begin{equation}
    \mathcal{G}_{\mu\nu}^a = \partial_\mu \mathcal{G}_\nu^a - \partial_\nu \mathcal{G}_\mu^a + g_s f^{abc} \mathcal{G}_\mu^b \mathcal{G}_\nu^c.
\end{equation}

The matching of the UV complete Lagrangian $\mathcal{L}_\text{UV} = \mathcal{L}_{\text{SM}} + \Delta \mathcal{L}_\mathcal{G}$, where $\Delta \mathcal{L}_\mathcal{G}$ is defined in Eq.~\ref{eq:coloron-lagrangian}, to the corresponding dimension-6 SMEFT Lagrangian, produces the Warsaw basis operators in Tab.~\ref{tab:WilsonOperatorsCombined}, where we distinguished the operators
\begin{equation*}
    \left(\mathcal{O}_{qq}^{(1)}\right)_{ij} = (\bar{q}_i \gamma_\mu q_i) (\bar{q}_j \gamma^\mu q_j), \quad  \quad \left(\mathcal{O}_{qq}^{\prime (1)}\right)_{ij}= (\bar{q}_i \gamma_\mu q_j) (\bar{q}_j \gamma^\mu q_i),
\end{equation*}
and
\begin{equation*}
    \left(\mathcal{O}_{qq}^{(3)}\right)_{ij} = (\bar{q}_i \tau^a \gamma_\mu q_i) (\bar{q}_j \tau^a \gamma^\mu q_j) \quad, \quad \left(\mathcal{O}_{qq}^{\prime (3)}\right)_{ij}= (\bar{q}_i \tau^a \gamma_\mu q_j) (\bar{q}_j \tau^a \gamma^\mu q_i),
\end{equation*}
where $i,j$ are generation indices, and $\tau^a$ are the Pauli matrices, i.e. the generators of the $\text{SU}(2)_L$ group. In the literature, these operators are also respectively named \textit{Direct} and \textit{Exchange} \cite{Greljo:2023adz}.

\begin{table}[ht!]
    \centering
    \renewcommand{\arraystretch}{1.4}
    \begin{tabular}{|l|l|l|}
        \hline
        \textbf{4-fermion Operator} & \textbf{Warsaw basis} & \textbf{Top basis} \\
        \hline\hline
        $(\bar{L} L)(\bar{L} L)$ &
        $\mathcal{O}_{qq}^{(1)}$, $\mathcal{O}_{qq}^{\prime (1)}$, $\mathcal{O}_{qq}^{\prime (3)}$ &
        $\mathcal{O}_{Qq}^{(1,8)}$ \\
        \hline
        $(\bar{R} R)(\bar{R} R)$ &
        $\mathcal{O}_{uu}$, $\mathcal{O}_{uu}^{\prime}$, $\mathcal{O}_{ud}^{(8)}$ &
        $\mathcal{O}_{tu}^{(8)}$, $\mathcal{O}_{td}^{(8)}$ \\
        \hline
        $(\bar{L} L)(\bar{R} R)$ &
        $\mathcal{O}_{qu}^{(8)}$, $\mathcal{O}_{qd}^{(8)}$ &
        $\mathcal{O}_{qt}^{(8)}$, $\mathcal{O}_{Qu}^{(8)}$, $\mathcal{O}_{Qd}^{(8)}$ \\
        \hline
    \end{tabular}
    \caption{$\hat{Z}$ model Wilson operators generated by matching Eq.~\ref{eq:coloron-lagrangian} to the dimension-6 SMEFT, shown in the Warsaw and top bases.}
    \label{tab:WilsonOperatorsCombined}
\end{table}

Since we are interested in the top sector measurements, we restrict to a model with $\text{U}(2)^3 = \text{U}(2)_q \times \text{U}(2)_u \times \text{U}(2)_d$ 
flavour symmetry, under which only the light quarks transform:
\begin{equation}
    q \mapsto \text{U}_q \, q, \quad u \mapsto \text{U}_u \, u, \quad d \mapsto \text{U}_d \, d, \quad Q \mapsto Q, \quad t \mapsto t, \quad b \mapsto b
\end{equation}
where $q$ is $\text{U}(2)_q$ doublet containing the first two generations left-handed doublets, $u$ is the $\text{U}(2)_u$ doublet containing the first two up-type right-handed singlets, $d$ is the $\text{U}(2)_d$ doublet containing the first two  down-type right-handed singlets, $Q$ is the $\text{U}(2)^3$ singlet which contains the heavy quark left-handed doublet, while $t$ and $b$ are the two $\text{U}(2)^3$ singlets which contain the $\text{SU}(2)_L$ singlets. Note that we neglect potential flavour mixing effects for simplicity.

We neglect purely light–light and heavy–heavy four-quark operators, which do not contribute to the observables considered in this analysis.
The matching to the SMEFT then creates a point-like 
interaction between one heavy and one light quark $\text{SU}(3)_c$ currents:
\begin{equation}
    \mathcal{O}_{VV} = \sum_{i=1}^2 (\bar{Q} \gamma^\mu T^a Q + \bar{t} \gamma^\mu T^a t) (\bar{q}_i \gamma_\mu T^a q_i + \bar{u}_i \gamma_\mu T^a u_i + \bar{d}_i \gamma_\mu T^a d_i),
    \label{eq:Ovv}
\end{equation}
Therefore, the restriction to the smaller set of Wilson operators 
\cite{Ellis:2020unq, Aguilar-Saavedra:2018ksv} shown in Tab.~\ref{tab:WilsonOperatorsCombined} leads to the following definition of the SMEFT 
Lagrangian
\begin{equation}
    \mathcal{L}^{\mathcal{G}}_\text{SMEFT} = \mathcal{L}_\text{SM} - \frac{g_s^2 \hat{Z}}{2 m_W^2} \qty[\mathcal{O}_{Qq}^{(1,8)} + \mathcal{O}_{Qu}^{(8)}+ \mathcal{O}_{Qd}^{(8)} + \mathcal{O}_{tu}^{(8)} + \mathcal{O}_{td}^{(8)} + \mathcal{O}_{qt}^{(8)}],  \quad \hat{Z} = \frac{g_\mathcal{G}^2}{g_s^2} \frac{m_W^2}{M_\mathcal{G}^2},
    \label{eq:smeft-lagrangian}
\end{equation}
where $\hat{Z}$ is the parameter which tunes the intensity of the new physics signature.

%% file: appendices/app3_data.tex
\section{Presentation of the dataset}
\label{app:data}
In this appendix, we present the real measurements used in the analysis of this project,
mostly coming from the NNPDF \cite{NNPDF:2021njg} and  \simunet{}~\cite{Costantini:2024xae} implementations. 
The HL-LHC projections have already been described in the main body of this paper in Tab.~\ref{tab:hmdy_hllhc}, 
\ref{tab:lhcb_dy_hllhc} and \ref{tab:ttbar_hllhc}.

\input{tables/data.tex}

%% file: tables/data.tex
\begin{table}[H]
    \centering
    \footnotesize
    \begin{tabularx}{\textwidth}{Xcccc}
    \toprule
    Dataset & N$_\text{dat}$ & $x$ & $Q$ [GeV] & Ref. \\
    \midrule
    NMC $d/p$                               & 121 & $[0.002000, 0.675]$ & $[0.400, 9.951]$ & \cite{Arneodo:1996kd} \\
    NMC $p$                                 & 204 & $[0.004000, 0.479]$ & $[0.894, 7.895]$ & \cite{Arneodo:1996qe} \\
    SLAC $p$                                &  33 & $[0.070000, 0.850]$ & $[0.762, 5.400]$ & \cite{Whitlow:1990gk,Whitlow:1991uw} \\
    SLAC $d$                                &  34 & $[0.070000, 0.850]$ & $[0.762, 5.393]$ & \cite{Whitlow:1990gk,Whitlow:1991uw} \\
    BCDMS $p$                               & 333 & $[0.070000, 0.750]$ & $[2.739, 15.16]$ & \cite{Benvenuti:1989rh} \\
    BCDMS $d$                               & 248 & $[0.070000, 0.750]$ & $[2.958, 15.16]$ & \cite{Benvenuti:1989fm} \\
    CHORUS $\sigma_{CC}^{\nu}$              & 416 & $[0.020000, 0.650]$ & $[0.513, 9.757]$ & \cite{Onengut:2005kv} \\
    CHORUS $\sigma_{CC}^{\bar{\nu}}$        & 416 & $[0.020000, 0.650]$ & $[0.513, 9.757]$ & \cite{Onengut:2005kv} \\
    NuTeV $\sigma_{c}^{\nu}$                &  39 & $[0.021000, 0.326]$ & $[1.062, 10.79]$ & \cite{Goncharov:2001qe} \\
    NuTeV $\sigma_{c}^{\bar{\nu}}$          &  37 & $[0.015000, 0.207]$ & $[0.875, 8.266]$ & \cite{Goncharov:2001qe} \\
    HERA I+II incl. NC $e^-p$               & 159 & $[0.000800, 0.650]$ & $[7.746, 223.6]$ & \cite{Abramowicz:2015mha} \\
    HERA I+II incl. NC $e^+p$ 460 GeV       & 204 & $[0.000046, 0.650]$ & $[1.414, 28.28]$ & \cite{Abramowicz:2015mha} \\
    HERA I+II incl. NC $e^+p$ 575 GeV       & 254 & $[0.000028, 0.650]$ & $[1.225, 28.28]$ & \cite{Abramowicz:2015mha} \\
    HERA I+II incl. NC $e^+p$ 820 GeV       &  70 & $[0.000001, 0.400]$ & $[0.212, 173.2]$ & \cite{Abramowicz:2015mha} \\
    HERA I+II incl. NC $e^+p$ 920 GeV       & 377 & $[0.000005, 0.650]$ & $[0.387, 173.2]$ & \cite{Abramowicz:2015mha} \\
    HERA I+II incl. CC $e^-p$               &  42 & $[0.008000, 0.650]$ & $[17.32, 173.2]$ & \cite{Abramowicz:2015mha} \\
    HERA I+II incl. CC $e^+p$               &  39 & $[0.008000, 0.400]$ & $[17.32, 173.2]$ & \cite{Abramowicz:2015mha} \\
    HERA comb. $\sigma_{c\bar c}^{\rm red}$ &  37 & $[0.000000, 0.050]$ & $[1.581, 141.4]$ & \cite{H1:2018flt} \\
    HERA comb. $\sigma_{b\bar b}^{\rm red}$ &  26 & $[0.000000, 0.050]$ & $[1.581, 141.4]$ & \cite{H1:2018flt} \\
    \midrule
    Total & 3089 & & & \\
    \bottomrule
    \end{tabularx}
    \caption{Deep inelastic scattering datasets. $F_2$ structure function, neutral- and charged-current cross-sections.}
    \label{tab:dis_datasets}
\end{table}

\begin{table}[H]
    \centering
    \footnotesize
    \begin{tabularx}{\textwidth}{Xcccc}
    \toprule
    Dataset & N$_\text{dat}$ & $-$ & $p_T$ [GeV] & Ref. \\
    \midrule
    ATLAS $W^+$+jet 8 TeV & 15 & $-$ &$[12.5, 700]$ & \cite{Aaboud:2017soa} \\
    ATLAS $W^-$+jet 8 TeV & 15 & $-$ &$[12.5, 700]$ & \cite{Aaboud:2017soa} \\
    \midrule
    Total & 30 & & & \\
    \bottomrule
    \end{tabularx}
    \caption{Associated $W^\pm$ vector bosons and jet production cross-section differential in the transverse momentum of the produced vector boson $p_T^{W^\pm}$.}
    \label{tab:W_jet_datasets}
\end{table}

\begin{table}[H]
    \centering
    \footnotesize
    \begin{tabularx}{\textwidth}{Xcccc}
    \toprule
    Dataset & N$_\text{dat}$ & $y$ & $Q$ [GeV] & Ref. \\
    \midrule
    DYE 866 $\sigma^d_{\rm DY}/\sigma^p_{\rm DY}$ & 15 & $[0.069, 1.530]$ & $[4.6, 12.9]$ & \cite{Towell:2001nh} \\
    DY E886 $\sigma^p_{\rm DY}$                   & 89 & $[0.000, 1.908]$ & $[4.4, 15.8]$ & \cite{Webb:2003ps} \\
    DY E605 $\sigma^p_{\rm DY}$                   & 85 & $[-0.20, 0.400]$ & $[7.1, 16.9]$ & \cite{Moreno:1990sf} \\
    DYE 906 $\sigma^d_{\rm DY}/\sigma^p_{\rm DY}$ &  6 & $[0.107, 0.772]$ & $[4.7, 6.36]$ & \cite{Dove:2021ejl} \\
    \midrule
    Total & 195 & & & \\
    \bottomrule
    \end{tabularx}
    \caption{Fixed target Drell-Yan process datasets.}
    \label{tab:fixed_target_dy_datasets}
\end{table}

\begin{table}[H]
    \centering
    \footnotesize
    \begin{tabularx}{\textwidth}{Xcccc}
    \toprule
    Dataset & N$_\text{dat}$ & $\eta_{ll}$ & $M_{ll}$ [GeV] & Ref. \\
    \midrule
    ATLAS low-mass DY 2011 $d\sigma_{Z/\gamma^{*}}/dM_{ll}$   & 6 & $[-2.10, 2.10]$ & $[14.5, 56.0]$ & \cite{Aad:2014qja} \\
    ATLAS high-mass DY 7 TeV $d\sigma_{Z/\gamma^{*}}/dM_{ll}$ & 5 & $[-2.10, 2.10]$ & $[123, 1250]$  & \cite{Aad:2013iua} \\
    \midrule
    Total & 11 & & & \\
    \bottomrule
\end{tabularx}
    \caption{Neutral-current Drell-Yan cross-section differential in the invariant mass $M_{ll}$ of the produced lepton-pair.}
    \label{tab:nc_dy_diff_mll_datasets}
\end{table}

\begin{table}[H]
    \centering
    \footnotesize
    \begin{tabularx}{\textwidth}{Xcccc}
    \toprule
    Dataset & N$_\text{dat}$ & $-$ & $p_T$ [GeV] & Ref. \\
    \midrule
    ATLAS $Z$ $p_T$ 8 TeV $(p_T^{ll},M_{ll})$ & 44 & $-$ & $[1, 550]$ & \cite{Aad:2015auj} \\
    \midrule
    Total & 44 & & & \\
    \bottomrule
    \end{tabularx}
    \caption{Neutral-current Drell-Yan cross-section differential in the transverse momentum of the produced vector boson $p_T^Z$, i.e. the transverse momentum of the outgoing lepton pair $p_T^{ll}$.}
    \label{tab:nc_dy_diff_pt_datasets}
\end{table}

\begin{table}[H]
    \centering
    \footnotesize
    \begin{tabularx}{\textwidth}{Xcccc}
    \toprule
    Dataset & N$_\text{dat}$ & $y_{ll}$ & $p_T^{ll}$ [GeV] & Ref. \\
    \midrule
    ATLAS $Z$ $p_T$ 8 TeV $(p_T^{ll},y_{ll})$ & 48 & $[0.2, 2.2]$ & $[1, 550]$  & \cite{Aad:2015auj} \\
    CMS $Z$ $p_T$ 8 TeV $(p_T^{ll},y_{ll})$   & 28 & $[0.2, 1.8]$ & $[10, 600]$ & \cite{Khachatryan:2015oqa} \\
    \midrule
    Total & 76 & & & \\
    \bottomrule
    \end{tabularx}
    \caption{Neutral-current Drell-Yan cross-section double differential in the transverse momentum $p_T^{ll}$ and the rapidity $y_{ll}$ of the produced lepton-pair.}
    \label{tab:nc_dy_diff_ptll_yll_datasets}
\end{table}

\begin{table}[H]
    \centering
    \footnotesize
    \begin{tabularx}{\textwidth}{Xcccc}
    \toprule
    Dataset & N$_\text{dat}$ & $y$ & $M$ [GeV] & Ref. \\
    \midrule
    CDF $Z$ rapidity                        &  28 & $[0.050, 2.800]$ & $M_Z$            & \cite{Aaltonen:2010zza} \\
    D0 $Z$ rapidity                         &  28 & $[0.050, 2.750]$ & $M_Z$            & \cite{Abazov:2007jy} \\
    D0 $W\to \mu\nu$ asymmetry              &   9 & $[0.100, 1.880]$ & $M_W$            & \cite{Abazov:2013rja} \\
    ATLAS $W,Z$ 7 TeV 2010                  &  30 & $[0.105, 3.200]$ & $[M_W, M_Z]$     & \cite{Aad:2011dm} \\
    ATLAS $W,Z$ 7 TeV 2011 Central sel.     &  46 & $[0.100, 2.340]$ & $[56, 133]$      & \cite{Aaboud:2016btc} \\
    ATLAS $W,Z$ 7 TeV 2011 Forward sel.     &  15 & $[1.300, 3.400]$ & $[91, 133]$      & \cite{Aaboud:2016btc} \\
    ATLAS DY 2D 8 TeV high mass             &  48 & $[0.100, 2.300]$ & $[133, 1000]$    & \cite{Aad:2016zzw} \\
    ATLAS DY 2D 8 TeV low mass              &  60 & $[0.100, 2.300]$ & $[56, 175]$      & \cite{Aaboud:2017ffb} \\
    CMS $W$ asymmetry 840 pb                &  11 & $[0.100, 2.300]$ & $M_W$            & \cite{Chatrchyan:2012xt} \\
    CMS $W$ asymmetry 4.7 fb                &  11 & $[0.100, 2.250]$ & $M_W$            & \cite{Chatrchyan:2013mza} \\
    CMS Drell-Yan 2D 7 TeV 2011             & 110 & $[0.050, 2.350]$ & $[25, 850]$      & \cite{Chatrchyan:2013tia} \\
    CMS $W$ rapidity 8 TeV                  &  22 & $[0.100, 2.250]$ & $M_W$            & \cite{Khachatryan:2016pev} \\
    LHCb $Z$ 940 pb                         &   9 & $[2.125, 4.125]$ & $M_Z$            & \cite{Aaij:2012mda} \\
    LHCb $Z\to ee$ 2 fb                     &  17 & $[2.062, 4.125]$ & $M_Z$            & \cite{Aaij:2015vua} \\
    LHCb $W,Z \to \mu$ 7 TeV                &  29 & $[2.062, 4.250]$ & $[M_W, M_Z]$     & \cite{Aaij:2015gna} \\
    LHCb $W,Z \to \mu$ 8 TeV                &  30 & $[2.062, 4.375]$ & $[M_W, M_Z]$     & \cite{Aaij:2015zlq} \\
    LHCb $Z\to \mu\mu$                      &  16 & $[2.062, 4.375]$ & $M_Z$            & \cite{Aaij:2016mgv} \\
    LHCb $Z\to ee$                          &  15 & $[2.062, 4.125]$ & $M_Z$            & \cite{Aaij:2015vua} \\
    \midrule
    Total & 534 & & & \\
    \bottomrule
    \end{tabularx}
    \caption{Drell-Yan datasets.}
    \label{tab:dy_datasets}
\end{table}

\begin{table}[H]
    \centering
    \footnotesize
    \begin{tabularx}{\textwidth}{Xcccc}
        \toprule
        Dataset & N$_\text{dat}$ & $\eta$ & $p_T$ [GeV] & Ref. \\
        \midrule
        ATLAS jets 8 TeV, R=0.6     & 171 & $[0.250, 2.750]$ & $[77.50, 2246]$ & \cite{Aaboud:2017dvo} \\
        CMS jets 8 TeV              & 185 & $[0.250, 2.750]$ & $[22.50, 2308]$ & \cite{Khachatryan:2016mlc} \\
        \midrule
        Total & 611 & & & \\
        \bottomrule
    \end{tabularx}
    \caption{Inclusive-jet cross-section double differential in the jet transverse momentum $p_T$ and the jet absolute rapidity $\abs{y}$.}
    \label{tab:inclusive_jet_datasets}
\end{table}

\begin{table}[H]
    \centering
    \footnotesize
    \begin{tabularx}{\textwidth}{Xcccc}
        \toprule
        Dataset & N$_\text{dat}$ & $\eta$ & $M_\text{jj}$ [GeV] & Ref. \\
        \midrule
        ATLAS dijets 7 TeV, R=0.6   & 90 & $[0.25, 2.75]$ & $[285.0, 4485]$ & \cite{Aad:2013tea} \\
        CMS dijets 7 8TeV           & 54 & $[0.25, 2.25]$ & $[246.5, 4381]$ & \cite{Chatrchyan:2012bja} \\
        \midrule
        Total & 144 & & & \\
        \bottomrule
        \end{tabularx}
    \caption{Dijets cross-section differential in the invariant mass of the jet-pair $M_\text{jj}$ and the absolute rapidity separation of the jets $\abs{y^*}$.}
    \label{tab:dijets_datasets}
\end{table}

\begin{table}[H]
    \centering
    \footnotesize
    \begin{tabularx}{\textwidth}{Xcccc}
        \toprule
        Dataset & N$_\text{dat}$ & $\eta_\gamma$ & $E_T^\gamma$ [GeV] & Ref. \\
        \midrule
        ATLAS direct photon production 13 TeV & 53 & $[0.125, 0.385]$ & $[137.5, 1300]$ & \cite{ATLAS:2017nah} \\
        \midrule
        Total & 53 & & & \\
        \bottomrule
    \end{tabularx}
    \caption{Direct photon production cross-section differential in the photon transverse energy $E_T^\gamma$.}
    \label{tab:direct_photon_datasets}
\end{table}

\begin{table}[H]
    \centering
    \footnotesize
    \begin{tabularx}{\textwidth}{Xcccc}
    \toprule
    Dataset & N$_\text{dat}$ & $-$ & $M_{t\bar{t}}$ [GeV] & Ref. \\
    \midrule
    ATLAS $t\bar{t}$ 8 TeV $m_{t\bar{t}}$ normalised    &  5 & $-$ & $[350.0, 1850.0]$ & \cite{Aaboud:2016iot} \\
    ATLAS $t\bar{t}$ 13 TeV $m_{t\bar{t}}$ normalised   &  8 & $-$ & $[362.5, 1750.0]$ & \cite{Aad:2019ntk} \\
    CMS $t\bar{t}$ 8 TeV $m_{t\bar{t}}$ normalised      &  5 & $-$ & $[340.0, 1300.0]$ & \cite{Sirunyan:2018ucr} \\
    CMS $t\bar{t}$ 13 TeV $m_{t\bar{t}}$ normalised     & 14 & $-$ & $[325.0, 2900.0]$ & \cite{CMS:2021vhb} \\
    \midrule
    ATLAS 8TeV $t\bar{t}$ asymmetry                     &  1 & $-$ & $m_t$             & \cite{Aad:2016ove} \\
    ATLAS 13TeV $t\bar{t}$ asymmetry                    &  5 & $-$ & $[250.0, 1750.0]$ & \cite{ATLAS:2022waa} \\
    CMS 8TeV $t\bar{t}$ asymmetry                       &  3 & $-$ & $[215.0, 5264.5]$ & \cite{Khachatryan:2016ysn} \\
    CMS 13TeV $t\bar{t}$ asymmetry                      &  3 & $-$ & $[350.0, 1350.0]$ & \cite{CMS-PAS-TOP-21-014} \\
    ATLAS/CMS 8TeV $t\bar{t}$ asymmetry                 &  6 & $-$ & $[210.0,  975.0]$ & \cite{Sirunyan:2017lvd} \\
    \midrule
    Total & 50 & & & \\
    \bottomrule
    \end{tabularx}
    \caption{$t\bar{t}$ cross-section differential in the invariant mass of the produced top-pair $M_{t\bar{t}}$, and $t\bar{t}$ charge asymmetry inclusive at 8 TeV and differential in the top-pair invariant mass $M_{t\bar{t}}$.}
    \label{tab:ttbar_xs_asy_diff_datasets}
\end{table}

\begin{table}[H]
    \centering
    \footnotesize
    \begin{tabularx}{\textwidth}{Xcccc}
        \toprule
        Dataset & N$_\text{dat}$ & $-$ & $p_T$ [GeV] & Ref. \\
        \midrule
        ATLAS 13TeV $t\bar{t}Z$ & 5 & $-$ & $[20.0, 255]$ & \cite{ATLAS:2021fzm} \\
        CMS 13TeV $t\bar{t}Z$   & 3 & $-$ & $[37.5, 375]$ & \cite{CMS:2019too} \\
        \midrule
        Total & 8 & & & \\
        \bottomrule
    \end{tabularx}
    \caption{Associated $t\bar{t}$ and $Z$ boson production differential in the $Z$ boson transverse momentum $p_T^Z$ datasets.}
    \label{tab:ttbar_Z_xs_diff_pt_datasets}
\end{table}

\begin{table}[H]
    \centering
    \footnotesize
    \begin{tabularx}{\textwidth}{Xcccc}
        \toprule
        Dataset & N$_\text{dat}$ & $y_t$ & $Q$ [GeV] & Ref. \\
        \midrule
        ATLAS single top 7 TeV $y_t$ normalised             & 3 & $[0.100, 0.850]$ & $m_t$ & \cite{Aad:2014fwa} \\
        ATLAS single antitop 7 TeV $y_{\bar{t}}$ normalised & 3 & $[0.100, 0.850]$ & $m_t$ & \cite{Aad:2014fwa} \\
        ATLAS single top 8 TeV $y_t$ normalised             & 3 & $[0.150, 1.000]$ & $m_t$ & \cite{Aaboud:2017pdi} \\
        ATLAS single antitop 8 TeV $y_{\bar{t}}$ normalised & 3 & $[0.150, 1.000]$ & $m_t$ & \cite{Aaboud:2017pdi} \\
        CMS 13TeV single-top rapidity $y_t$                 & 4 & $[0.100, 1.050]$ & $m_t$ & \cite{Sirunyan:2019hqb} \\
        \midrule
        Total & 16 & & & \\
        \bottomrule
    \end{tabularx}
    \caption{Single top/antitop production distributions differential in the quark rapidity $y_{t / \bar{t}}$.}
    \label{tab:single_top_diff_datasets}
\end{table}

\begin{table}[H]
    \centering
    \footnotesize
    \begin{tabularx}{\textwidth}{Xcccc}
        \toprule
        Dataset & N$_\text{dat}$ & $y_{t\bar{t}}$ & $Q$ [GeV] & Ref. \\
        \midrule
        ATLAS $t\bar{t}$ 8 TeV $y_t$ normalised                 &  4 & $[0.200, 2.050]$ & $m_t$ & \cite{Aad:2015mbv}\\
        \midrule
        ATLAS $t\bar{t}$ 8 TeV $y_{t\bar{t}}$                   &  4 & $[0.150, 1.900]$  & $m_t$         & \cite{Aad:2015mbv} \\
        ATLAS $t\bar{t}$ 13 TeV $(m_{t\bar{t}},y_{t\bar{t}})$   & 10 & $[0.165, 2.025]$  & $[350, 1985]$ & \cite{ATLAS:2020ccu} \\
        CMS $t\bar{t}$ 8 TeV $y_{t\bar{t}}$                     &  9 & $[-1.820, 1.820]$ & $m_t$         & \cite{Khachatryan:2015oqa} \\
        CMS $t\bar{t}$ 13 TeV $(m_{t\bar{t}},y_{t\bar{t}})$     & 16 & $[0.175, 1.825]$  & $[370, 1075]$ & \cite{Sirunyan:2017azo} \\
        \midrule
        Total & 43 & & & \\
        \bottomrule
    \end{tabularx}
    \caption{$t\bar{t}$ production cross-section differential in the rapidity of the top quark $y_t$, and the top-pair $y_{t\bar{t}}$.}
    \label{tab:ttbar_xs_diff_y_datasets}
\end{table}

\begin{table}[H]
    \centering
    \footnotesize
    \begin{tabularx}{\textwidth}{Xcccc}
        \toprule
        Dataset & N$_\text{dat}$ & $-$ & $Q$ [GeV] & Ref.\\
        \midrule
        ATLAS $W,Z$ incl. 13 TeV $\sigma^{fid}$                                                   & 3 & $-$ & $[M_W, M_Z]$ & \cite{Aad:2016naf} \\
        ATLAS $\sigma_{tt}^{\rm tot}$ 7 TeV                                                       & 1 & $-$ & $m_t$        & \cite{Aad:2014kva} \\
        ATLAS $\sigma_{tt}^{\rm tot}$ 8 TeV                                                       & 1 & $-$ & $m_t$        & \cite{Aad:2014kva} \\
        ATLAS $\sigma_{t\bar{t}}^{\rm tot}$ (l+jets) 8 TeV                                        & 1 & $-$ & $m_t$        & \cite{ATLAS:2017wvi} \\
        ATLAS $\sigma_{t\bar{t}}^{\rm tot}$ (dilepton) 13 TeV                                     & 1 & $-$ & $m_t$        & \cite{ATLAS:2019hau} \\
        ATLAS $\sigma_{t\bar{t}}^{\rm tot}$ (hadronic) 13 TeV                                     & 1 & $-$ & $m_t$        & \cite{ATLAS:2020ccu} \\
        ATLAS $\sigma_{t\bar{t}}^{\rm tot}$ (l+jets) 13 TeV                                       & 1 & $-$ & $m_t$        & \cite{Aad:2020tmz} \\
        CMS $\sigma_{tt}^{\rm tot}$ 5 TeV                                                         & 1 & $-$ & $m_t$        & \cite{Sirunyan:2017ule} \\
        CMS $\sigma_{tt}^{\rm tot}$ 7 TeV                                                         & 1 & $-$ & $m_t$        & \cite{Spannagel:2016cqt} \\
        CMS $\sigma_{tt}^{\rm tot}$ 8 TeV                                                         & 1 & $-$ & $m_t$        & \cite{Spannagel:2016cqt} \\
        CMS $\sigma_{tt}^{\rm tot}$ 13 TeV                                                        & 1 & $-$ & $m_t$        & \cite{Khachatryan:2015uqb} \\
        CMS $\sigma_{t\bar{t}}^{\rm tot}$ (l+jets) 13 TeV                                         & 1 & $-$ & $m_t$        & \cite{CMS:2021vhb} \\
        ATLAS $\sigma_{ttZ}^{\rm tot}$ 8 TeV                                                      & 1 & $-$ & $m_t$        & \cite{Aad:2015eua} \\
        ATLAS $\sigma_{ttW}^{\rm tot}$ 8 TeV                                                      & 1 & $-$ & $m_t$        & \cite{Aad:2015eua} \\
        ATLAS $\sigma_{ttZ}^{\rm tot}$ 13 TeV                                                     & 1 & $-$ & $m_t$        & \cite{Aaboud:2019njj} \\
        ATLAS $\sigma_{ttW}^{\rm tot}$ 13 TeV                                                     & 1 & $-$ & $m_t$        & \cite{Aaboud:2019njj} \\
        CMS $\sigma_{ttZ}^{\rm tot}$ 8 TeV                                                        & 1 & $-$ & $m_t$        & \cite{Khachatryan:2015sha} \\
        CMS $\sigma_{ttW}^{\rm tot}$ 8 TeV                                                        & 1 & $-$ & $m_t$        & \cite{Khachatryan:2015sha} \\
        CMS $\sigma_{ttZ}^{\rm tot}$ 13 TeV                                                       & 1 & $-$ & $m_t$        & \cite{Sirunyan:2017uzs} \\
        CMS $\sigma_{ttW}^{\rm tot}$ 13 TeV                                                       & 1 & $-$ & $m_t$        & \cite{Sirunyan:2017uzs} \\
        ATLAS $\sigma_{t}^{\rm tot}$ 7 TeV                                                        & 1 & $-$ & $m_t$        & \cite{Aad:2014fwa} \\
        ATLAS $\sigma_{\bar t}^{\rm tot}$ 7 TeV                                                   & 1 & $-$ & $m_t$        & \cite{Aad:2014fwa} \\
        ATLAS $\sigma_{t}^{\rm tot}$ 8 TeV                                                        & 1 & $-$ & $m_t$        & \cite{Aaboud:2017pdi} \\
        ATLAS $\sigma_{\bar t}^{\rm tot}$ 8 TeV                                                   & 1 & $-$ & $m_t$        & \cite{Aaboud:2017pdi} \\
        ATLAS $\sigma_{t}^{\rm tot}$ in s-channel, 8 TeV                                          & 1 & $-$ & $m_t$        & \cite{Aad:2015upn} \\
        ATLAS $\sigma_{t}^{\rm tot}$ 13 TeV                                                       & 1 & $-$ & $m_t$        & \cite{Aaboud:2016ymp} \\
        ATLAS $\sigma_{\bar t}^{\rm tot}$ 13 TeV                                                  & 1 & $-$ & $m_t$        & \cite{Aaboud:2016ymp} \\
        ATLAS $\sigma_{t}^{\rm tot}$ in s-channel, 13 TeV                                         & 1 & $-$ & $m_t$        & \cite{ATLAS:2022wfk} \\
        CMS single top $\sigma_{t}+\sigma_{\bar{t}}$ 7 TeV                                        & 1 & $-$ & $m_t$        & \cite{Chatrchyan:2012ep} \\
        CMS $\sigma_{t}^{\rm tot}$ 8 TeV                                                          & 1 & $-$ & $m_t$        & \cite{Khachatryan:2014iya} \\
        CMS $\sigma_{\bar t}^{\rm tot}$ 8 TeV                                                     & 1 & $-$ & $m_t$        & \cite{Khachatryan:2014iya} \\
        CMS $\sigma_{t}^{\rm tot}$ in s-channel, 8 TeV                                            & 1 & $-$ & $m_t$        & \cite{Khachatryan:2016ewo} \\
        CMS $\sigma_{t}^{\rm tot}$ 13 TeV                                                         & 1 & $-$ & $m_t$        & \cite{Sirunyan:2016cdg} \\
        CMS $\sigma_{\bar t}^{\rm tot}$ 13 TeV                                                    & 1 & $-$ & $m_t$        & \cite{Sirunyan:2016cdg} \\
        ATLAS $\sigma_{tW}^{\rm tot}$ 8 TeV                                                       & 1 & $-$ & $m_t$        & \cite{Aad:2015eto} \\
        ATLAS $\sigma_{tW}^{\rm tot}$ (single lepton) 8 TeV                                       & 1 & $-$ & $m_t$        & \cite{Aad:2020zhd} \\
        ATLAS $\sigma_{tW}^{\rm tot}$ 13 TeV                                                      & 1 & $-$ & $m_t$        & \cite{Aaboud:2016lpj} \\
        CMS $\sigma_{tW}^{\rm tot}$ 8 TeV                                                         & 1 & $-$ & $m_t$        & \cite{Chatrchyan:2014tua} \\
        CMS $\sigma_{tW}^{\rm tot}$ 13 TeV                                                        & 1 & $-$ & $m_t$        & \cite{Sirunyan:2018lcp} \\
        CMS $\sigma_{tW}^{\rm tot}$ (single lepton) 13 TeV                                        & 1 & $-$ & $m_t$        & \cite{CMS:2021vqm} \\
        \midrule
        Total & 42 & & & \\
        \bottomrule
    \end{tabularx}
    \caption{Inclusive $t\bar{t}$ production cross-section datasets.}
    \label{tab:incl_ttbar_xs_datasets}
\end{table}

%% file: appendices/app4_fit_quality.tex
\section{Evaluation of fit qualities}

In this appendix, we present the quality of the different fits performed in this
study. We show their $\Delta \chi^2$ dataset-per-dataset as well as the global 
$\Delta \chi^2$ (taking into account the correlations between the datasets) 
with respect to a baseline SM-like fit, specified in each case.

In Tab.~\ref{tab:chi2_DY_fits} we display the $\Delta \chi ^2$ of the fits performed on the Drell-Yan
sector in the presence of both a $W'$ ($\hat{W} = 8 \times 10^{-5}$) and a $Z'$
($\hat{Y} = 1.5 \times 10^{-4}$) displayed in
Fig.~\ref{fig:DY_SMEFT_comparison} for the SMEFT and in Fig.~\ref{fig:DY_PDF_lumi_comparison}
for the PDFs. We use the separate fit using the "True PDF" as the 
baseline and compute with respect to it the $\Delta \chi^2$ of the conservative fit, 
the simultaneous fit and the BSM-biased fit dataset-per-dataset.

In Tab.~\ref{tab:chi2_top_fits} we show the same comparison for the fits on the $t\bar{t}$
sector in the presence of a heavy gluon with $\hat{Z} = 4 \times 10^{-4}$. The fit 
results were displayed in Fig.~\ref{fig:ttbar_smeft_fits} for the SMEFT and on 
Fig.~\ref{fig:gg_lumi_Z4_comp} for the PDF marginal distributions.

Finally, in Tab.~\ref{tab:chi2_energycut_fits} we compare the separate PDF and SMEFT fits 
performed with various energy cuts on the PDF dataset presented in Fig.~\ref{fig:energy_cuts_cont}. Here, the SMEFT fit obtained using PDFs fitted to the full set of SM data is taken as the reference, and the $\Delta \chi^2$ values of the other fits are computed relative to this baseline.

\input{tables/chi2_table_DY_long}

\input{tables/chi2_table_TOP}

\input{tables/chi2_table_energy_cuts}

%% file: tables/chi2_table_DY_long.tex
\renewcommand{\arraystretch}{1.20}
\tiny

\begin{longtable}{ l c C{1.4cm} C{1.4cm} C{1.4cm} }

\label{tab:chi2_DY_fits} \\

\toprule
Dataset & $n_{\rm dat}$ & \multicolumn{3}{c}{$\Delta \chi^2$} \\
\cmidrule(lr){3-5}
 &  & SMEFT Fit (BSM biased) & Conservative Fit & Simultaneous Fit \\
\midrule
\endfirsthead

\midrule
\multicolumn{5}{r}{\emph{Continued on next page}}
\endfoot

\endlastfoot
NMC $d/p$ & 121 & +0.912 & +0.020 & -0.025 \\
NMC $p$ & 204 & +0.178 & -0.021 & -0.025 \\
SLAC $p$ & 33 & +0.012 & -0.001 & -0.004 \\
SLAC $d$ & 34 & +0.040 & -0.004 & -0.006 \\
BCDMS $p$ & 333 & +0.075 & -0.042 & -0.085 \\
BCDMS $d$ & 248 & +0.325 & -0.011 & -0.028 \\
CHORUS $\sigma\_{CC}^{\nu}$ & 416 & +0.307 & -0.014 & -0.035 \\
CHORUS $\sigma_{CC}^{\bar{\nu}}$ & 416 & +1.041 & -0.005 & -0.025 \\
NuTeV $\sigma_{c}^{\nu}$ & 39 & +0.250 & -0.012 & -0.013 \\
NuTeV $\sigma_{c}^{\bar{\nu}}$ & 37 & +3.029 & +0.030 & +0.016 \\
HERA I+II inclusive NC $e^-p$ & 159 & +0.131 & -0.006 & -0.018 \\
HERA I+II inclusive NC $e^+p$ 460 GeV & 204 & +0.021 & -0.007 & -0.007 \\
HERA I+II inclusive NC $e^+p$ 575 GeV & 254 & +0.031 & -0.005 & -0.005 \\
HERA I+II inclusive NC $e^+p$ 820 GeV & 70 & +0.080 & -0.003 & -0.003 \\
HERA I+II inclusive NC $e^+p$ 920 GeV & 377 & +0.584 & -0.052 & -0.055 \\
HERA I+II inclusive CC $e^-p$ & 42 & +0.030 & -0.001 & -0.002 \\
HERA I+II inclusive CC $e^+p$ & 39 & +0.344 & -0.003 & -0.010 \\
HERA comb. $\sigma_{c\bar c}^{\rm red}$ & 37 & +0.239 & -0.015 & -0.013 \\
HERA comb. $\sigma_{b\bar b}^{\rm red}$ & 26 & +0.001 & -0.000 & -0.000 \\
CDF $Z$ rapidity (new) & 28 & +0.224 & -0.014 & -0.020 \\
D0 $Z$ rapidity & 28 & +0.062 & -0.003 & -0.003 \\
D0 $W\to\mu\nu$ asymmetry & 9 & +0.307 & -0.069 & -0.070 \\
ATLAS $W,Z$ inclusive 13 TeV & 3 & +0.002 & -0.009 & -0.007 \\
ATLAS $W^+$+jet 8 TeV & 15 & +0.032 & +0.000 & +0.000 \\
ATLAS $W^-$+jet 8 TeV & 15 & +0.074 & -0.000 & -0.001 \\
ATLAS direct photon production 13 TeV & 53 & +0.118 & -0.002 & -0.002 \\
ATLAS $W,Z$ 7 TeV 2011 Central selection & 46 & +0.330 & -0.002 & -0.014 \\
ATLAS $W,Z$ 7 TeV 2010 & 30 & +0.035 & +0.000 & +0.000 \\
ATLAS $W,Z$ 7 TeV 2011 Forward selection & 15 & +0.015 & -0.001 & -0.002 \\
ATLAS $Z$ $p_T$ 8 TeV $(p_T^{ll},M_{ll})$ & 44 & +0.061 & -0.000 & -0.001 \\
ATLAS $Z$ $p_T$ 8 TeV $(p_T^{ll},y_{ll})$ & 48 & +0.195 & -0.017 & -0.012 \\
CMSZDIFF12 & 28 & +0.030 & -0.003 & -0.002 \\
CMS $W$ asymmetry 840 pb & 11 & +0.232 & +0.008 & +0.004 \\
CMS $W$ asymmetry 4.7 fb & 11 & +0.048 & -0.017 & -0.013 \\
CMS $W$ rapidity 8 TeV & 22 & +0.086 & +0.025 & -0.011 \\
LHCb $Z$ 940 pb & 9 & +0.008 & -0.001 & -0.001 \\
LHCb $Z\to ee$ 2 fb & 17 & +0.026 & -0.004 & -0.004 \\
LHCb $W,Z\to\mu$ 7 TeV & 29 & +1.232 & -0.039 & -0.049 \\
LHCb $W,Z\to\mu$ 8 TeV & 30 & +1.490 & -0.057 & -0.053 \\
LHCb $Z\to\mu\mu$ & 16 & +0.004 & -0.000 & -0.001 \\
LHCb $Z\to ee$ & 15 & +0.001 & -0.000 & -0.000 \\
DYE 866 $\sigma^d_{\rm DY}/\sigma^p_{\rm DY}$ & 15 & +0.324 & -0.009 & +0.005 \\
DY E886 $\sigma^p_{\rm DY}$ & 89 & +8.849 & -0.025 & -0.064 \\
DY E605 $\sigma^p_{\rm DY}$ & 85 & +0.283 & +0.004 & -0.006 \\
DYE 906 $\sigma^d_{\rm DY}/\sigma^p_{\rm DY}$ & 6 & +0.417 & -0.006 & -0.029 \\
ATLAS DY 2D 8 TeV low mass & 60 & +0.058 & -0.006 & -0.014 \\
HL-LHC Forward Drell-Yan  14 TeV - $W^-$ -  & 8 & +0.376 & +0.008 & -0.013 \\
HL-LHC Forward Drell-Yan  14 TeV - $W^+$ & 8 & +0.522 & +0.002 & -0.015 \\
ATLAS DY 2D 8 TeV high mass & 48 & +0.947 & -0.009 & -0.031 \\
ATLAS low-mass DY 2011 & 6 & +0.001 & +0.000 & -0.000 \\
CMS DY 2D 7 TeV  & 110 & +0.089 & +0.014 & -0.001 \\
CMS DY 1D 8 TeV & 41 & +0.566 & +0.014 & -0.012 \\
ATLAS HM DY 7 TeV & 5 & +0.020 & +0.002 & -0.000 \\
CMS HM DY 13 TeV - combined channel & 43 & +0.057 & +0.003 & -0.002 \\
HL-LHC HM DY 14 TeV - neutral current - electron channel & 12 & +1.852 & +0.104 & +0.035 \\
HL-LHC HM DY 14 TeV - neutral current - muon channel & 12 & +1.856 & +0.104 & +0.035 \\
HL-LHC HM DY 14 TeV - charged current - electron channel & 16 & +3.104 & +0.083 & +0.036 \\
\midrule
GLOBAL & 4363 & +31.964 & -0.117 & -0.728 \\
\midrule
\caption{\small
The total dataset $\Delta \chi^2$ for datasets entering the SMEFT and PDF fits, presented respectively in
Fig.~\ref{fig:DY_SMEFT_comparison} and in
Fig.~\ref{fig:DY_PDF_lumi_comparison} for the DY sector analysis.
For each dataset, we indicate the number of data points $n_{\rm dat}$ and the corresponding
$\Delta \chi^2$ with respect to the fit using the true PDF, chosen as a reference.
}
\end{longtable}

%% file: tables/chi2_table_TOP.tex
\renewcommand{\arraystretch}{1.20}

\begin{longtable}{ l c C{1.4cm} C{1.4cm} C{1.4cm} }

\label{tab:chi2_top_fits} \\

\toprule
Dataset & $n_{\rm dat}$ & \multicolumn{3}{c}{$\Delta \chi^2$} \\
\cmidrule(lr){3-5}
 &  & SMEFT Fit (BSM biased) & Conservative Fit & Simultaneous Fit \\
\midrule
\endfirsthead

\toprule
Dataset & $n_{\rm dat}$ & \multicolumn{3}{c}{$\Delta \chi^2$} \\
\cmidrule(lr){3-5}
 &  & SMEFT Fit (BSM biased) & Conservative Fit & Simultaneous Fit \\
\midrule
\endhead

\midrule
\multicolumn{5}{r}{\emph{Continued on next page}}
\endfoot

\endlastfoot

\midrule
ATLAS $\sigma_{tt}^{\rm tot}$ 7 TeV & 1 & +0.021 & +0.000 & +0.000  \\
ATLAS $\sigma_{tt}^{\rm tot}$ 8 TeV & 1 & +0.012 & +0.000 & +0.000 \\
ATLAS $t\bar{t}$ 8 TeV $y_t$ normalised & 4 & +0.447 & +0.017 & -0.006 \\
ATLAS $t\bar{t}$ 8 TeV $m_{t\bar{t}}$ normalised & 5 & +0.015 & +0.000 & +0.000 \\
ATLAS $\sigma_{t\bar{t}}^{\rm tot}$ (l+jets) 8 TeV & 1 & +0.006 & +0.000 & +0.000 \\
ATLAS $t\bar{t}$ 13 TeV $m_{t\bar{t}}$ normalised & 8 & +0.018 & +0.000 & +0.000 \\
ATLAS $t\bar{t}$ 13 TeV $(m_{t\bar{t}},y_{t\bar{t}})$ & 10 & +0.538 & +0.009 & -0.001 \\
ATLAS $\sigma_{t\bar{t}}^{\rm tot}$ (l+jets) 13 TeV & 1 & +0.000 & +0.000 & +0.000 \\
ATLAS $\sigma_{t\bar{t}}^{\rm tot}$ (dilepton) 13 TeV & 1 & -0.000 & +0.000 & +0.000 \\
ATLAS $\sigma_{t\bar{t}}^{\rm tot}$ (hadronic) 13 TeV & 1 & +0.000 & +0.000 & +0.000 \\
CMS $\sigma_{tt}^{\rm tot}$ 5 TeV & 1 & +0.007 & +0.000 & +0.000 \\
CMS $\sigma_{tt}^{\rm tot}$ 7 TeV & 1 & +0.020 & +0.000 & +0.000 \\
CMS $\sigma_{tt}^{\rm tot}$ 8 TeV & 1 & +0.016 & +0.000 & +0.000 \\
CMS $\sigma_{tt}^{\rm tot}$ 13 TeV & 1 & +0.000 & +0.000 & +0.000 \\
CMS $t\bar{t}$ 8 TeV $y_{t\bar{t}}$ & 9 & +1.142 & +0.017 & -0.005  \\
CMS $t\bar{t}$ 13 TeV $(m_{t\bar{t}},y_{t\bar{t}})$ & 16 & +0.631 & +0.010 & -0.002  \\
CMS $t\bar{t}$ 8 TeV $m_{t\bar{t}}$ normalised & 5 & +0.015 & -0.000 & -0.000  \\
CMS $t\bar{t}$ 13 TeV $m_{t\bar{t}}$ normalised & 14 & +0.150 & +0.000 & -0.000 \\
CMS $\sigma_{t\bar{t}}^{\rm tot}$ (l+jets) 13 TeV & 1 & +0.000 & +0.000 & +0.000 \\
HL-LHC high-mass $t\bar t$ production & 18 & +1.080 & +0.009 & -0.004 \\
\midrule
GLOBAL & 4363 & +5.536 & +0.107 & +0.034  \\
\midrule
\caption{\small
The total dataset $\chi^2$ for datasets entering the SMEFT and PDF fits in the top sector analysis, presented respectively in Fig.~\ref{fig:ttbar_smeft_fits}
and in Fig.~\ref{fig:gg_lumi_Z4_comp}.
For each dataset, we indicate the number of data points $n_{\rm dat}$ and the corresponding
$\Delta \chi^2$ with respect to the fit using the true PDF, chosen as a reference.
}
\end{longtable}

%% file: tables/chi2_table_energy_cuts.tex
\renewcommand{\arraystretch}{1.20}

\begin{longtable}{ l c C{1.4cm} C{1.4cm} C{1.4cm} C{1.4cm}  }

\label{tab:chi2_energycut_fits} \\

\toprule
Dataset & $n_{\rm dat}$ & \multicolumn{4}{c}{$\Delta \chi^2$} \\
\cmidrule(lr){3-6}
 &  & Full Data Contaminated & 1500GeV Cut Contaminated & 1000GeV Cut Contaminated & 500GeV Cut Contaminated   \\
\midrule
\endfirsthead

\toprule
Dataset & $n_{\rm dat}$ & \multicolumn{4}{c}{$\Delta \chi^2$} \\
\cmidrule(lr){3-6}
 &  &Full Data Contaminated & 1500GeV Cut Contaminated & 1000GeV Cut Contaminated & 500GeV Cut Contaminated \\
\midrule
\endhead

\midrule
\multicolumn{5}{r}{\emph{Continued on next page}}
\endfoot

\endlastfoot

ATLAS $W,Z$ inclusive 13 TeV & 3 & -0.003 & -0.010 & -0.007 & -0.013 \\
ATLAS $W^+$+jet 8 TeV & 15 & +0.034 & +0.023 & +0.013 & -0.000 \\
ATLAS $W^-$+jet 8 TeV & 15 & +0.051 & +0.018 & +0.006 & +0.000 \\
ATLAS direct photon production 13 TeV & 53 & +0.035 & +0.021 & +0.007 & -0.002 \\
ATLAS $W,Z$ 7 TeV 2011 Central selection & 46 & +0.216 & +0.030 & -0.021 & -0.047 \\
ATLAS $W,Z$ 7 TeV 2010 & 30 & +0.023 & +0.012 & +0.003 & -0.004 \\
ATLAS $W,Z$ 7 TeV 2011 Forward selection & 15 & +0.011 & -0.003 & -0.004 & -0.002 \\
ATLAS $Z$ $p_T$ 8 TeV $(p_T^{ll},M_{ll})$ & 44 & +0.018 & +0.013 & +0.006 & -0.001 \\
ATLAS $Z$ $p_T$ 8 TeV $(p_T^{ll},y_{ll})$ & 48 & +0.206 & +0.047 & +0.014 & -0.023 \\
CMS $Z$ $p_T$ 8 TeV $(p_T^{ll},y_{ll})$ & 28 & +0.028 & +0.018 & +0.007 & -0.004 \\
CMS $W$ asymmetry 840 pb & 11 & +0.142 & +0.229 & +0.187 & +0.002 \\
CMS $W$ asymmetry 4.7 fb & 11 & -0.027 & -0.008 & +0.002 & -0.024 \\
CMS $W$ rapidity 8 TeV & 22 & +0.160 & +0.081 & +0.143 & -0.023 \\
LHCb $Z$ 940 pb & 9 & +0.007 & -0.001 & -0.001 & -0.002 \\
LHCb $Z\to ee$ 2 fb & 17 & +0.021 & -0.000 & -0.002 & -0.006 \\
LHCb $W,Z\to\mu$ 7 TeV & 29 & +0.863 & +0.235 & +0.142 & -0.066 \\
LHCb $W,Z\to\mu$ 8 TeV & 30 & +1.003 & +0.291 & +0.160 & -0.085 \\
LHCb $Z\to\mu\mu$ & 16 & +0.005 & +0.002 & -0.000 & -0.002 \\
LHCb $Z\to ee$ & 15 & +0.003 & +0.001 & +0.000 & -0.001 \\
DYE 866 $\sigma^d_{\rm DY}/\sigma^p_{\rm DY}$ & 15 & +1.183 & +0.574 & +0.332 & -0.011 \\
DY E886 $\sigma^p_{\rm DY}$ & 89 & +3.511 & +0.855 & +0.224 & -0.036 \\
DY E605 $\sigma^p_{\rm DY}$ & 85 & +0.147 & +0.072 & +0.042 & -0.002 \\
DYE 906 $\sigma^d_{\rm DY}/\sigma^p_{\rm DY}$ & 6 & +0.490 & +0.060 & +0.032 & -0.019 \\
ATLAS DY 2D 8 TeV low mass & 60 & +0.039 & -0.012 & -0.013 & -0.012 \\
HL-LHC Forward Drell-Yan 14 TeV (CC) $\mu+$  & 8 & +0.265 & +0.070 & +0.049 & -0.034 \\
HL-LHC Forward Drell-Yan 14 TeV (CC) $\mu$ & 8 & +0.419 & +0.019 & -0.035 & -0.008 \\
HL-LHC Forward Drell-Yan 14 TeV (NC) & 18 & +0.161 & +0.013 & -0.011 & -0.036 \\
ATLAS DY 2D 8 TeV high mass & 48 & +0.362 & +0.331 & +0.197 & -0.044 \\
ATLAS low-mass DY 2011 & 6 & -0.000 & -0.001 & -0.001 & +0.000 \\
CMS Drell-Yan 2D 7 TeV 2011 & 110 & +0.031 & +0.019 & +0.015 & +0.007 \\
CMS DY 1D 8 TeV & 41 & +0.359 & +0.220 & +0.155 & -0.014 \\
ATLAS high-mass DY 7 TeV $d\sigma_{Z/\gamma^{*}}/dM_{ll}$ & 5 & +0.007 & +0.007 & +0.006 & +0.000 \\
CMS HM DY 13 TeV - combined channel & 43 & +0.053 & +0.025 & +0.019 & -0.003 \\
HL-LHC High-mass Drell--Yan (NC) (electron) & 12 & +0.258 & +0.826 & +0.680 & +0.056 \\
HL-LHC High-mass Drell--Yan (NC) (muon) & 12 & +0.252 & +0.812 & +0.668 & +0.056 \\
HL-LHC High-mass Drell--Yan (CC) (electron) & 16 & +3.158 & +6.101 & +4.301 & +0.012 \\
\midrule
GLOBAL & 4363 & +20.857 & +13.241 & +8.215 & -1.154 \\
\midrule
\caption{\small
The total dataset $\chi^2$ for datasets entering the SMEFT and PDF fits for the energy cut-off analysis, presented in Fig.~\ref{fig:energy_cuts_cont}. 
Only the DY sector has been presented.
For each dataset, we indicate the number of data points $n_{\rm dat}$ and the corresponding
$\Delta \chi^2$ with respect to the fit using the true PDF, chosen as a reference.
}
\end{longtable}